\documentclass[11pt]{article}
\usepackage{jheppub}

\usepackage{amssymb}
\usepackage{url}
\usepackage{calc}
\usepackage{graphicx}
\usepackage{amsfonts}
\usepackage{amsbsy}
\usepackage[all,2cell,cmtip]{xy}
\UseTwocells
\usepackage{amsmath}
\usepackage{dsfont}
\usepackage{bbm}
\usepackage{fixmath}
\usepackage{upgreek}
\usepackage{amssymb,amscd}
\usepackage{mathrsfs}
\usepackage{amsmath,amsthm}
\usepackage{MnSymbol}
\usepackage{slashed}
\usepackage{dsfont}
\usepackage{tikz}
\usepackage{physics}

\usetikzlibrary{positioning}
\usetikzlibrary{arrows.meta}
\usetikzlibrary{shapes.geometric, arrows}

\tikzstyle{hex} = [regular polygon,regular polygon sides=6, draw,
    text width=2em, text centered]
\tikzstyle{hexs}= [regular polygon,regular polygon sides=5, draw,
    text width=1.7em, text centered]
\tikzstyle{line} = [draw, -latex']
\newcommand{\eq}[1]{\begin{align}\begin{split}#1\end{split}\end{align}}
\newcommand{\twopi}[1]{{\frac{#1}{2\pi}}}

\def\be{ \begin{equation} }
\def\ee{ \end{equation}}

\def\exp{{\rm exp}}

\renewcommand{\Re}{{\rm Re }}

\def\Tr{{\rm Tr}}

\def\half{\frac{1}{2}}

\def\tG{{\widetilde{G}}}
\def\tJ{{\tilde{J}}}
\def\tB{{\tilde{B}}}

\def\one{{\hbox{ 1\kern-.8mm l}}}

\def\vx{{\vec{x}}}

\def\CA{{\cal A}}

\def\CC {{\cal C}}
\def\CD {{\cal D}}

\def\CF {{\cal F}}

\def\CH {{\cal H}}
\def\CI {{\cal I}}

\def\CL {{\cal L}}

\def\CN {{\cal N}}
\def\CO {{\cal O}}
\def\CP {{\cal P}}

\def\CW {{\cal W}}

\def\CO {{\cal O}}

\def\CH {{\cal H}}
\def\CI {{{\cal I}}}

\def\CT {{\cal T}}
\def\CU {{\cal U}}

\def\IG{\mathbb{G}}

\def\IQ{\mathbb{Q}}
\def\IR{{\mathbb{R}}}
\def\IS{{\mathbb{S}}}

\def\IZ{{\mathbb{Z}}}
\def\IZN{{\mathbb{Z}_N}}
\def\hatG{{\widehat{G}}}
\def\hatF{{\widehat{F}}}
\def\tildeG{{\tilde{G}}}

\def\fg{\mathfrak{g}}

\def\ft{\mathfrak{t}}

\def\ft{\mathfrak{t}}

\def\rmk#1{\bigskip\noindent{\bf Remark} }
\def\aside#1{\bigskip\noindent{\bf Aside} }

\def\beq{\begin{equation}}
\def\eeq{\end{equation}}
\def\bea{\begin{eqnarray}}
\def\eea{\end{eqnarray}}

\title{Introduction to Generalized Global Symmetries in QFT 
and Particle Physics}
\author[a]{T. Daniel Brennan} 
\author[b]{, Sungwoo Hong} 
\affiliation[a]{Department of Physics, University of California, San Diego, CA, 92093, USA}
\affiliation[b]{Department of Physics, KAIST, Daejeon, 34141, Korea}

\emailAdd{tbrennan@ucsd.edu, sungwooh@kaist.ac.kr}

\abstract{
Generalized symmetries (also known as categorical symmetries) is a newly developing technique for studying quantum field theories. It has given us new insights into the structure of QFT and many new powerful tools that can be applied to the study of particle phenomenology. 
In these notes we give an exposition to the topic of generalized/categorical symmetries for  high energy phenomenologists although the topics covered may be useful to the broader physics community. Here we describe generalized symmetries without the use of category theory and pay particular attention to the introduction of discrete symmetries and their gauging. 
}

\preprint{}
	
\begin{document}

\maketitle

\section{Introduction}

Symmetry has been one of the guiding tenets of theoretical physics for the past century. Recently, many different types of symmetries have been organized into the common language of higher form global symmetries \cite{Gaiotto:2014kfa,Aharony:2013hda}. This consolidation has made it easier to analyze the symmetries of general quantum field theories and has led to  new results including conjectures on the IR phase structure of interacting gauge theories \cite{Gaiotto:2017yup,Komargodski:2017keh,Gaiotto:2017tne,Gomis:2017ixy,Cordova:2017kue,Cordova:2018acb,Cordova:2019bsd}. 

Even more recently, it has been realized that higher form global symmetries fit into the more general structure of categorical global symmetries \cite{Baez:2004in,Baez:2010ya,Gukov:2013zka,Kapustin:2013uxa,Cordova:2018cvg,Benini:2018reh,Brennan:2020ehu,Bhardwaj:2023wzd,Bartsch:2022mpm,Bartsch:2022ytj,Bartsch:2023pzl,Bhardwaj:2022yxj,Bhardwaj:2022lsg,Bhardwaj:2023ayw,Bartsch:2023wvv,Tachikawa:2017gyf,Choi:2022jqy,Choi:2022zal,Choi:2022fgx,Choi:2022rfe,Cordova:2022ieu,Bhardwaj:2022kot,Bhardwaj:2022maz} which is described a more general framework than groups: category theory. Of particular note in regards to the application to particle phenomenology are the special case of higher group and non-invertible global symmetries. 
Higher group global symmetries describe the case when higher form global symmetries of different degrees mix together in a way that is reminiscent of group extensions 
while non-invertible symmetries are purely categorical, although they are sometimes associated to certain classical group-like symmetries. The new framework of categorical symmetries  provides more tools to study quantum field theories and in particular can be used to provide new constraints on RG flows. 

In this review, we provide a pedagogical introduction to some aspects of categorical symmetry that are relevant for the study of particle phenomenology. This has a strong focus on higher form, higher group, and Lagrangian non-invertible global symmetries. It is our intent to make this review accessible to both formal and phenomenology high energy theorists. 
 Here we assume a graduate student level understanding of quantum field theory as well as some understanding of group theory, differential forms, cohomology, and basic bundle theory. We provide some basic review of some of these concepts in Appendix \ref{app:differentialforms} but for a more complete review, we refer the reader to \cite{Nakahara:2003nw}. 

The order of topics are as follows. 
In Section \ref{sec:continuous higher-form} we discuss continuous higher form global symmetries and their spontaneous breaking. Here we focus on using the current algebra to explicitly demonstrate the action of symmetry defect operators on charged operators. In Section \ref{sec:discrete} we discuss discrete global symmetries. There we show how to use discrete background gauge fields with a particular focus on the application to center symmetry in non-abelian gauge theory. Throughout our discussion of symmetries, we will additionally discuss  `t Hooft anomalies of higher form global symmetries. In addition to spelling out a general procedure for computing anomalies, we examine some important examples such as the mixed anomaly between time reversal symmetry and center symmetry in non-abelian Yang-Mills theory. In Section \ref{sec:highergroup} we then give an introduction to higher group global symmetries with a focus on 2-group and 3-group global symmetry in $4d$ and examine their implications on RG flows in phenomenological models. We then conclude with Section \ref{sec:NIS} in which we give an introduction to Lagrangian non-invertible symmetries and their application to chiral symmetries and axions. 

We would like to emphasize that all results contained in this review are found in the literature. Since these notes are generally constructive, we will only include references in the main text when referring the reader for further material. 

We would additionally like to point out that in this review we will not discuss any category theory explicitly. Along with this, we will not discuss any other newer ideas such as the SymmTFT \cite{Freed:2022qnc,Freed:2022iao,Kaidi:2022cpf,Apruzzi:2021nmk} or anomalies of more general categorical symmetries \cite{Kaidi:2023maf,Zhang:2023wlu,Choi:2023xjw}. See \cite{Schafer-Nameki:2023jdn} for complementary review using this approach.

\bigskip
\noindent \emph{Note to  the reader:}~
We hope that these lecture notes are helpful to those who are interested in learning the subject. Please feel free to reach out with any comments or suggestions for improvement. 

\section{Continuous Generalized Global Symmetries}
\label{sec:continuous higher-form}

In this section, we introduce higher-form global symmetries. 
Our main focus will be continuous higher-form symmetries for which conserved currents exit.
The discrete analog will be described in Section~\ref{sec:discrete}. 
We will begin by describing ``ordinary symmetries'' (i.e. 0-form symmetries) in the language that can be easily generalized to higher-form symmetries.

\subsection{Ordinary symmetry}
\label{subsed:0-form}

Let us first recall the case of ``ordinary'' continuous global symmetries  in $d$ space time dimensions. We will also assume that our QFT has a Lagrangian although it is not essential in our discussion. 

Consider a QFT with an action $S [\phi]$ ($\phi$ here denotes collectively all  dynamical fields) which is invariant under an infinitesimal transformation $\phi (x) \mapsto \phi' (x) \approx \phi (x) + \alpha\, \delta \phi (x)$. Under a position-dependent transformation with $\alpha (x)$, the action shifts as
\beq
\delta S = \int \partial_\mu \alpha(x)\, j^\mu (x)\,d^dx~,
\eeq
implying that the Noether current is conserved
\eq{\label{eq:0form_current_conservation}
\partial_\mu j^\mu(x)=0~.
}
Since these ``ordinary'' symmetries act on 0-dimensional objects (i.e.~local operators which excite particles when operated on the vacuum), in the language of higher-form symmetry, they are called 0-form symmetries.

We say that a quantum field theory has such a continuous symmetry with group $G$ if there is a $G$ action on some collection of local operators and we have an associated conserved current $j^\mu$. For future reference, we note that this current can be dualized to a 1-form current $j_1 = j_\mu dx^\mu$. In the notation of differential forms, the conservation of the current is expressed as $j_1$ being  (co-)closed:\footnote{We will frequently use differential form notations in this note. The \emph{conservation} equation (\ref{eq:0form_current_conservation}) is written as $d \ast  j _1= 0$ while $d j_1 = 0$ means the 1-form current is closed. Sometimes, we will say that $j_1$ is co-closed when $d \ast  j _1= 0$ because the co-differential (adjoint of the exterior derivative $d$) acting on $p$-form in $d$-dimensional space time dimension is $\delta =(-1)^{pd+d+1}\ast d\ast $. ``$\ast $'' is the Hodge dual operation. See Appendix~\ref{app:differentialforms} for a short review on differential forms. }
\eq{
\partial_\mu j^\mu(x)=0\quad\Longrightarrow \quad d\ast j_1=0~. 
}

In general, one  can couple the conserved current to a background gauge field by adding 
\eq{
S=...+i \int A_\mu(x) \,j^\mu(x)d^dx~ = \dots + i \int A_1 \wedge \ast  j_1,
}
to the action. This makes it clear that 0-form symmetry current couples to a 1-form background gauge field. Due to the conservation of the current, the action is invariant under gauge transformations of the background gauge field:
\eq{
\delta S_\lambda &=i\int \partial_\mu \lambda(x) j^\mu (x) d^d x=-i\int \lambda(x) \partial_\mu j^\mu(x)\, d^dx=0~,\\
\delta S_\lambda & =i \int d\lambda(x) \wedge \ast j_1=-i \int \lambda(x)\, d\ast j_1=0~,\\
\delta_\lambda A_\mu&=[\lambda(x),A_\mu]+\partial_\mu \lambda(x)\quad, \quad \delta_\lambda A_1=d\lambda(x)~.
}
Introducing a background gauge field turns out to be a very powerful method for studying symmetries because asking whether or not $j^\mu(x)$ is conserved is translated  into asking whether or not the path integral is invariant under (background) gauge transformations of $A_\mu$.  
In addition to being a potent tool for studying anomalies, this technique also extends very simply to  general symmetries. Utilizing this method to study symmetries is one of the central themes of these lectures. 

Now let us pick a local operator  $\CO_R(x)$ that transforms under $G$ with representation $R$:
\eq{
g\cdot \CO_R(x)=R(g)^i_{~j}\CO_R^j(x)~.
} 
From the above variation of the background gauge field we can derive the Ward identity:\footnote{
In particular, the fact that $G$ is a symmetry implies that 
\eq{
\delta_\lambda\big\langle \CO_R(x)\big\rangle=0\quad, \quad \big\langle \delta_{\lambda} \CO_R(x)-i\lambda(y) \partial_\mu j^\mu(y) \CO_R(x)\big\rangle=0~.
}
Then from the variation of $\CO_R(x)$:
\eq{
\delta_{\lambda}\CO_R(x)=i\delta(x-y)\,R(\lambda(y)) \CO_R(x))~,
}
we arrive at equation \eqref{0formward}. 
}
\be\label{0formward}
\partial_\mu j^\mu(x) \,\CO_R(y)=\delta^{(d)}(x-y)\,\, R(T^a)\, \CO_R(y)~,
\ee
which is simply the quantum (operator) version of the conservation equation in the classical field theory.

Since the symmetry $G$ has an associated current, we can integrate it over all of space ($X_{d-1}$) to define the associated conserved charge:
\eq{
Q(X_{d-1})=\int_{X_{d-1}} j^\mu(x)\,\hat{n}_\mu\, d^{d-1}x~,
}
where $\hat{n}$ is a unit normal vector to $X_{d-1}$. Alternatively, we can define the charge operator in terms of differential forms as
\eq{
Q(X_{d-1})=\int_{X_{d-1}}\ast j_1~. 
}

In a Hamiltonian picture, where we quantize on $X_{d-1}$, we can then define the $G$-action on the associated Hilbert space $\CH_{X_{d-1}}$ by the unitary operator
\eq{\label{Hilb0form}
U_g=e^{i \lambda\, Q(X_{d-1})}\quad, \quad g=e^{i \lambda}~. 
}
Since states in the Hilbert space can be prepared by acting on the vacuum with local operators $\CO(x)$, the above action of $U_g$ on $\CH_{X_{d-1}}$ translates into an action of the unitary operator $U_g$ on $\CO(x)$. In a Euclidean theory, this action can be realized by a more general operator:\footnote{Here it is sufficient to think of symmetries in terms of Euclidean field theories because states in the Lorentzian field theory can be prepared by the Euclidean path integral. }
\eq{
U_g(\Sigma_{d-1})=\exp\left\{i \lambda \int_{\Sigma} j^\mu(x) \hat{n}_\mu d^{d-1}x\right\} = \exp\left\{i \lambda \int_{\Sigma} \ast j_1\right\}~,
}
where here $\Sigma_{d-1}$ is any $(d-1)$-dimensional manifold without boundary\footnote{Here we allow   symmetry defect operators whose boundary coincides with the boundary of $X_{d-1}$.} and $\hat{n}^\mu$ is its normal vector. This operator is called a \emph{symmetry defect operator} (SDO).  

In order for $U_g(\Sigma_{d-1})$ to enact the symmetry action, we require that the symmetry defect operators satisfy a group multiplication law. Indeed, one can check explicitly, using the fact that the currents satisfy the Operator Product Expansion (OPE) 
\eq{
\partial_\mu j_a^\mu(x) j^\nu_b(y)=f_{ab}^{~c}\ j_c^\nu(x)\, \delta^{(d)}(x-y)~,
}
which follows from the Ward identity. Here $f_{ab}^{~c}$ are the $G$-structure constants so 
that the 0-form symmetry defect operators satisfy the group multiplication law
\eq{
U_{g_1}(\Sigma)\cdot U_{g_2}(\Sigma)=U_{g_1g_2}(\Sigma)~.
} 
\begin{figure}
\center
\includegraphics[scale=0.8,trim=6cm 15cm 6cm 5.5cm,clip]{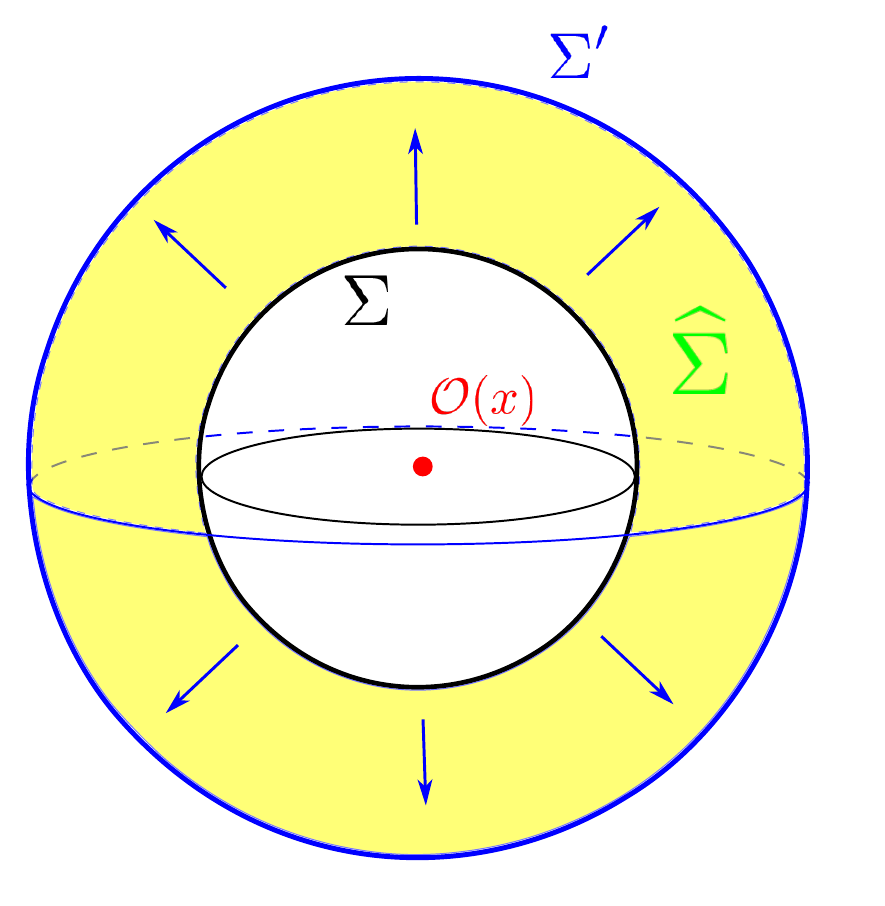}
\caption{This figure shows the deformation of a symmetry defect operator $U_g$ wrapped on a surface $\Sigma$ to a homotopically equivalent $\Sigma'$. Due to the conservation of the associated current, the symmetry operators are equivalent $U_g(\Sigma)\cong U_g(\Sigma')$. }
\label{fig:0formtop}
\end{figure}
Additionally, note that the conservation of $j^\mu(x)$ (or equivalently closed-ness of the dual current $\ast j_1$) implies that the $U_g(\Sigma)$ are topological. To be precise, the conservation of $j^\mu$ implies that $U_g(\Sigma)$ only depends on the choice of $\Sigma_{d-1}$ up to homotopy. To show this, let us pick a $\Sigma_{d-1}$ and a homotopic $\Sigma_{d-1}'$ (i.e.~$\Sigma_{d-1}'$ can be obtained as a smooth deformation of $\Sigma_{d-1}$).  
We can now compute
\eq{
U_g(\Sigma_{d-1})\cdot U_{g^{-1}}(\Sigma'_{d-1}) &=\exp\left\{i \lambda \left(\int_{\Sigma}j^\mu(x)\hat{n}_\mu d^{d-1}x-\int_{\Sigma'}j^\mu(x)\hat{n}_\mu d^{d-1}x\right)\right\} \\
&=\exp \left\{i \lambda \int_{\widehat{\Sigma}} \partial_\mu j^\mu(x)d^{d}x\right\}~,
}
where $\widehat\Sigma$ is the 4-volume bounded by $\Sigma_{d-1},\Sigma'_{d-1}$ : $\partial \widehat\Sigma=\Sigma_{d-1}\cup \Sigma_{d-1}^{\prime\vee}$ (where $\Sigma^\vee$ is $\Sigma$ with opposite orientation). 
Now using the fact that
\eq{
\partial_\mu j^\mu (x)=0~,
}
in the presence of no charged operators, we see that 
\eq{
U_g(\Sigma_{d-1})\cdot U_{g^{-1}}(\Sigma'_{d-1})=\mathds{1}~,
} 
when $\Sigma_{d-1}$ is smoothly deformable (homotopic) to $\Sigma_{d-1}^\prime$ 
and hence $U_{g^{-1}}(\Sigma'_{d-1})\cong U_{g^{-1}}(\Sigma_{d-1})$ which implies topological invariance of the $U_g(\Sigma_{d-1})$. 
See Figure \ref{fig:0formtop}.  

\begin{figure}
\begin{center}
\includegraphics[scale=0.9,trim=1.5cm 16.5cm 5cm 6cm,clip]{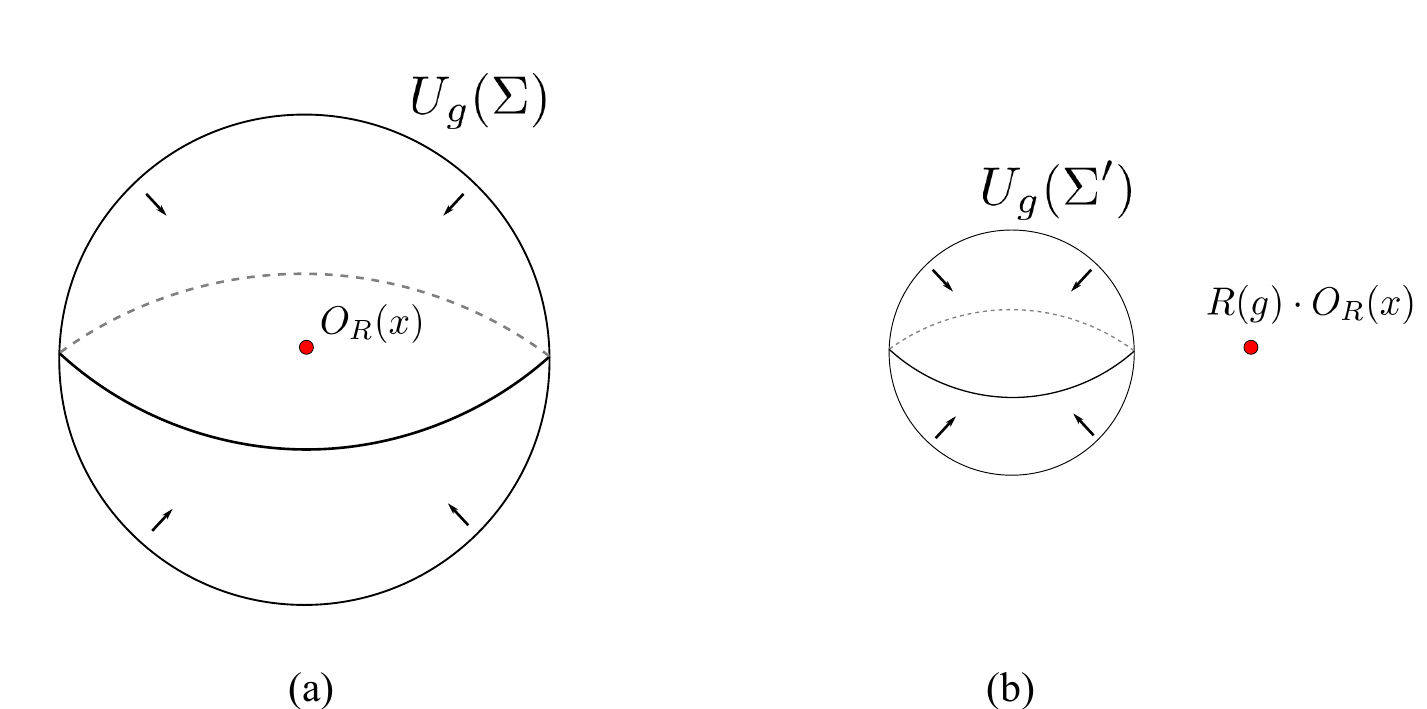}
\end{center}
\caption{This figure illustrates how symmetry defect operators $U_g$ can act on a local operator $O(x)$ in $3d$. (a) is an initial configuration in which $U_g(\Sigma)$ wraps an $S^2$ that links $O_R(x)$.  (b) shows the action of $U_g$ on $O_R(x)$ by contracting the $S^2$ through the point $x$ to $U_g(\Sigma')$.}
\label{fig:0form}
\end{figure}

It is now straightforward to show that $U_g(\Sigma_{d-1})$ acts on charged operators as 
\eq{
U_g(\Sigma_{d-1}) \CO_R(x)=R(g) \cdot\CO_R(x)U_g(\Sigma'_{d-1})~,
}
where $\Sigma_{d-1}$ is smoothly deformable to $\Sigma'_{d-1}$ by passing through the point  $x$. See Figure \ref{fig:0form}. 

Physically, the operator $U_g(\Sigma)$ implements the group action on the charged operator $\CO_R(x)$ when a smooth deformation of $\Sigma$ intersects $x$. This follows from the Ward identity  \eqref{0formward}. We can compute this action of $U_g(\Sigma_{d-1})$ explicitly. By expanding the exponential, we find
\eq{
U_g(\Sigma_{d-1})\cdot &\CO_R(x)\cdot U_{g^{-1}}(\Sigma_{d-1}')=e^{i \lambda^a \int_{\widehat\Sigma}\partial_\mu j_a^\mu(y)d^dy }\CO_R(x)\\
&=\sum_{n=0}^\infty \frac{(i \lambda^a)^n}{n!}\left(\int_{\widehat\Sigma} \partial_\mu j_a^\mu(y)d^dy\right)^n \CO_R(x)\\
&=\sum_{n=0}^\infty \frac{(i \lambda^aR(T^a))^n}{n!}\left(\int_{\widehat\Sigma} \delta^{(d)}(x-y)d^dy\right)^n \CO_R(x)\\
&=\sum_{n=0}\frac{(i \lambda^a R(T^a))^n}{n!}\CO_R(x)=R(g) \cdot \CO_R(x)~,
}
where $\widehat\Sigma$ is the 4-manifold bounded by $\Sigma_{d-1}$ and $\Sigma_{d-1}'$ and $g=e^{i \lambda^a T^a}$. 
In summary, these symmetry defect operators enact symmetry transformations on charged operators.

We would like to emphasize the relationship between symmetry defect operators and an associated background gauge field. The connection between the two viewpoints is that symmetry defect operators implement a background gauge transformation along its world volume:
\eq{
U_{g_0}(\Sigma_{d-1})\quad\Longleftrightarrow \quad g(x)=\exp\left\{i \lambda_0 \,\Theta(x;\Sigma)\right\}~,
}
where $g_0=e^{i \lambda_0}$ is a group element and $\Theta(x;\Sigma)$ is the Heaviside step function that is 0 to one side of $\Sigma_{d-1}$ and 1 to the other.\footnote{To be precise, we take $\Sigma_{d-1}$ to be orientable and we pick a global vector field $\hat{n}$ on $\Sigma_{d-1}$ defining an orientation of the normal bundle of $\Sigma_{d-1}\subset M_d$ in Euclidean space time. Then we say that $\Theta(x;\Sigma)=0$ if there exists a path $\gamma:[0,1]\to M_d$ with end points $\gamma(0)=x$ and $\gamma(1)\in \Sigma_{d-1}$ whose tangent vector $\hat{v}_1=\lim_{t\to 1}\gamma'(t)$ has negative inner product with $\hat{n}$ at $\gamma(1)$. Similarly, $\Theta(x;\Sigma)=1$ if there exists a similar path such that $\hat{v}_1$ has positive inner product with $\hat{n}$.} This leads to a background gauge field  
\eq{\label{backgroundgaugeSDO}
g(x)=\exp\left\{i \lambda_0 \,\Theta(x;\Sigma)\right\}\quad\Rightarrow\quad A_\mu=\lambda_0 \,\delta(x\in \Sigma_{d-1})\hat{n}_\mu~,
}
where $\hat{n}_\mu$ is the normal vector-field to $\Sigma_{d-1}$. Due to the fact that the SDO gives rise to a gauge field that goes like a $\delta$-function, we can think of any smooth gauge field as a collection of ``smoothed out'' symmetry defect operators. 

We can also use this viewpoint to have a understand the action of the symmetry defect operators on local operators. For simplicity let us consider the case of a $U(1)$ symmetry and pick a background gauge field generated by the symmetry defect operator $U_{g_0}(\Sigma_{d-1})$ as above. 
 Insert a charged operator $\CO_q(x)$ at a point $x\notin \Sigma_{d-1}$. Now, let us consider moving $\CO_q$ along a path $\gamma:x\longmapsto y$ that passes through $\Sigma_{d-1}$. With regard to the $U(1)$ symmetry, charged operator $\CO_q$ acquires a phase 
\eq{
\CO_q(x)\longmapsto e^{iq \int_\gamma A_\mu dx^\mu }\CO_q(y)~.
}
Since $\gamma$ passes through $\Sigma_{d-1}$ the integral 
\eq{
\int_\gamma A_\mu(x)\,d x^\mu=\int_{\gamma}\lambda_0\delta(x\in \Sigma_{d-1})\hat{n}_\mu dx^\mu=\lambda_0~.
} 
Therefore, we see that $\CO_q$ acquires a phase 
\eq{
\CO_q(x)\longmapsto e^{iq \lambda_0}\CO_q(y)~,
}
which matches the action we previously derived of a charged operator passes through a symmetry defect operator. 
 
\subsection{1-Form Global Symmetries}

Now let us consider the first example of higher form global symmetries called a ``1-form'' global symmetry. A 1-form global symmetry is an \emph{abelian} symmetry group $G^{(1)}$ that acts on 1-dimensional operators (i.e.~line operators). Since we have assumed  $G^{(1)}$ to be abelian and continuous, $G^{(1)}\cong U(1)^N$. However, for simplicity let us consider the case $G^{(1)}=U(1)$. Later, we will show that all higher-form global symmetries are necessarily abelian, but for now it can be taken as a simplifying assumption.

To a continuous 1-form global symmetry, we can associate a \emph{conserved} 2-tensor antisymmetric current:
\eq{
\partial_\mu J^{[\mu \nu]}(x)=0 \;\;\;\;\; \text{or} \;\;\;\;\; d \ast  J_2 = 0~.
}
Here the conservation of the current $J^{\mu\nu}(x)$ corresponds to the associated 2-form current being (co)-closed. 

As in the case of 0-form global symmetries, we can couple this current to a 2-form background gauge field
\eq{
S=...+i \int B_{\mu\nu}J^{\mu\nu} d^dx=...+i \int B_2\wedge 
\ast J_{2}~. 
}
The fact that $\ast J_{2}$ is closed implies that the theory is invariant under 1-form background gauge transformations of $B_2$:
\eq{
\delta B_{\mu\nu}=\partial_{[\mu}\Lambda_{\nu]}\quad,\quad 
\delta_\Lambda B_2=d\Lambda_1~,
}
where $\Lambda_1$ is a 1-form transformation parameter. Again, we can see that the coupling to the current in the action is invariant under the background gauge transformation 
\eq{
\delta_\Lambda S=i \int d\Lambda_1 \wedge \ast J_2=i\int \Lambda_1\wedge d \ast J_2=0~,
}
due to the fact that $\ast J_2$ is closed.

As before, we can define a symmetry defect operator by exponentiating the associated charge operator associated to the integral of the dual current along a $(d-2)$-manifold $\Sigma_{d-2}$:

\eq{
U_g(\Sigma_{d-2})=e^{i \lambda \oint_\Sigma \ast J_2}\quad, \quad g=e^{i \lambda},~~~ \lambda \in S^1~.
}
Again, we would like these operators to (i) obey a group multiplication law and (ii) be topological. 

We first need to show  that the SDOs obey a group multiplication law. This can also be achieved as before by first noting that the currents satisfy the OPE 
\eq{
\partial_\mu J^{\mu\nu}(x)J^{\alpha\beta}(y)=0~,
}
which again can be derived from the Ward identity. 
This results from the fact that we have restricted $G^{(1)}$ to be abelian. We can then compute the product
\eq{
 U_{g_1}(\Sigma_{d-2}) \cdot U_{g_2}(\Sigma_{d-2})=e^{i (\lambda_1+\lambda_2)\oint_\Sigma J}~,
}
by using the Baker-Campbell-Hausdorff formula. Therefore, we indeed see that the SDOs constructed in this way obey a group multiplication law.

Now we can show that  SDOs are topological by computing the product of $ U_g(\Sigma_{d-2})$ with $ U_{g^{-1}}(\Sigma_{d-2}')$ where $\Sigma\sim \Sigma'$ is homotopic:
\eq{
 U_{g}(\Sigma_{d-2})\cdot U_{g^{-1}}(\Sigma'_{d-2})={\rm exp}\left\{i \lambda \left(\int_\Sigma J_{d-2}-\int_{\Sigma'}J_{d-2}\right)\right\}=\exp\left\{i \lambda \int _{\widehat{\Sigma}}dJ_{d-2}\right\}=\mathds{1}~,
}
where $\widehat{\Sigma}$ is the $(d-1)$-dimensional surface swept out by the deformation of $\Sigma_{d-2}$ to $\Sigma_{d-2}'$ : $\partial \widehat\Sigma=\Sigma_{d-2}\cup \overline{\Sigma_{d-2}'}$.

Now let us consider the action of the 1-form symmetry defect operators on charged operators. 
The operators that are charged under 1-form global symmetries are 1-dimensional operators (i.e.~line operators) $L_q(\gamma)$. Because these operators are charged with respect to the current $J$, they obey the Ward identity:
\eq{
\partial_\mu J^{\mu\nu}(x)\, L_q(\gamma)=q \,\delta^{(d-1)}(x\in \gamma)\,\hat{n}^\nu \, L_q(\gamma)~,
}
 where $\hat{n}$ is the tangent vector to $\gamma$. Alternatively, this above expression can be written more cleanly using differential forms as
\eq{\label{Ward1form}
d\ast J_{2}(x)\, L_q(\gamma)=q\,\delta^{(d-1)}(x\in \gamma)\, L_q(\gamma)~,
}
where  $\delta^{(d-1)}(x\in \gamma)$ is the $(d-1)$-form that integrates to 1 on any manifold transversely intersecting $\gamma$ and zero otherwise. To be exact, $\delta^{(d-1)}(x\in \gamma)$ is the Thom class of the normal bundle of $\gamma \hookrightarrow M_d$ where $M_d$ is the $d$-dimensional space time manifold. Physically, the Thom class works as described above, but for a more precise discussion see \cite{Harvey:1998bx,Harvey:2005it}.

Since the concepts of linking and intersection appear frequently, we will take a moment to briefly explain them. Consider two submanifolds, $U_q$ of dimension $q$ and $V_r$ of dimension $r$, of a given finite dimensional smooth manifold $M$ of dimension $d$. The two submanifolds $U_q$ and $V_r$ are said to \emph{intersect transversally}  if at every intersection point $\forall p \in U_q \cap V_r$, their separate tangent spaces at that point, $T_p U_q$ and $T_p V_r$, together generate a $(q+r)$-dimensional subspace of the tangent space of $M$ at $p$: $ T_p U_q \oplus T_p V_r\subseteq T_p M$. 

\begin{figure}
\begin{center}
\includegraphics[scale=0.85,trim=1.5cm 23cm 0cm 1.5cm,clip]{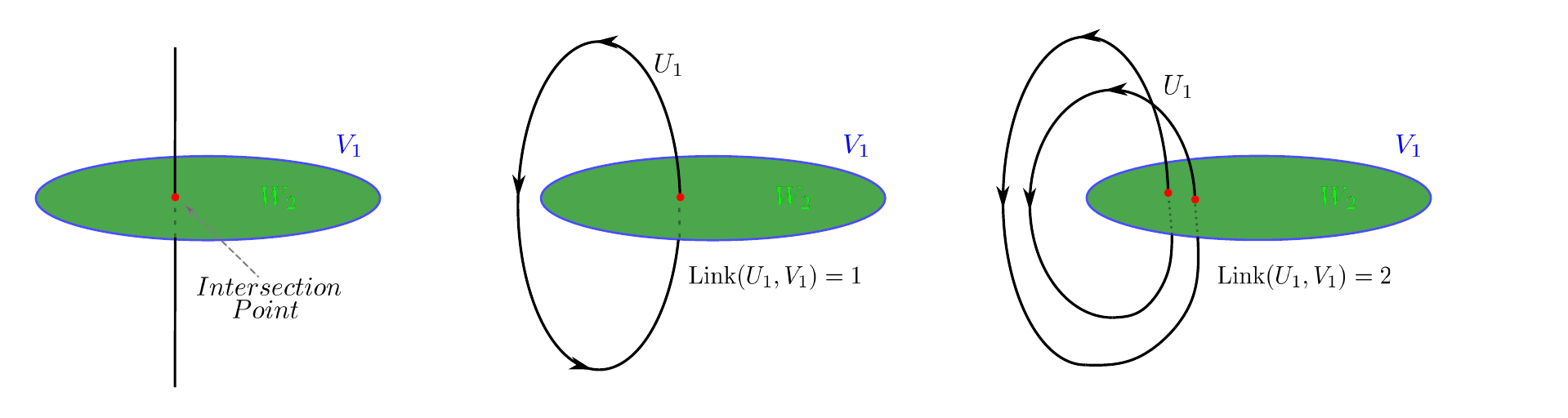}
\end{center}
\caption{In this figure we illustrate the idea of linking manifolds in $3d$. Here, $U_1$ and $V_1$ are linking manifolds which are described by summing over the signs of the intersection points of $W_2$, which fills in $V_1$, with $U_1$. 
}
\label{fig:intersection_linking_3d}
\end{figure}

The linking number is defined as follows. Let $M$ be a compact, oriented $d$ dimensional manifold without boundary, i.e. closed $\partial M = \emptyset$. Let $U_q$ and $V_r$ be two oriented submanifold of dimension $q$ and $r$ with $q+r=d-1$. We further assume that $U_q$ and $V_r$ are non-intersecting and they are homotopically trivial, which means that $U_q$ and $V_r$ can be thought to be boundaries of manifolds of one higher dimension. This allows us to introduce an $(r+1)$-dimensional manifold with boundary $W_{r+1}\subset M$ such that $V_r = \partial W_{r+1}$. Then generically $U_q$ and $W_{r+1}$ intersect in a finite number of points $\lbrace p_i \rbrace$. Since $q+r+1=d$, the tangent space at each intersection point $p_i$ generates the entire tangent space of $M$: $T_{p_i}M=T_{p_i}U_q\oplus T_{p_i}W_{r+1}$. Thus, the orientations on $U_q$ and $W_{r+1}$ define an orientation on $M$ at each $p_i$. We can then define $\text{sign} (p_i) = \pm 1$ if such an induced orientation of $M$ at $p_i$ is the same as (opposite to) the original orientation of $M$, the \emph{linking number} of $U_q$ and $V_{r=d-q-1}$ can be defined as 
\beq
\text{Link} \left( U_q, V_{d-q-1} \right) \equiv \sum_i \text{sign} (p_i).
\eeq
Note that while the number of intersection points is not invariant under the choice of $W_{r+1}$ that ``fills in'' $V_r$, the total linking number is independent of the choice of $W_{r+1}$. An illustration of linking of two lines in $3d$ is shown in Figure~\ref{fig:intersection_linking_3d}.

\begin{figure}
\begin{center}
\includegraphics[scale=0.9,trim=3.7cm 18.2cm 8cm 3.5cm,clip]{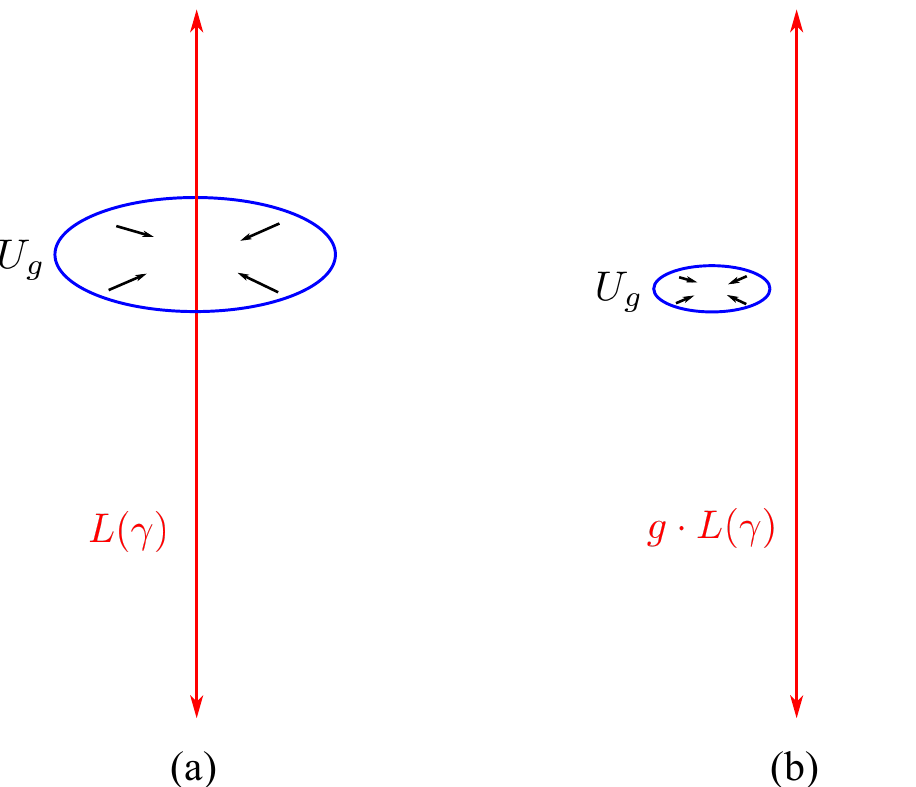}
\end{center}
\caption{This figure illustrates how 1-form symmetry defect operators $U_g$ act on charged line operators $L(\gamma)$. (a) shows a 1-form symmetry defect operator $U_g$ wrapping a charged line operator $L(\gamma)$. (b) then shows the effect of contracting $U_g$ through the line operator. }
\label{fig:1formaction}
\end{figure}

Going back to the discussion of 1-form symmetry, from the Ward identity \eqref{Ward1form}, one can show that the 1-form symmetry defect operator acts on line operators in the following way. Consider a line operator of charge $q$ wrapped on a curve $\gamma$: $L_q(\gamma)$. By counting arguments, we can see that the curve $\gamma$ can be linked by a $d-2$-dimensional manifold. Let us then wrap a SDO 
on such a linking manifold $\Sigma_{d-2}$ : $ U_g(\Sigma_{d-2})$. Then we can deform $\Sigma_{d-2}\to \Sigma'_{d-2}$ so that it crosses $\gamma$. Due to the contact term in the Ward identity, when the SDO intersects $L_q(\gamma)$, it generates a phase: 
\eq{
\big\langle U_g(\Sigma_{d-2})L_q(\gamma)\big\rangle=e^{i \lambda q \,\text{Link} \left(\Sigma_{d-2}, \gamma \right)} \big\langle L_q(\gamma)  U_g(\Sigma'_{d-2})\big\rangle~.
}
This discussion makes it clear why the current for a 1-form symmetry needs to be a 2-form conserved current: a line in $d$-dimension can only link with a $(d-2)$ dimensional manifold which corresponds to conserved $2$-tensor currents. 
See Figure \ref{fig:1formaction}. 

\begin{figure}
\begin{center}
\includegraphics[scale=0.9,trim=1.5cm 7.5cm 2cm 16cm,clip]{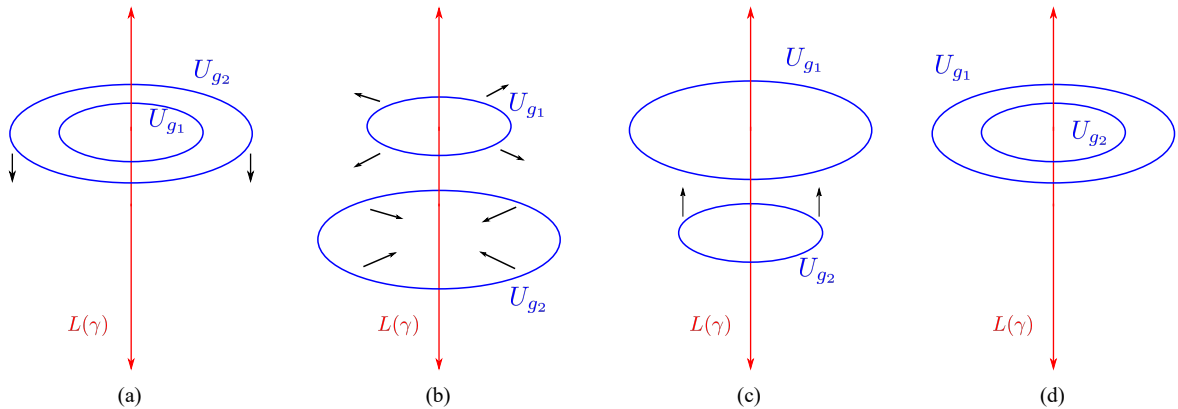}
\end{center}
\caption{This figure illustrates why 1-form symmetries must be abelian. (a) shows a configuration of parallel 1-form symmetry defect operators $U_{g_1},U_{g_2}$ wrapping a charged line operator $L(\gamma)$. The symmetry defect operators here have a natural radial ordering. (b)-(d) show a series of topological deformations of the symmetry defect operators that allows one to exchange the order of these operators. Hence, consistency with topological invariance implies that the product of 1-form symmetry defect operators must be abelian. }
\label{fig:1formbraiding}
\end{figure}

Now we would like to address the fact that we have restricted $G^{(1)}$ to be abelian. In order to have a well defined notion of a non-abelian symmetry group, we must have a well defined notion of ordering in the product of the SDOs. For the case of 0-form symmetries, the SDOs are $(d-1)$-dimensional so that there is only a single transverse direction and there is a well defined (local) notion of ordering. However, for a 1-form symmetry, the SDOs are $(d-2)$-dimensional so that there is locally a transverse 2-plane. The dimensionality of this transverse space allows one to exchange symmetry defect operators by a smooth deformation. 
Because of this, there is no well defined notion of ordering:
\eq{
 U_{g_1}\cdot  U_{g_2} \;\; \underset{\text{top. deform.}}{\simeq} \;\; U_{g_2}\cdot  U_{g_1}~.
}
See Figure \ref{fig:1formbraiding}. This lack of a notion of ordering implies that all 1-form symmetry groups must be abelian.

\subsubsection{Example: $U(1)$ Maxwell Theory}
\label{subsubsec:Maxwell Theory}

Let us illustrate the structure of 1-form global symmetries with an example. Consider $U(1)$ pure Maxwell theory in $4d$ with gauge field $A$ and coupling $g$. 
The action is given by  
\beq
S = \frac{1}{2g^2} \int F \wedge \ast  F = - \frac{1}{4g^2} \int F_{\mu\nu} F^{\mu\nu},
\eeq
where $F = dA$ is the 2-form field strength. The equation of motion of $A$ is 
\eq{
d \ast  F = 0~.
}
This equation can be viewed as a conservation of 2-form current (up to a choice of normalization) $J_2^e = \frac{1}{g^2} F$.   It implies that the pure $U(1)$ gauge theory has a $U(1)$ 1-form symmetry, often called the \emph{electric 1-form symmetry} or \emph{1-form center symmetry}. 

The corresponding symmetry defect operator is obtained by the exponential of the dual 2-form current $\ast J_2^e$ integrated over a closed 2-manifold (or 2-cycle): 
\eq{
 U^e_g(\Sigma_2)=\exp\left\{i \lambda \oint_{\Sigma_2} \ast J_2^e\right\}=\exp\left\{\frac{i \lambda }{g^2}\oint_{\Sigma_2} \ast F\right\}~.
}
Notice that $\frac{1}{g^2}\int_{\Sigma_2} \ast F$ measures the charge enclosed by $\Sigma_2$. This expression also clarifies that the symmetry under consideration is $U(1)$ since we have the charge quantization $\oint_{\Sigma_2} \ast  F = 2\pi \IZ$. 

The corresponding charged operator $W (q, \gamma)$ is the Wilson line:
\eq{
W(q, \gamma)=e^{i q\int_\gamma A}, \;\;\; q \in \IZ~.
}
The charge of the Wilson line is measured by the SDO via a non-trivial linking:
\eq{\label{eq:1-form electric_linking}
\big\langle U^e_g(\Sigma_{2})W(q, \gamma)\big\rangle=e^{i   q \lambda\, \text{Link} (\Sigma_2, \gamma)} \big\langle W (q, \gamma)  U^e_g(\Sigma'_{2})\big\rangle~,
}
where $\Sigma'_2$ is homotopic to $\Sigma_2$ and it does not link $\gamma$.
Here the phase $e^{i  q \lambda}$ can be understood as the fact that in deforming $\Sigma\to \Sigma'$ the electric charge enclosed by $ U^e_g$ goes from $q\to 0$. Thus $ U^e_g(\Sigma_2)\longmapsto e^{i q\lambda}\,  U^e_g(\Sigma'_2)$.    

This action can be derived as follows. 
Physically, a Wilson line can be thought of as a world line of super massive, stable (probe) particle which carries electric charge. The presence of such a particle will source an electric flux, modifying Maxwell's equations. Concretely, we can insert a Wilson line in the path integral as
\bea
&& \int [ d A ] e^{i q \int_\gamma A} e^{\frac{1}{2g^2} \int_{M_4} F \wedge \ast  F} = \int [ d A ] e^{i q \int_{M_4} \delta^3 (\gamma) \wedge A + \frac{1}{2g^2} F \wedge \ast  F}~,
\eea
where $\delta^3 (\gamma)$ is a 3-form delta function defined as 
\beq
\int_{M_3^T} \delta^3 (\gamma) = 1~,
\eeq
and $M_3^T$ is a 3-manifold that transversely intersects $\gamma$ once.       
This makes it clear that the Wilson line acts as an electric \emph{source}, and in its presence, the equation of motion for $A$ becomes
\beq
d \ast  F = q\, g^2 \delta^3 (x\in \gamma)~.
\eeq
Now suppose we integrate the above equation over a 3-manifold $\Sigma_3$ with boundary:
\beq
\oint_{\Sigma_2} \ast  F = q g^2 \text{Link} (\Sigma_2, \gamma)~.
\eeq
The integral of the left-hand-side will compute the asymptotic electric flux coming out of $\partial \Sigma_3$, while the integral of the right-hand-side will be proportional to the linking of the $\gamma$ with $\partial \Sigma_3$. 
This reproduces the action of the SDO on a Wilson line in \eqref{eq:1-form electric_linking} by exponentiation. 

We can also see the action on the Wilson line by noting that the associated current $\ast J_2^e$ is the conjugate momentum operator to the dynamical gauge field. This means that the symmetry defect operator acts on the dynamical gauge field by shifting:
\beq\label{Ashift}
A \to A + \lambda_1~,
\eeq
where $\lambda_1$ is a 1-form transformation parameter satisfying $\oint \lambda_1 \in U(1)$. One can easily check that under this transformation, the Wilson line indeed can pick up a $U(1)$ phase consistent with \eqref{eq:1-form electric_linking}. 

There also exists a dual 1-form symmetry called \emph{magnetic 1-form symmetry}. This follows from the Bianchi identity.
\beq
d F =0~.
\eeq
We view this as a conservation equation for a 2-form current $J_2^m = \frac{1}{2\pi} \ast  F$. The SDO of magnetic 1-form symmetry is constructed by integrating the dual current over a 2-cycle:  
\beq
U_g^m (\Sigma_2) = \exp\left\{i \lambda \oint_{\Sigma_2} \ast  J_2^m\right\} = \exp\left\{ i\lambda \oint_{\Sigma_2} \frac{F}{2\pi}\right\}~,
\eeq
We note that $\oint_{\Sigma_2} F$ measures the magnetic flux passing through $\Sigma_2$. From this, we learn that the charged object is 't Hooft line operators%
\beq
T_1 (m, \gamma) = e^{i m  \int_\gamma \tilde{A}}
\eeq
where $\tilde{A}$ is the dual photon defined by $\ast  F = d \tilde{A}$. 

Alternatively, we can also define the `t Hooft line operator intrinsically in terms of the photon field $A$. 
Let us fix a curve $\gamma$ and charge $m$ we wish to insert the `t Hooft line operator on. To define the operator, we need to cut out an infinitesimal tube surrounding the curve $\gamma$. In the transverse space\footnote{Here by transverse we simply mean if you zoom in close enough to the line the space looks like $\IR_\gamma\times \IR_T^3$ where $\IR_\gamma$ is the direction along the curve $\gamma$. The transverse space is the $\IR_T^3$ factor. } this will locally look like cutting out an infinitesimal 3-ball surrounding a point. On the boundary of this 3-ball  in the transverse space (i.e. a transverse infinitesimal 2-sphere) we impose the boundary condition on the gauge field 
\eq{
A_T=m A_{\rm Dirac}~. 
}
Here $A_{\rm Dirac}$ is the Dirac monopole background which has a smooth solution everywhere except for at the origin. The boundary condition (i.e. $A_{\rm Dirac}$) can be constructed by gluing together a smooth solution in the Northern and Southern hemispheres
\eq{
A_{\rm Dirac}^{(N/S)}=\half\big(\pm1-\cos(\theta)\big)d\phi~,
}
by the gauge transformation $g=e^{ i \phi}$.\footnote{Note that sometimes the Dirac monopole includes a singular Dirac string. However, this is not physical when gauge symmetry is preserved and is simply a reflection of the fact that not all smooth gauge fields can be written in a single coordinate patch. }

The action of the 1-form magnetic symmetry group on 't Hooft lines is identical to the action of the 1-form electric symmetry on the Wilson lines:
\eq{\label{eq:1-form magnetic_linking}
\big\langle U^m_g(\Sigma_{2})T_1(m, \gamma)\big\rangle=e^{i   m \lambda\, \text{Link} (\Sigma_2, \gamma)} \big\langle T_1 (m, \gamma)  U^m_g(\Sigma'_{2})\big\rangle~.
}

We can couple the theory to 2-form background gauge fields for both of the 1-form symmetries. These appear in the action as
\beq
S = \frac{1}{2g^2} \int \left( F - B_2^e \right) \wedge \ast  \left( F - B_2^e \right) + \frac{i}{2\pi} \int B_2^m \wedge F~,
\label{eq:Maxwell_with_BGF}
\eeq
 where $B_2^e$ ($B_2^m)$ is the 2-form background gauge field of 1-form electric (magnetic) symmetry. The second term clearly takes the standard form of coupling between the current and its background gauge field $i\int B \wedge \ast  J$.\footnote{In fact, this provides a justification for the normalization of the current. Namely, it is with particular normalization of $\frac{i}{2\pi}$ that the coupling term is invariant under the 1-form background gauge transformation $B_2^m \to B_2^m + d \lambda_1$ once the Dirac quantization condition $\oint F = 2\pi \IZ$ is taken into account. } The first term however, is a bit different. Here we have supplemented the usual coupling of a background gauge field to its associated current by an additional local counter term that is only dependent on the background gauge fields: 
 \eq{
 S_{c.t.}=\frac{1}{2g^2}\int B_2^e\wedge B_2^e~.
}
This term is necessary to make the kinetic term invariant under background gauge transformations since the dynamical (electric) gauge field is also shifted under these background gauge transformations as in \eqref{Ashift}. Since this counter term only depends on background gauge fields, it does not effect any of the dynamics of the theory. 

In fact, turning on couplings to background gauge fields makes it straightforward to analyze 't Hooft anomalies among global symmetries. Concretely, we now show that 1-form electric and magnetic symmetries have a mixed 't Hooft anomaly. The 't Hooft anomalies of a QFT can be probed by the non-invariance of the partition function under background gauge transformations with all  background gauge fields activated. In particular, to probe the anomalies of a 1-form global symmetry $G^{(1)}$ with background gauge field $B$, one needs compute 
\beq
Z [ B + d \lambda, \lbrace C \rbrace ] = e^{i \int \CA [\lambda, B, \lbrace C \rbrace ]} Z [ B , \lbrace C \rbrace ]~,
\eeq
where $\lbrace C \rbrace$ is the set of other background gauge fields and $\CA$ is  the anomalous phase.

In the case of Maxwell's theory, the action \eqref{eq:Maxwell_with_BGF} is not invariant under the 1-form electric background gauge transformations $B_2^e \to B_2^e + d \lambda_1^e$:
\beq
\delta S = \frac{i}{2\pi} \int B_2^m \wedge d \lambda_1^e~,
\label{eq:Maxwell_tHooft_anom1}
\eeq
because the dynamical gauge field shifts under 1-form electric symmetry transformations. 

Of course, such a non-invariance does not signal a violation of the symmetry. it instead captures the obstruction to gauging both of the 1-form global symmetries simultaneously. However, there is no obstruction to just gauging the one of the 1-form symmetries. With this choice of counter terms, we see that the action is invariant under magnetic 1-form transformations and thus the 1-form magnetic symmetry can be consistently gauged while the 1-form electric symmetry cannot. 
 
In order to gauge the electric 1-form symmetry, we can add an addition local counter-term (made only of background gauge fields) to make the action manifestly invariant under the electric symmetry:
\eq{
S_{c.t.}=-\frac{i}{2\pi}\int B_2^m\wedge B_2^e~.
}
Now the action reads 
\beq
S = \frac{1}{2g^2} \int \left( F - B_2^e \right) \wedge \ast  \left( F - B_2^e \right) + \frac{i}{2\pi} \int B_2^m \wedge \left( F - B_2^e \right)~,
\label{eq:Maxwell_with_BGF_2}
\eeq
which is invariant under the electric 1-form symmetry transformations, but   not  under the magnetic 1-form background transformations $B_2^m \to B_2^m + d \lambda_1^m$:
\beq
\delta S = \frac{i}{2\pi} \int - d \lambda_1^m \wedge B_2^e~.
\label{eq:Maxwell_tHooft_anom2}
\eeq
In fact, one can show that there exists no choice of local counter-terms that make the theory invariant under both symmetries. This is the statement that there is anomaly between the two symmetries and how it obstructs us from simultaneously gauging both symmetries. 

This mixed 't Hooft anomaly can be written concisely in terms of anomaly inflow action in five dimensions:
\beq
S_{\scriptscriptstyle \text{inflow}} = -\frac{i}{2\pi}\int_{N_5}  B_2^m \wedge d B_2^e
\eeq
and $\partial N_5 = M_4$. 
Here we use the idea of ``anomaly inflow'' (which we discuss in more detail in Section \ref{sec:anomalies}) 
to describe the anomalous variation of the partition function. 
 It is a straightforward exercise to show that background gauge transformations of the inflow action indeed reproduces the anomalous variation \eqref{eq:Maxwell_tHooft_anom2} on the $4d$ boundary $M_4$ of the $5d$ manifold $N_5$, while the anomalous variation of \eqref{eq:Maxwell_tHooft_anom1} is reproduced by switching the labels $e\leftrightarrow m$. We will discuss anomalies and the idea of anomaly inflow below in more detail in Section \ref{sec:anomalies}.

\subsection{$p$-Form Global Symmetry}

The structure of ``0-form'' and ``1-form'' global symmetries fit into a more general family of symmetries called ``higher form global symmetries'' which are indexed by an integer $p$ (sometimes called the ``degree''). 

A \emph{$p$-form global symmetry}, denoted $G^{(p)}$ is a symmetry that acts on $p$-dimensional charged operators. It has a corresponding $(p+1)$-index anti-symmetric tensor (or simply $(p+1)$-form) conserved current which can be recast as a closed $(d-p-1)$-form dual current $\ast J_{p+1}$ :
\eq{
\partial_{\mu_1}J^{[\mu_1...\mu_{p+1}]}(x)=0\quad \Longrightarrow\quad d\ast J_{p+1}(x)=0~.
}
Again, the conservation of $J^{\mu_1...\mu_{p+1}}(x)$ corresponds to $\ast J_{p+1}$ being a closed form.

A theory with $p$-form global symmetry can then be coupled to an abelian $(p+1)$-form background gauge field  $B_{p+1}$ by the term
 \eq{
 S_p = i \int B_{p+1}\wedge \ast J_{p+1}~,  
 }
 where $\Lambda_p$ is a $p$-form background gauge transformation parameter. Again, the fact that $\ast J_{p+1}$ is closed means that the action is invariant under the background gauge transformation 
\eq{
\delta B_{p+1}=d\Lambda_{p}~.
}
Now, as before, we can define a $(d-p-1)$-dimensional topological symmetry defect operator  
\eq{
 U_g(\Sigma_{d-p-1})=\exp\left\{i \lambda \int_\Sigma \ast J_{p+1}\right\}\quad, \quad g=e^{i \lambda}~.
}
Again, the fact that $\ast J_{p+1}$ is closed  implies that the symmetry defect operator is topological. It can be proven by following the same steps as the 0- and 1-form symmetry cases discussed above.

\begin{figure}
\begin{center}
\includegraphics[scale=1.2,trim=5.5cm 16.35cm 3cm 5cm,clip]{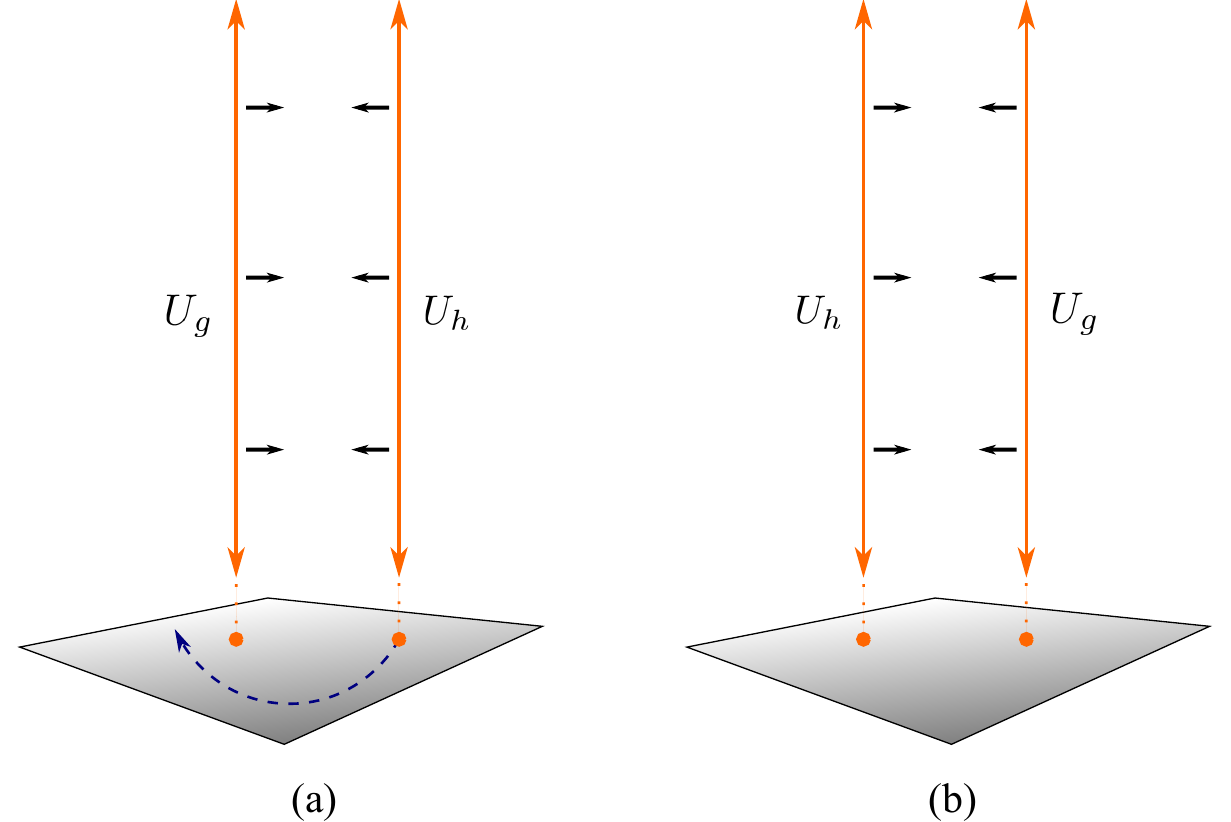}
\end{center}
\caption{This figure illustrates why $p$-form symmetries must be abelian for $p\geq1$. (a) shows a configuration of parallel $p$-form symmetry defect operators $U_g,U_h$ in which $U_g$ is to the ``left'' of $U_h$. They are projected onto the transverse 2-plane  below. Since these operators are topological, they can be exchanged in their transverse 2-plane by a smooth deformation so that $U_h$ is to the left of $U_g$. This means that there is no natural ordering of the product of operators and therefore that the multiplication must be abelian. This is the higher dimensional analog of Figure \ref{fig:1formbraiding}.}
\label{fig:pformbraiding}
\end{figure}

As in the case of 1-form global symmetries, the dimensionality of the symmetry defect operators for $p\geq1$ obstruct a well defined notion of ordering of the $ U_g(\Sigma)$ and therefore $G^{(p)}$ must be abelian for $p\geq 1$.  
See Figure \ref{fig:pformbraiding}.

Again, since $G^{(p)}$ is abelian, the currents $\ast J_{p+1}$ satisfy
\eq{
d\ast J_{p+1}(x) \,\ast J_{p+1}(y)=0~,
}
and therefore the symmetry defect operators satisfy the group relation 
\eq{
 U_{g_1}(\Sigma_{d-p-1})\cdot  U_{g_2}(\Sigma_{d-p-1})= U_{g_1g_2}(\Sigma_{d-p-1})~. 
}

Recall that $p$-form global symmetries are defined as those that act on $p$-dimensional charged obects. This implies OPE coming from the Ward identity:
\eq{\label{pformWard}
d\ast J_{p+1}(x) W_q(\Gamma_p)=q\,\delta^{(d-p)}(x\in \Gamma_p)\, W_q(\Gamma_p)~,
}
where $W_q(\Gamma_p)$ is a charge $q$, $p$-dimensional operator. From this Ward identity, we can derive the action of the SDO on $W_q(\Gamma_p)$. Let us fix a particular $\Gamma_p$ and pick a linking $\Sigma_{d-p-1}$ manifold that is homotopic to $\Sigma'_{d-p-1}$ that does not link $\Gamma_p$. Then, from the Ward identity \eqref{pformWard} we can derive 
\eq{
\big\langle U_g(\Sigma_{d-p-1})\,W_q(\Gamma_p)\big\rangle=e^{iq\lambda \,\text{Link} (\Sigma, \Gamma) }\big\langle  W_q(\Gamma_p)\,U_g(\Sigma'_{d-p-1})\big\rangle\quad, \quad g=e^{i \lambda}~.
}

As in the case of the 0-form global symmetry, we can also understand the action of the symmetry defect operators as implementing background gauge transformations on charged operators. In particular, to $ U_g(\Sigma_{d-p-1})$, we can associate the background gauge transformation 
\eq{
\Lambda_p=\lambda\, \Omega_p~,
}
where here $\Omega_p$ is a $p$-form that is known as the global angular form of the normal bundle of $\Sigma_{d-p-1}\subset M_d$ \cite{Harvey:2005it}. Physically, this global angular form is the gauge transformation that gives the $p$-form gauge field unit winding in the $(p+1)$-dimensional transverse space. For example, $\Omega_2=dA_{\rm Dirac}$. 
 This leads to the background gauge field  
\eq{
B_{p+1}=d\Lambda_p=\lambda \,\delta^{(p+1)}(x\in \Sigma_{d-p-1})~,
}
where $\delta(x\in \Sigma_{d-p-1})$ is the Dirac delta $(p+1)$-form on the transverse space (formally the Thom class of the normal bundle of $\Sigma_{d-p-1}$).

\subsubsection{Example: Periodic Scalar Field}
\label{periodicscalar}

An example of a theory with higher form global symmetry is that of the periodic scalar field in $4d$. This theory occurs naturally as the Goldstone boson of spontaneously broken $U(1)$ global symmetry. However, the following discussion does not need any specific UV completion.\footnote{The following discussion applies to axion theory without a coupling to gauge bosons. The presence of couplings to gauge bosons introduces interesting effects on the global symmetries of the theory. We refer to \cite{Hidaka:2020iaz,Hidaka:2020izy,Brennan:2020ehu,Brennan:2023kpw} for detailed discussion.}

Consider the action 
\eq{
S=\half\int  d\phi \wedge \ast d\phi~,
}
where $\phi\sim \phi+2\pi f$ with a dimension one parameter $f$ related to the symmetry breaking scale. 
The equation of motion and Bianchi identity are 
\beq
d \ast  d \phi = 0 \;, \quad \quad d^2 \phi = 0.
\eeq
These two equations can be seen as conservation of two currents, or equivalently closed-ness of 3-form and 1-form dual currents, respectively:
\eq{
j_3=\ast d \phi \; ,\quad \quad J_1=\frac{d\phi}{2\pi}~.
} 
Here $j_3\leftrightarrow j^\mu=\partial^\mu\phi$ is the current associated to the (0-form) shift symmetry of $\phi$. Recall that the Goldstone boson has the shift symmetry, $\phi \to \phi + c$ which is captured by the fact that its equation of motion $\square \phi = 0$ does not include terms polynomial in $\phi$. 
The second $J_1\leftrightarrow J^{\mu\nu\alpha}=\epsilon^{\mu\nu\alpha\beta}\partial_\beta\phi $ is the current for the 2-form global symmetry associated to winding modes of $\phi$. This may be seen by looking at the charge operator $Q = \oint \frac{d \phi}{2\pi}$. For smooth field configurations \underline{on flat space}, the integral evaluates to zero. 
However, this is not true if we are on a space time with non-trivial 1-cycles (i.e. non-contractible loops). Additionally, if we allow for singular field configurations, then $\phi$ can have non-trivial winding around string-like singularities. These string-like singularities often arise naturally in UV completions as cosmic strings.  

\begin{figure}
\begin{center}
\includegraphics[scale=0.8,trim=4cm 19.5cm 4cm 2.5cm,clip]{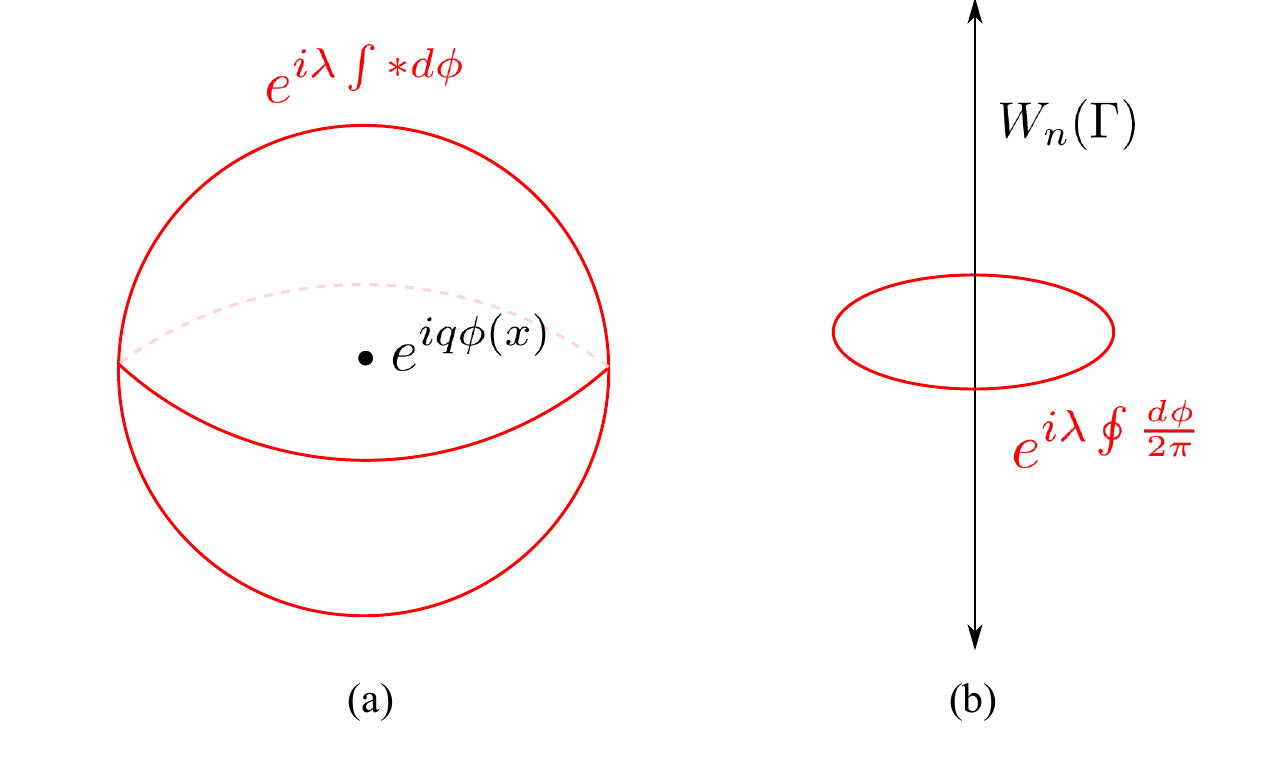}
\end{center}
\caption{This figure illustrates the charged operators and symmetry defect operators for the periodic scalar field along a spatial slice. (a) displays the features of the 0-form symmetry: the charged operators (black) are the local operators $\CO_q(x)=e^{i q\phi(x)}$ and the symmetry defect operators (red) $U_g=e^{ i\lambda \int \ast d\phi}$ wrap linking 3-manifolds. (b.) displays the features of the  dual 2-form symmetry: the charged operators (black) are the two-dimensional string operators $W_n(\Gamma_2)$  and the symmetry defect operators (red) $\CU_g=e^{i \lambda\oint \frac{d\phi}{2\pi} }$ wrap linking circles. }
\label{fig:axionstringdef}
\end{figure}

We can couple these currents to background gauge fields $A_1$ and $C_3$ as 
\eq{
S=...+i \int A_1\wedge j_3+i \int C_3\wedge J_1~, 
}
where $A_1,C_3$ are 1- and 3-form $U(1)$ gauge fields.
Here the charged objects  with respect to the 0-form and 2-form global symmetries are 
\eq{
\text{0\text{-}form}:~~ &\CO_q(x)=e^{i q \phi(x)}\quad, \quad q\in \IZ~,\\
\text{2\text{-}form}:~~&W_n(\Gamma_2) \hspace{15mm},\quad n\in \IZ~,
}
where $\CO_q(x)$ is a local operator (note the  charge $q\in \IZ$ due to gauge invariance under $\phi \sim \phi + 2\pi$)  and $W_n(\Gamma_2)$ is a string/vortex operator wrapped on the 2D\ cycle $\Gamma_2$ around which $\phi$ winds $n$-times.\footnote{
The string/vortex operator on $\Gamma_2$ can be defined more precisely in analogy with the monopole line. First, one must cut out a tube around $\Gamma_2$ which has a transverse space which is $\IR^2$ with an infinitesimal disc cut out around the origin. For the operator of charge $n$, one then imposes boundary conditions $\phi$ so that it winds $n$-times around the excised disc. 
} The corresponding symmetry defect operators are
\eq{
\text{0\text{-}form}:~~ U_g(\Sigma_3)&=\exp\left\{ i\lambda \int_{\Sigma_3} \ast d \phi\right\}~,\\ 
\text{2\text{-}form}:~~ \CU_g(\widetilde\Sigma_1)&=\exp\left\{i
 \lambda \int_{\widetilde\Sigma_1} \frac{d\phi}{2\pi}\right\}~.
}
Here $U_g(\Sigma_3)$ is the SDO for the 0-form symmetry which acts on the charged operator $e^{i q\phi(x)}$ by shifting $\phi$ since $\ast j_3(x)$ is the canonically conjugate momentum operator to $\phi(x)$.

Similarly,  $\CU_g(\widetilde\Sigma_1)$ is the SDO for the 2-form global symmetry. It acts on the strings by measuring the winding number of $\phi$ around the curve $\widetilde\Sigma_1$ just as the electric and magnetic symmetry defect operators in $U(1)$ Maxwell theory act on  charged Wilson and monopole lines by measuring the enclosed electric and magnetic charges.  See Figure \ref{fig:axionstringdef}. 
  
\subsection{Spontaneous Symmetry Breaking}

As is the case for ordinary global symmetries, spontaneously broken higher form global symmetries also lead to massless fields \cite{Gaiotto:2014kfa,Lake:2018dqm}. 

The standard story is that a spontaneously broken continuous 0-form global symmetry $G\to H$  has an associated Goldstone boson $\pi(x)$ that is valued in (the Lie algebra of) the coset space $G/H$. The Goldstone field exhibits a non-linear realization of the symmetry -- i.e.~  it shifts under background gauge transformations along broken directions. Consequently,  the current operator for a broken generator  sources the Goldstone boson field.  

For example, for a spontaneously broken 0-form $U(1)$ symmetry , the background gauge transformation 
\eq{
\delta_\lambda A =d\lambda~,
}
acts on the Goldstone boson by a shift
\eq{
 \delta_\lambda \pi(x)= \lambda(x)~.
} 

Now recall that a  $p$-form global symmetry has a $(p+1)$-form background gauge field with a  $p$-form gauge parameter
\eq{
B_{p+1} \longmapsto B_{p+1}+d\Lambda_p~.
}
In the case when this symmetry is spontaneously broken, the background gauge transformation will shift the Goldstone boson in analogy with the case of the Goldstone boson associated to spontaneously broken 0-form global symmetries. However, since the gauge transformation parameter is a $p$-form field, we see that the associated Goldstone field will also be a $p$-form gauge field. 

As with 0-form global symmetry, spontaneous symmetry breaking of higher form  global symmetries can be measured by the vacuum expectation value of a charged operator (i.e.~an order parameter). The reason is that a symmetry is preserved iff the charge operator acts trivially on the vacuum state. It then follows that any charged operator should have trivial vacuum expectation value unless the symmetry is spontaneously broken:
\eq{
0=\langle0|[Q,\CO(x)]|0\rangle=\langle0|\CO(x)|0\rangle~.
} 

In the case of a $p$-form global symmetry, the charged operators are $p$-dimensional operators $W(\Gamma_p)$. In general, we can express the expectation value of such an operator as 
\eq{
\big\langle W(\Gamma_p)\big\rangle\sim e^{-F(\Gamma_p)}~,
}
where $F(\Gamma_p)$ is some function with $\Gamma_p$-dependence. 

Let us denote $\operatorname{Vol}(\Gamma_p)$ to be the volume of $\Gamma_p$. Note that this is not the volume enclosed by the manifold $\Gamma_p$ -- in the case where $p=1$, ${\rm Vol}(\Gamma_1)$ is the perimeter of the curve $\Gamma_1$. 
We can now differentiate between two types of behaviors:
\eq{
\lim_{\operatorname{Vol}(\Gamma_p)\to \infty}\Re\left[\frac{F(\Gamma_p)}{\operatorname{Vol}(\Gamma_p)}\right]=\begin{cases}
{\rm finite}& \text{Spontaneous Symmetry Breaking}\\
\infty& \text{Symmetry Unbroken}
\end{cases}
}
Here the case where the above limit is finite corresponds to spontaneous symmetry breaking (even though $F(\Gamma_p) \to \infty$) because, given a $W(\Gamma_p)$ with the above behavior, we can define a renormalized operator which is  shifted by a local counter term 
\eq{
\widehat{W}(\Gamma_p)=\exp\left\{-c \,\oint_{\Gamma_p}dV\right\}\times W(\Gamma_p)~, }
where $dV$ is the volume form of $\Gamma_p$, that has behavior 
\eq{
\lim_{\operatorname{Vol}(\Gamma_p)\to \infty}\big\langle \widehat{W}(\Gamma_p)\big\rangle\neq 0~,
}
which is finite. This non-zero expectation value implies the vacuum state is charged with respect to the generalized global symmetry because there is a charged operator with non-trivial vacuum expectation value. Conversely, if the above limit of $\Re[F(\Gamma_p)/\operatorname{Vol}(\Gamma_p)]$ is infinite, there is no local counter term with which to renormalize the operator so that it has finite expectation value in the infinite volume limit. In this case, the vacuum state is not charged and there is no spontaneous symmetry breaking. \\

\noindent \textbf{Example: $4d$ $SU(N)$ Yang Mills}

Our discussion of how the scaling of the expectation value of of Wilson-type operators indicates the spontaneous breaking (or preservation) of a higher form global symmetry is reminiscent of the Wilson and `t Hooft prescription for confinement \cite{Wilson:1974sk,tHooft:1977nqb,tHooft:1978pzo}. Indeed, these papers lay the foundation for what we now call 1-form global symmetries \cite{tHooft:1977nqb,tHooft:1978pzo}. In fact, the modern statement that a theory confines is to say that 1-form center symmetry is conserved. 

To connect our discussion to those of Wilson and `t Hooft, we can apply our above discussion to the case of 1-form global symmetries in $4d$ pure $SU(N)$ Yang-Mills theory. Consider a Wilson operator in the fundamental representation $R_f$ of $SU(N)$:
\eq{
\langle W_{R_f}(\gamma)\rangle=\langle \Tr_{R_f} \CP\,e^{i \oint_\gamma A}\rangle=e^{F({\rm Per}(\gamma))}~,
}
where here ${\rm Per}(\gamma)$ is the perimeter of the curve $\gamma$. We now say that the associated center symmetry (which for $SU(N)$ is actually the discrete group $\IZ_N$) is spontaneously broken if 
\eq{
\lim_{{\rm Per}(\gamma)\to \infty} \frac{F({\rm Per}(\gamma))}{{\rm Per}(\gamma)}=c\quad, \quad |c|<\infty~.
}
Here we see explicitly that when $c\neq 0$ that we can can create a charged operator \eq{
\widetilde{W}_{R_f}=\Tr_{R_f} \CP\,e^{i \oint_\gamma A-c \oint_\gamma ds}
}
which has non-trivial expectation value in the vacuum state even when ${\rm Per}(\gamma)\to \infty$. \\

\noindent \textbf {Example: $U(1)$ Gauge Theory}

Another  nice example of this spontaneous symmetry breaking is in $U(1)$ Maxwell theory. Here there is a 1-form global symmetry for which $\ast J_2=-\frac{i}{g^2}\ast F$. This symmetry is spontaneously broken by the vacuum and the corresponding Goldstone boson is the massless photon.  
This follows from the fact that the photon is the Goldstone boson for this 1-form gauge symmetry by noting that the 1-form background gauge transformation shifts the photon field. This is because $\ast \frac{F}{g^2}$ is the conjugate momentum to the gauge field $A$ and hence the canonical commutation relations imply that  $\ast J_2$ generates a shift in the dynamical gauge field. Because of this, we see that $\ast J_2$ acts on the vacuum state to source the photon field and the 1-form global symmetry (not gauge invariance) is spontaneously broken.

\subsubsection{$p$-Form Coleman-Mermin-Wagner Theorem}

As in the case of ordinary global symmetries, one can rule out spontaneous breaking of a higher form symmetry based on the space time dimension of the theory  \cite{Gaiotto:2014kfa,Lake:2018dqm}. In particular, consider a theory with a  $p$-form global symmetry in $d$-space time dimensions. If this symmetry is spontaneously broken, then the resulting Goldstone field is a (massless) $p$-form field. 

Now consider the theory of a free, massless $p$-form field in $d$-space time dimensions:
\eq{
S=\half\int_{M_d} d B_p \wedge \ast d B_p~.
}
The theory is invariant under the $(p-1)$-form gauge transformation 
\eq{
\delta B_p= d\Lambda_{p-1}~,
}
and the equation of motion for $B_p$ is given by 
\eq{
d\ast dB_p=0~.
}

In order to derive a bound on the spontaneous breaking of continuous higher form symmetries, we can study the large distance behavior of the 2-point function of the candidate Goldstone field. If the 2-point function blows up at large separation, it  indicates that the symmetry in question cannot be spontaneously broken because it leads to an ill-defined Goldstone theory. 

Therefore, we would like to study the large distance behavior of 2-point function of the Goldstone boson $B_p$ in Euclidean signature. 
This is controlled by 
the Green's function for the equation of motion. The large $r$ dependence is controlled by the terms with radial derivatives  which corresponds to the equations of motion for the purely angular components of $B_p$. Here we parametrize the angular components as 
\eq{
B_p^{(ang)}=B_{\mu_1,...,\mu_p}r^p d\theta^{\mu_1}\wedge... \wedge d\theta^{\mu_p}~,
}
where here we include the $r^p$ to emphasize the difference between differential form and the tensor field $B_{\mu_1...\mu_p}$.  Here we mean that the angular differential forms $d\theta^{\mu_a}$ are singular whereas $r \,d\theta^{\mu_a}$ are smooth, and thus it only really makes sense to talk about the tensor field $B_{\mu_1...\mu_p}$ as defined above so that any singularities of $B_p^{(ang)}$ are reflected by the $B_{\mu_1...\mu_p}$. 

Now from the equation of motion we can read off that the large distance behavior of the Green's function 
\eq{
 d \ast d B_p^{(ang)}=0\quad\Longrightarrow \quad \frac{r^{d-1}}{r^{2(d-p-1)}}\partial_r\left(\frac{r^{d-1}}{r^{2p}}\partial_r\left(G(\vx,\vx')r^p\right)\right)=0~.
} 
Let us then take the ansatz that $G(\vx,\vx')\sim |\vx-\vx'|^{-s}$. We then find that the behavior of the solution away from $r=0$ is $s=d-p-2$. Therefore, we find that the Green's function parametrically goes like $G(\vx,\vx')\sim \frac{1}{|\vx-\vx'|^{d-p-2}}$. This means that the 
 Green's function is bounded at infinity if and only if
\eq{
s>0\quad\Longrightarrow\quad  d-p-2>0~,
}
and conversely that the Green's function $G(x,x')$ diverges as $|x-x'|\to \infty$ if $d-p\leq 2$. 

Again, since the Green's function encodes the behavior of the 2-point function of the associated putative Goldstone mode, we see that for $d-p\leq 2$ that the 2-point function diverges. We take this to mean that a $p$-form global symmetry cannot be spontaneously broken when the Green's function of the would-be Goldstone field diverges in analogy with the Coleman-Mermin-Wagner theorem regarding continuous 0-form global symmetries in $2d$ \cite{Mermin:1966fe,Coleman:1973ci}. In $4d$, this theorem tells that $p$-form global symmetry with $p \geq 2$ can not be spontaneously broken.

\subsection{Anomalies}
\label{sec:anomalies}
Symmetries are an incredibly powerful tool for studying the renormalization group flow of quantum field theories. A conserved global symmetry will protect terms that explicitly break the global symmetry upon flowing to the IR. However, there are many different ways symmetries can be matched between the UV and IR: a UV symmetry can be spontaneously broken, trivially conserved, or can be fractionalized to name possibilities.  One tool that further constrains the ways symmetries are matched between the UV and IR is by use of anomalies. 

Anomalies (here by which we mean `t Hooft anomalies) are RG invariant quantities that give information about how symmetries are matched between the UV and IR.   The standard example of this is a chiral symmetry in a theory with free fermions. In this case, the action is invariant under a chiral $U(1)$ rotation of fermion fields, but we see that the 3-point function of associated currents does not respect the associated conservation equation. This can be computed directly from the 1-loop triangle diagram following \cite{Adler:1969gk,Bell:1969ts,Fujikawa:1983bg}. One way we can see that the anomaly is matched along RG flow is an old argument due to `t Hooft in which we can add free, decoupled fermions to the theory that transform under the symmetry so as to cancel the anomaly. We then have a good symmetry that we can track along the RG flow and fermions in the IR that transform anomalously under the symmetry. This implies that the IR theory without the free fermions must also transform anomalously under the original symmetry so as to cancel the anomaly of the decoupled fermions. 
However, this argument only really works for a small class of anomalies. 

The modern perspective on anomalies uses the idea of inflow from TQFTs to describe anomalies and show their invariance along RG flows. The idea is the following. Consider a theory with a (0-form) classical, continuous symmetry group $G$. To this classical symmetry, we can identify an associated classically conserved current and can couple the theory to a classical gauge field $A_1$ .\footnote{Here by coupling a symmetry to a classical gauge field we mean that we turn on a fixed gauge field in the action as if we had coupled to a dynamical gauge field without a kinetic term. This gauge field couples to the associated current at linear order in the usual way. 
} In particular, since we are not summing over the gauge field in the path integral, the partition function is a functional of the ``background gauge field'' $A_1$:
\eq{
Z[A_1]=\int[d\Phi]\,e^{-S[\Phi;A_1]}~. 
}
The symmetry $G$ is then anomalous if the partition function is not invariant under gauge transformations of the background gauge field $A_1$: 
\eq{
Z[A_1+d\lambda_0]=e^{ i \int \CA [A_1,\lambda_0]}\times Z[A_1]~. 
}
Here the phase $\CA[A_1,\lambda_0]$ is the ``anomalous phase'' which vanishes when $\lambda_0=0$:
\eq{
\CA [A_1,\lambda_0=0]=0~. 
} 
This anomalous phase can be understood as the boundary variation of a $d+1$-dimensional TQFT. The idea is similar to that of `t Hooft except that we can couple our $d$-dimensional theory on $M_d$ to a decoupled TQFT living on a $d+1$-dimensional manifold $N_{d+1}$ that bounds $M_d$: $\partial N_{d+1}=M_d$. Again, we can treat this coupled system as invariant under RG flows and again allows us to match the anomaly of our $d$-dimensional theory along an RG flow.

More concretely, let us consider a $d$-dimensional theory on space time $M_d$ with anomalous phase $\CA [A_1,\lambda_0]$. We can write the total derivative of the anomalous phase in terms of the variation of some $(d+1)$-dimensional form depending on $A_1$:
\eq{
d\CA[A_1,\lambda_0]=\widehat\CA[A_1+d\lambda_0]~.
}
This means that we can cancel the anomalous variation of the action by defining the modified path integral
\eq{
\widehat{Z}[A_1]:=Z[A_1]\times {\rm exp}\left\{ i \int_{N_{d+1}}\widehat\CA[A_1]\right\}~.
}
We can interpret the $(d+1)$-dimensional phase as describing an (invertible) topological quantum field theory that is often referred to as a \emph{symmetry protected phase} or SPT. Since this SPT is a decoupled theory, we can track it along the RG flow to infer that the associated `t Hooft anomaly is RG invariant. This is the paradigm of ``anomaly inflow.''

For completeness, let us consider the example of the standard cubic anomaly of a chiral symmetry. Using the Fujikawa method, we can see that the anomalous variation of the action is given by 
\eq{
\widehat\CA[A_1,\lambda_0]=\kappa \lambda_0\frac{F_2\wedge F_2}{8\pi^2}~, 
}
where $F_2=dA_1$ is the field strength for the chiral symmetry. This anomalous variation can be described as the boundary variation of the $5d$ Chern-Simons action 
\eq{
\CA[A_1]=\kappa A_1\wedge \frac{F_2\wedge F_2}{8\pi^2}~. 
}
Note that this is related to the index of a Dirac operator in $6d$ and fits into the ``descent procedure'' for anomalies. See \cite{Callan:1984sa, Witten:2019bou, Harvey:2005it, Weinberg:1996kr, Hong:2020bvq} for more details.

\subsubsection{Anomalies of Higher Form Global Symmetries}

As in the case of ``ordinary'' global symmetries, higher form global symmetries can also have anomalies. An anomaly can be diagnosed by turning on background gauge fields $\{A_{p_i}\}$ and studying the variation of the path integral under background gauge transformations. We say that the theory has an anomaly if the path integral varies by an overall phase:
\eq{
Z[A_{p_i}+d\Lambda_{i}]=Z[A_{p_i}]\times e^{i \int \CA(\Lambda_i,A_{p_i})}~.
}
In general, the anomaly $\CA$, can be any local functional involving the background gauge fields, their field strengths, and the background gauge transformation parameters $\Lambda_i$. In the case when turning off all of the $A_{p_i}$ eliminates the phase, we say that there is an 't Hooft anomaly -- otherwise there is an ABJ-type anomaly.\footnote{In the modern vernacular, when we refer to ``anomalies'' we do not refer to  ABJ-type anomalies since they are not preserved along RG flows in general. } Note that this formalism allows for the existence of  mixed anomalies between higher form global symmetries of different degree.

We will now demonstrate two important classes of mixed anomalies involving higher form global symmetries. 
The first of these arises in $p$-form electrodynamics, but more generally appears for any pair of dual symmetries.   The second class of examples arises from the fact that discrete center symmetry affects the quantization of the instanton number in non-abelian gauge theories.

\subsubsection{Anomaly of $p$-form Electrodynamics}

Let us consider $p$-form electrodynamics.\footnote{For a more detailed discussion of $p$-form electromagnetism see \cite{Freed:2006yc,Freed:2006ya}.} This is the theory of a $U(1)$ $p$-form gauge field $A_p$ with action 
\eq{
S=\int \frac{1}{2g^2}F_{p+1}\wedge \ast F_{p+1}\quad, \quad F_{p+1}=dA_p~. 
}
Here the $U(1)$ gauge symmetry acts as 
\eq{
A_p\longmapsto A_p+d\lambda_{p-1}~.
}
And the equations of motion are given by 
\eq{
d\ast d A_p=0~,
}
with the additional constraint $d F_{p+1}=d^2 A_p=0$ called the Bianchi identity. Because of this we can define two conserved currents 
\eq{
\ast J_e=\frac{i}{g^2}\ast F_{p+1}\quad, \quad \ast J_m= \twopi{F_{p+1}}~,
}
corresponding to center and magnetic 1-form  symmetry. The first of these acts by shifting the gauge field by 
\eq{
A_p\longmapsto A_p+\alpha \lambda_p~,
}
where $\lambda_p\in \Omega^p_\IZ(M_d)$ and $\alpha \in  \IS^1  $. 
This corresponds to shifting $F_{p+1}$ by a closed $(p+1)$-form $\Lambda^{(e)}_{p+1}=\alpha d\lambda_p$. 
The magnetic 1-form  symmetry does not shift the gauge field $A_p$. 

We can now couple the theory  to background gauge fields $B_{p+1}^{(e)}\,,\,B_{d-p-1}^{(m)}$ for the center and magnetic 1-form  gauge field respectively. We are also free to add local counter terms to the theory that are only dependent on background gauge fields. With this freedom, we can choose counter terms so that the action is independent of the variation $F_{p+1}\longmapsto F_{p+1}+\Lambda^{(e)}_{p+1}$:
\eq{
S=\int \frac{1}{g^2}(F_{p+1}-B_{p+1}^{(e)})\wedge \ast(F_{p+1}-B_{p+1}^{(e)})~. 
}
Next, we can also add the coupling to the magnetic background gauge field:
\eq{
S=\int \frac{1}{g^2}(F_{p+1}-B_{p+1}^{(e)})\wedge \ast (F_{p+1}-B_{p+1}^{(e)})+\twopi{i} F_{p+1}\wedge B_{d-p-1}^{(m)}~. 
}
Now we see that the center background gauge transformation
\eq{
\delta  B_{p+1}^{(e)}=\Lambda^{(e)}_{p+1}\quad, \quad\Lambda^{(e)}_{p+1}\in H^{p+1}(M_d;\IZ)
}
 shifts the action 
\eq{
\delta_e S=i\int \Lambda^{(e)}_{p+1}\wedge \twopi{B^{(m)}_{d-p-1}}~.
}
This term implies the existence of a mixed $(p+1)$-form and $(d-p-1)$-form anomaly. This is because the above transformation cannot be canceled by any local counter term but can be canceled by inflow from the 5D anomaly term 
\eq{\label{mixedEManom}
\CI=\twopi{i} \int B_{p+1}^{(e)}\wedge dB^{(m)}_{d-p-1}~. 
}
This mixed anomaly physically implies that we cannot physically gauge a symmetry and  its magnetic dual simultaneously \cite{Gaiotto:2014kfa}. \\

\section{Discrete Global Symmetries}
\label{sec:discrete}

Discrete global symmetries are a bit more subtle than their continuous counterparts. The reason is that discrete symmetries generically do not have an associated conserved current (except when descending from a broken continuous symmetry). However, there is still a notion of background gauge fields and topological symmetry defect operators for discrete symmetries.

As in the case of a continuous $p$-form global symmetry, a theory with a discrete $p$-form global symmetry has a collection of $(d-p-1)$-dimensional symmetry defect operators $ U_g(\Sigma_{d-p-1})$ that enact the discrete symmetry transformations on $p$-dimensional charged operators $W_q(\Gamma_p)$. The existence of the SDOs can be argued from the consistency of the action of the global symmetry in a Euclidean formulation of the theory where there is no preferred time coordinate. 

Again, the symmetry defect operators obey a group multiplication law
\eq{
 U_{g_1}(\Sigma_{d-p-1})\cdot U_{g_2}(\Sigma_{d-p-1})= U_{g_1g_2}(\Sigma_{d-p-1})~,
}
which implies that a discrete $p$-form global symmetry must be abelian for $p\geq1$, 
and they act on charged operators by linking
\eq{
\big\langle\,  U_{g}(\Sigma_{d-p-1})\, W_q(\Gamma_p)\,\big\rangle=R_q(g)\cdot \big\langle W_q(\Gamma_p)\,U_g(\Sigma_{d-p-1}^\prime)\big\rangle~,
}
where $\Sigma_{d-p-1}$ links $\Gamma_p$ (while $\Sigma_{d-p-1}^\prime$ does not). 

A question that one might have is why discrete symmetries should be described by \emph{topological} operators since we have no current with which to derive this condition. There are several ways to see this fact. 

One way is the following.  Consider a (Euclidean) space time which looks like $\IR_t\times M$ and consider the (defect) Hilbert space at a fixed time $t\in \IR_t$ where we have inserted a charged operator along the time direction (plus any necessary spatial directions). Because the charged operator transforms under a global symmetry, the Hilbert space decomposes into superselection sectors corresponding to the charge of the inserted operator. In particular, the action of the symmetry group should be independent of any operation that preserves the charge. For example, even in a non-conformal theory, the action of the symmetry defect operator should be invariant under any dilatation in the space transverse to the charged operator because such a transformation (which can be enacted by insertions of the stress tensor) should act within a given super selection sector of the Hilbert space due to the conserved (discrete) global symmetry. 
 More generally, 
even in non-topological theories, the action of a symmetry defect operator should be invariant under any (nice) diffeomorphism in the transverse plane for the same reasons. The conclusion is that any operator that commutes with such an action of the stress tensor must be topological and therefore the symmetry defect operators for a discrete symmetry must be topological. 

Another way to see the topological nature of the symmetry defect operators is the following. The insertion of (networks of) symmetry defect operators corresponds to turning on a background gauge field as described in \eqref{backgroundgaugeSDO}. As we will show in the next section, discrete gauge fields -- which are defined in terms of their holonomy -- are always trivial holonomies along contractible cycles. In the language of symmetry defect operators, this is equivalent to the statement that a topologically trivial network is equivalent to the trivial network which implies that the symmetry defect operators are themselves topological. 

\bigskip While the technology of symmetry defect operators is useful to study discrete global symmetries, there are many cases where we can also make use of the paradigm of coupling to background gauge fields. 
However, the background gauge fields for discrete symmetries must themselves correspond to discrete gauge groups. So, before we continue with the study of discrete global symmetries in QFT we will first take a quick detour to briefly discuss  discrete gauge theory.

\subsection{A First Look at Discrete Gauge Theory}

Gauge theory is the theory of principal $G$-bundles with connection. Principal $G$-bundles ($P$) are continuous spaces that can be realized as a copy of $G$ fibered above a base space $X$ -- i.e.~to every point $x\in X$ we can associate a copy of $G$ (the fiber). This is often written
\eq{
\xymatrix{
G\ar[r]&P\ar[d]\\
&X
}
}
Locally, (that is in some local coordinate patch $U\subset X$), 
 we can realize the total space as a trivial product $P\big{|}_U \sim U\times G$. However, globally the bundle $P$ can be non-trivial in that the fiber can ``rotate'' as we move along a path between coordinate patches in the base space. 
 
The rotation of the fiber is captured by the connection $A$ which tells us how to connect the fiber $G$ across different coordinate patches. Given a good covering $\{U_\alpha\}$ of $X$, we can understand the connection $A$ as a collection of gauge transformations $g_{\alpha\beta}$ 
 \eq{
 g_{\alpha\beta}:U_\alpha\cap U_\beta\to G~,
}
that relates the different trivializations $P\big{|}_{U_\alpha},P\big{|}_{U_\beta}$ on the intersection $U_\alpha\cap U_\beta$.

Explicitly, when $G$ is a 0-form symmetry group and continuous the connection $A$, which is locally a 1-form, is related between different patches by $g_{\alpha\beta}$:
\eq{\label{localconn}
A_\beta=g_{\alpha\beta}^{-1}A_\alpha g_{\alpha\beta}+i g_{\alpha\beta}^{-1}dg_{\alpha\beta}~,
}
on $U_\alpha\cap U_\beta$ where $A_\alpha$ is the local expression for $A$ on $U_\alpha$. It is critical that, while the local expression for $A$ allows us to work in different coordinate patches, the definition of $A$ is independent of a choice of covering $\{U_\alpha\}$. 

Now let us try to apply this construction to discrete gauge theory. Since $G$ is discrete, $g_{\alpha\beta}$ must also be valued in a discrete group. Here, we immediately run into a problem: the patching condition on the connection \eqref{localconn} does not appear to make sense. 

Mathematically, there is a well defined construction of discrete gauge theory. However, it is a bit abstract so we will not go down this route -- instead we will take a more physical, constructive approach. For the interested reader the canonical reference for physicists   is \cite{Dijkgraaf:1989pz}.

\subsubsection* {$\mathbb{Z}_N$ Gauge Theory}

In physical situations, the most common discrete gauge theories (both dynamical and background gauge fields corresponding to discrete global symmetries) are $\IZ_N$ gauge theories. Even though $\IZ_N$ is perhaps commonplace in physics, it is also very special. The reason is that $\IZ_N\subset U(1)$ as groups. Because of this, we can actually embed $\IZ_N$ gauge theory into $U(1)$ gauge theory.

Abstractly, we can take the defining features of $\IZ_N$ connections to be the following. A $\IZ_N$ connection is defined as having holonomies valued in $\IZ_N$
\eq{
{\rm exp}\left\{i \oint_\gamma A\right\}=e^{2\pi i n/N}~,~n=0,1,...,N-1~.
}
Because the holonomy is only defined mod$_N$, $\IZ_N$ gauge theory has a $\IZ_N$ classification of Wilson lines. 

Oftentimes, the more ``mathematical'' literature uses integer quantized fields. If $A_{p+1}$ is a $(p+1)$-form gauge field, then one often works with the associated finite cohomology class  $w_{p+1}\in H^{p+1}(M;\IZ_N)$ which has the property that 
\eq{
\oint_\gamma w_{p+1}=0,1,...,N-1~.
}
In the case of a 1-form gauge field $w_{1}$ is related to $A$ by a renormalization $A=\frac{2\pi}{N} w_1\in H^1(M;2\pi \IZ_N)$. 
See Appendix \ref{app:differentialforms} for some discussion on cohomology classes and \cite{Nakahara:2003nw} for a more detailed discussion.

The restriction on the holonomy implies that that connection is (locally) flat.\footnote{Note that even though the gauge field is flat, it is not trivial. This should be apparent by the fact that the holonomy is in general non-trivial if we are considering $\IZ_N$ gauge theory on a space with non-contractible loops.}  To see this, first note that the holonomy is invariant under a continuous deformation of $\gamma$. The reason is that the holonomy is valued in a discrete group and consequently would need to jump discontinuously.  This implies that for a discrete group, the holonomy around any contractible path is trivial because any such a path can be smoothly contracted to a point where integral evaluates to zero. However, for  any contractible path $\gamma$, which we can realize  as the boundary of a disk $D_2$ ($\gamma=\partial D_2$), we can write the holonomy as an integral of the field strength 
\eq{
{\rm exp}\left\{i \oint_\gamma A\right\}={\rm exp}\left\{i \int_{D_2}F\right\}~,
} 
which implies that the curvature for any discrete gauge theory must be locally trivial.  Note that  this only shows that the field strength is trivial along contractible cycles. In general, the ``field strength'' of a discrete gauge theory is non-trivial. 

The ``field strength'' of a discrete gauge field is given by what is called the Bockstein map. The Bockstein map is an operation in discrete cohomology:\footnote{
To be precise, this is the Bockstein map associated to the short exact sequence \eq{1\longrightarrow \IZ_N \overset{\times N}{\longrightarrow} \IZ_{N^2} \overset{{\rm mod}_N}{\longrightarrow}\IZ_N\longrightarrow 1~.} 
The reason why the Bockstein can be defined in terms of the derivative of an integral lift mod$_N$ is that this short exact sequence can be lifted:
\eq{
\xymatrix{
1\ar[r]&\IZ \ar[r]^{\times N}\ar[d]_{{\rm mod}_N}&\IZ \ar[r]^{{\rm mod}_N}\ar[d]^{{\rm mod}_{N^2}}&\IZ_N\ar[r]\ar[d]&1\\
1\ar[r]&\IZ_N\ar[r]^{\times N}&\IZ_{N^2}\ar[r]^{{\rm mod}_N}&\IZ_N\ar[r]&1
}
}
so that the standard $\IZ_N$ Bockstein can be defined in terms of the ``integral Bockstein'' mod$_N$.  
See \cite{Benini:2018reh} for more details.}
\eq{
{\rm Bock}:H^p(M_d;\IZ_N)\to H^{p+1}(M_d;\IZ_N)~.
}
  However, As physicists, is perhaps easier to describe $\IZ_N$ gauge fields and their field strength as the restriction of $U(1)$ gauge theory. This embedding works as follows.

As we discussed above, the discrete gauge field is defined by having holonomies valued in $\IZ_N$. To a $\IZ_N$ connection $A_{\IZ_N}$, we can pick an associated $U(1)$ connection $\widehat{A}_{\IZ_N}$, called an ``integral lift,'' which is required to have the same holonomies:
\eq{
{\rm exp}\left\{i \oint_\gamma A_{\IZ_N}\right\}={\rm exp}\left\{i\oint_\gamma\widehat{A}_{\IZ_N}\right\}\quad \Longrightarrow \quad \oint_\gamma \frac{A_{\IZ_N}}{2\pi}=\oint_\gamma\frac{\widehat{A}_{\IZ_N}}{2\pi}~{\rm mod}_\IZ~. 
}
Since lifting the $\IZ_N$ connection to the $U(1)$ connection only requires that the periods of  $\widehat{A}_{\IZ_N},A_{\IZ_N}$ match mod$_\IZ$, it is clear that there is no unique choice of $\widehat{A}_{\IZ_N}$. However, these choices are related by large $U(1)$ gauge transformation:
\eq{\widehat{A}_{\IZ_N} \longmapsto \widehat{A}_{\IZ_N} + \Lambda_1\quad, \quad  \oint \Lambda_1 = 2\pi \IZ~.
}
Now note that since $A_{\IZ_N}\in H^1(M;2\pi\IZ_N)$, we have $dA_{\IZ_N}=0$ in $\IZ_N$-valued cohomology. However, a general $U(1)$ gauge field need not be closed and so the choice of lift $\widehat{A}_{\IZ_N}$ is in general not closed. This means that we can define a notion of curvature for $A_{\IZ_N}$ in terms of the field strength of its integral lift:
\eq{
\frac{\widehat{F}_{\IZ_N}}{2\pi}:=\frac{d\widehat{A}_{\IZ_N}}{2\pi}\in H^2(M;\IZ)~.
}
This data is independent of the choice of lift because they are related by large $U(1)$ gauge transformations which does not affect the associated field strength. However, for a $\IZ_N$ gauge field, an integer lift has only a $\IZ_N$ amount of data, and so the field strength is only meaningful mod$_N$. This allows us to define the discrete field strength:
\eq{
\frac{F_{\IZ_N}}{2\pi}:=\frac{d\widehat{A}_{\IZ_N}}{2\pi}~{\rm mod}_N~. 
}
As it turns out, this is exactly the definition of the Bockstein map for the associated $w_1\in H^1(M;\IZ_N)$:
\eq{
{\rm Bock}(w_1)=\frac{F_{\IZ_N}}{2\pi}=\frac{1}{N}d\widehat{w}_1~{\rm mod}_N~,
}
where $\widehat{w}_1\in \Omega^1_\IZ(M)$ is an ``integral lift'' of $w_1$ which is a 1-form with integral periods on $M$. Note that the choice of $\widehat{w}_1$ is equivalent to the choice of $\widehat{A}_\IZN$ where the large $U(1)$ gauge transformations correspond to shifting the choice of $\widehat{w}_1\longmapsto \widehat{w}_1+N\Lambda_1$.

There are two more important parts of the literature on discrete gauge theory that we think is important to understand. The first is the cup product: $\smile$. This is the generalization of the wedge product of differential forms to discrete cohomology classes. It has the same anti-symmetry properties and can be treated as the wedge product for all intents and purposes in these notes
\eq{
\alpha_p\smile \beta_q\in H^{p+q}(M;\IZ_N)\quad, \quad \alpha_p\smile \beta_q=(-1)^{pq}\beta_q\smile \alpha_p~.
}
Another important construction in discrete gauge theory (and to understanding the literature) is the Pontryagin square. Loosely speaking, the Pontryagin square is a discrete version of operation that sends $F \mapsto F \wedge F$. More precisely, in the setting appropriate for discrete gauge theory, the Pontryagin square is a map:
\eq{
\CP:H^{2k}(X;\IZ_N)\longmapsto \begin{cases}
H^{4k}(X;\IZ_N) & N~{\rm odd}\\
H^{4k}(X; \IZ_{2N}) & N~{\rm even}
\end{cases}
}
For $N$ odd, 
\eq{\CP(v)=\widehat{v}\wedge \widehat{v}~~{\rm mod}_N~,}
and for $N$ even, 
\eq{
\CP(v)=\widehat{v}\wedge \widehat{v}~~ {\rm mod}_{2N}~,
}
where $\widehat{v}$ is an integer lift of $v\in H^{2k}(X;\IZ_N)$. Here there is an enhancement from $\IZ_N\to \IZ_{2N}$ for the case of $N$ even because $\widehat{v}\wedge \widehat{v}$ is  actually invariant under choice of integer lift mod$_{2N}$ as opposed to mod$_N$. This follows from the variation under the shift:
\eq{
\widehat{v}\longmapsto \widehat{v}+ N\Lambda_{2k}\quad, \quad \widehat{v}\wedge \widehat{v}\longmapsto \widehat{v}\wedge \widehat{v}+2N \widehat{v}\wedge \Lambda_{2k}+N^2\Lambda_{2k}\wedge \Lambda_{2k}~. 
}
Since $N\in 2\IZ$, we see that $2N,N^2\in 2N\IZ$ and therefore we see that $\widehat{v}\wedge \widehat{v}$ mod$_{2N}$ is invariant under the choice of lift and defines an element in $H^{4k}(X;\IZ_{2N})$. For a more detailed discussion of the Pontryagin square, see \cite{Kapustin:2013qsa,Hsin:2020nts}.\footnote{In particular, there exists a definition of the Pontryagin square operation without relying on picking integer lifts. This requires the use of cup-i products which we will not discuss here -- see \cite{Hsin:2020nts} for example for more discussion. }

The discussion of this section are summarized in the following table:

\begin{center}
\begin{tabular}{|c|c|}
\hline
$\IZ_N$ Gauge Theory & $\IZ_N\subset U(1)$ Gauge Theory\\
\hline 
$w_1=\frac{2\pi}{N} A_{\IZ_N}$ Connection & $\widehat{A}_{\IZ_N}$ $U(1)$ connection with\\
&$\oint_\gamma A_{\IZ_N}=\oint_\gamma\widehat{A}_\IZN~{\rm mod}_{2\pi}$\\
\vspace{-0.3cm}\\
Discrete Field strength: Bock$(w_1)\in H^2(M;\IZ_N)$ & Field strength $\widehat{F}_\IZN=d\widehat{A}_\IZN$ with \\
&Bock$(w_1)=\frac{\widehat{F}_\IZN}{2\pi}~{\rm mod}_N$\\\hline
\end{tabular}
\end{center}

\subsection{BF Gauge Theory}

We would now like to demonstrate one way to describe $\IZ_N$ $p$-form gauge theory called BF theory \cite{Horowitz:1989ng,Maldacena:2001ss,Banks:2010zn}. In words, BF gauge theory is a constrained $U(1)$ gauge theory so that on $\IZ_N$ gauge fields contribute to the path integral. Consequently, the field strengths are (generically) flat and Wilson lines only have a well defined charge mod$_N$. 

Let us consider a $d$-dimensional theory of a $U(1)$ $p$-form gauge field $A_p$ and a  $U(1)$ $(d-p-1)$-form gauge field $B_{d-p-1}$. We consider the action
\eq{\label{eq:BF theory}
S=\frac{iN}{2\pi} \int B_{d-p-1} \wedge dA_p~.
}
The action is invariant under $(p-1)$-form and $(d-p-2)$-form gauge transformations 
\eq{\label{eq:BF theory_action}
A_p\longmapsto A_p+d\Lambda_{p-1}\quad, \quad B_{d-p-1}\longmapsto B_{d-p-1}+d\Lambda_{d-p-2}~. 
}
Here, the gauge parameters $d\Lambda_{q}\in H^{q+1}(M_d;\IZ)$ for $q=(p-1)$ or $(d-p-2)$ have integral periods: $\oint_{\Sigma_{q+1}}\frac{d\Lambda_{q}}{2\pi}\in \IZ$ for any closed $(q+1)$-manifold $\Sigma_q$. 

Note that this theory is topological. This is clear because the action can be written in a way that is metric independent (i.e.~in differential form notation without using the Hodge star operation). This implies that there can be no propagating degrees of freedom. This follows from the constraints imposed by the equations of motion 
\eq{
N\,\frac{dA_p}{2\pi}=0\quad, \quad N\,\twopi{dB_{d-p-1}}=0~,
}
which restricts the gauge fields $A_p,B_{d-p-1}$ to be flat. 

Before we analyze higher-form (discrete) symmetries of the theory, we first explain how the action in \eqref{eq:BF theory_action} describes $\IZ_N$ gauge theory. We do this in two different ways. First, in Section~\ref{subsubsec:Wilson_ops_BF}, we will directly analyze the BF theory to show that the path integral restricts to a sum over $\IZ_N$ gauge fields. Then, in Section~\ref{subsubsec:BF from AHM} we derive the BF action (for $p=1$) from the low energy limit of an Abelian Higgs Model with a charge $N$ Higgs field. In this model, the gauge symmetry is spontaneously broken $U(1) \to \IZ_N$ which gives a simpler way to infer the $\IZ_N$ nature of BF theory.

\subsubsection{Direct Analysis of BF theory}
\label{subsubsec:Wilson_ops_BF}

While the equations of motion imply that gauge fields in BF theory are flat, this nevertheless does not mean that the theory is trivial. The reason is that BF theory admits non-trivial gauge invariant operators. To see this, we first note that this theory projects out all field configurations except for those for which the holonomies are $\IZ_N$-valued:
\eq{\label{flatholonomy}
N\oint \frac{A_p}{2\pi}~,~N\oint \twopi{B_{d-p-1}}\in \IZ~.
}
This follows from the fact that $\oint\frac{dA_p}{2\pi},\oint\frac{dB_{d-p-1}}{2\pi}\in \IZ$ (recall $A_p$ and $B_{d-p-1}$ are $U(1)$ gauge fields) so that the path integral acts as a (discrete) Fourier series transformation. The rough idea is captured by computing the sum+integral:
\eq{
Z[f]=\sum_{k\in \IZ} \int dxe^{ 2\pi i N\, k x}f(x)=\sum_{x\in \frac{1}{N}\IZ}f(x)~,
}
where $f(x)$ is a real function. Here we can think of $Z[f]$ is like a correlation function of $f$ which is dependent on one of the gauge fields $x$. The idea of this integral is similar to that of (a rescaled) Poisson resummation.

Similarly, in the BF partition function, the sum over $\frac{dA_p}{2\pi}$ fluxes cancels all contributions from field configurations from $B_{d-p-1}$ except those that satisfy \eqref{flatholonomy} -- thus localizing to the path integral over $\IZ_N$ gauge fields:
\eq{
Z&=\int[dA][dB]\,{\rm exp}\left\{\frac{iN}{2\pi}\int B_{d-p-1}\wedge dA_p\right\}\\
&=\sum_{[dA_p]\in H^{p+1}(M_d;\IZ)}\int [dA][dB]\, {\rm exp}\left\{2\pi i \int \frac{dA_p}{2\pi}\wedge \frac{NB_{d-p-1}}{2\pi}\right\}=\int_{\eqref{flatholonomy}} [dB]
}
This means that the theory has a collection of gauge invariant generalized Wilson operators
\eq{
W_n(\Gamma_p)=e^{i n\int_{\Gamma} A_p}\quad, \quad \CW_m(\widetilde\Gamma_{d-p-1})=e^{im \int_{\widetilde\Gamma }B_{d-p-1}}~,
}
that measure these non-trivial $\IZ_N$  gauge fields that are summed over in the path integral. 
Since $dA_p=dB_{d-p-1}=0$, these operators are topological - i.e.~correlation functions with operators $W_n(\Gamma),\CW_m(\widetilde\Gamma)$ are invariant under smooth deformations of $\Gamma,\widetilde\Gamma$. 

Let us consider the insertion of a $W_n(\Gamma_p)$ in the path integral.
\beq
\int [ dA ] [ dB] \; e^{i n \int_\Gamma A_p} \; e^{\frac{iN}{2\pi} \int A_p \wedge d B_{d-p-1}}
\eeq
Since $A_p$ is an abelian gauge field, we can exchange the insertion of $W_n$ in the path integral for a term in the action 
\eq{
S\longmapsto S+i n \int A_p\wedge \delta^{(d-p)}(x\in \Gamma_p)~,
}
where $\delta^{(d-p)}(x\in \Gamma_p)$ is the $(d-p)$-form delta function that is defined as \footnote{Again, this is the Thom class for the embedding $\Gamma_p\hookrightarrow M_d$.}
\eq{
\int_{M_d}\alpha_p\wedge \delta^{(d-p)}(\Gamma_p)=\int_{\Gamma_p}\alpha_p~,\quad \alpha_p\in \Omega^p(M_d)~.
}  
Alternatively, $\delta^{(d-p)}(\Gamma_p)$ satisfies
\beq
\int_{\Sigma_{d-p}} \delta^{(d-p)}(\Gamma_p) = \left\lbrace
\begin{array}{ll}
1 \;\;\; \text{if} \; \Sigma_{d-p} \; \text{transversely intersects} \; \Gamma_p \; \text{once} \\
0 \;\;\; \text{if} \; \Sigma_{d-p} \; \text{does not intersects} \; \Gamma_p.
\end{array}
\right.
\label{eq:Thom class delta}
\eeq
See Figure \ref{fig:intersection_linking_3d} and related discussion in Section~\ref{subsubsec:Maxwell Theory}. This additional term changes the equations of motion for $A_p$:
\eq{\label{eq:BF_change of eom by insertion}
N\twopi{dB_{d-p-1}}=n\ \delta^{(d-p)}(x\in \Gamma_p)\quad \longrightarrow \quad \twopi{dB_{d-p-1}}=\frac{n}{N}\ \delta^{(d-p)}(x\in \Gamma_p)~. 
}
We see that the operator $W_n(\Gamma_p)$ induces a fractional (magnetic) charge for $B_{d-p-1}$, hence non-trivial holonomy of the Wilson operator $\CW_m(\widetilde\Gamma_{d-p-1})=e^{im \int_{\widetilde\Gamma }B_{d-p-1}}$. However, when $n\geq N$ we can perform a large $U(1)$ gauge transformation of $B_{d-p-1}$ to reduce $n$ by multiples of $N$. 
 In particular, we can take
\eq{
\delta B_{d-p-1}=q\ \omega_{d-p-1}(\Gamma_p)\quad, \quad q\in \IZ~,
} 
where $\omega_{d-p-1}(\Gamma_p)$ is the global angular form surrounding $\Gamma_p$ which is heuristically the volume form on the (foliated) concentric $(d-p-1)$-spheres surrounding $\Gamma_p$.\footnote{ Here we do not need to worry about the fact that $\omega_{d-p-2}(\Gamma_p)$ is singular on $\Gamma_p$. The reason is that regularizing the Wilson operators requires excising a small tubular region $T_\epsilon$  around $\Gamma_p\subset T_\epsilon$  since the fields are singular there.   
In the space $M_d\backslash T_\epsilon$ the above gauge transformation is well defined and thus we find that the magnetic charge is only defined mod$_N$ \cite{Maldacena:2001ss}.} Alternatively, the charge $N$ Wilson operators do not contribute to correlation functions due to the localization of the path integral on $\IZ_N$ gauge field configurations and are consequently ``trivial'' line operators as they are ``invisible'' in the theory \cite{Banks:2010zn}.

Therefore, we see that BF-theory is indeed a $\IZ_N$ gauge theory by restricting the path integral to $\IZ_N$ gauge fields. This fact is also reflected in the fact that  the charge of Wilson operators can always be reduced mod$_N$.  

Before we end this section, we discuss the implication of \eqref{eq:BF_change of eom by insertion} on a correlation function of Wilson operators. This equation  combined with the defining property of $\delta^{(d-p)}(\Gamma_p)$  in \eqref{eq:Thom class delta} shows that the insertion of Wilson operator turns on non-trivial $\IZ_N$ charge for $B_{d-p-1}$ field if and only if the world volume of the inserted Wilson operator $\Gamma_p$ intersects non-trivially with $\Sigma_{d-p}$, or equivalently, if the world volume $\Gamma_p$ links non-trivially with the world volume $\partial \Sigma_{d-p} = \widetilde\Gamma_{d-p-1}$ of $B_{d-p-1}$. The exponentiation of this identity  therefore implies that  the correlation function is given by 
\beq
\left\langle W_n(\Gamma_p) \;  \CW_m(\widetilde\Gamma_{d-p-1}) \right\rangle = e^{\frac{2\pi i}{N} n m \text{Link} (\Gamma, \widetilde\Gamma)}~.
\eeq
This will be an important formula when we discuss global symmetries of BF theory in Section~\ref{subsubsec:BF_global symmetry}.

\subsubsection{$\IZ_N$ Gauge Theory from Abelian Higgs Model}
\label{subsubsec:BF from AHM}

The BF theory has a natural description as the IR limit of a particularly simple local quantum field theory: the charge $N$ Abelian Higgs model. Here our discussion will follow \cite{Banks:2010zn,Nakahara:2003nw}.

For concreteness, we consider a $4d$ $U(1)$ gauge theory with a complex scalar field $\Phi$ of charge $N$. This will lead us to the BF theory with $p=1$ and $d=4$.

The  4$d$ Abelian Higgs model we wish to study is described by the Lagrangian 
\eq{
\CL=(d\Phi-i N A_1\Phi)\wedge \ast (d\Phi-i N A_1\Phi)+\frac{1}{2g^2}F_2\wedge \ast F_2 -V(\Phi)~,
}
where $A_1$ is the $U(1)$ gauge field with field strength $F_2=dA_1$. 
Let us choose a potential $V(\Phi)$ such that $\Phi$ condenses $\langle \Phi\rangle=\Lambda$ and spontaneously breaks the $U(1)$ gauge symmetry. 

Because the Higgs field has charge $N$, the Higgs field transforms under $U(1)$ gauge transformations as 
\eq{
e^{i \phi \hat{Q}}\cdot \Phi= e^{ i N \phi}\Phi~,
}
where $\hat{Q}$ is the generator of gauge transformations. 
Consequently, the Higgs field is invariant under $\IZ_N \subset U(1)$ gauge transformations (i.e. $\phi=\frac{2\pi k}{N}$) and thus the condensation spontaneously breaks $U(1)$ while preserving a $\IZ_N\subset U(1)$ gauge symmetry. This in turn implies that in the deep IR the massive gauge boson and radial Higgs modes decouple and the theory is that of $\IZ_N$ gauge theory. Our task is to show that this ``remnant'' $\IZ_N$ gauge theory is nothing but the topological $4d$ BF theory \eqref{eq:BF theory_action} with $p=1$. 

Below the energy scale $\Lambda$, the radial mode of $\Phi$ is frozen out  and the low energy modes of $\Phi$ can be parametrized by $\Phi=\Lambda\, e^{ i \varphi}$ where now $\varphi$ is a periodic scalar field $\varphi\sim \varphi+2\pi$. If we substitute this into the Lagrangian we find 
\eq{
\CL=\Lambda^2(d\varphi-N A_1)\wedge \ast (d\varphi - N A_1)+\frac{1}{2g^2}F_2\wedge \ast F_2~.
}
Now the equation of motion for $\varphi$ is given by 
\beq
d \ast  \left( d \varphi - N A_1 \right) = 0~.
\eeq 

We now will proceed by dualizing $\varphi$. Morally, the idea of dualization is that of a Fourier transform. This is captured by the
 familiar Gaussian integral:
\beq
Z = \int d x\, e^{-a x^2 + ib x} = \int dx \,e^{-a \left( x -i \frac{b}{2a} \right)^2} e^{-\frac{b^2}{4a}} = \sqrt{\frac{\pi}{a}} \; e^{-\frac{b^2}{4a}}~.
\eeq
We view this as an analog to path integral where $x$ is a dynamical field with coupling strength $a$. We then add a term $ibx$, where $b$ is the dual field.\footnote{
More precisely, if we want to identify $x^2$ and $b^2$ as kinetic terms, we should think of $x,b$ as the momentum/field strength for some dual pair of dynamical fields. 
}
 Integrating out $x$  leads to the dual description with the action $S = \frac{b^2}{4a}$; it is again a Gaussian theory of the dual field $b$ with $a \to 1/a$.

With this preparation, we dualize by introducing the dual 2-form field $B_2$:
\beq
 \CL=\Lambda^2(d\varphi-N A_1)\wedge \ast (d\varphi - N A_1) + \frac{i}{2\pi} dB_2\wedge  d\varphi +  \frac{1}{2g^2}F_2\wedge \ast F_2~.
\eeq
Note that although $d\varphi$ is not gauge invariant, the coupling between $\varphi$ and $B_2$ is gauge invariant   due to the fact that $B_2$ is a $U(1)$ gauge field so that the field strength $\frac{H_3}{2\pi} =\frac{dB_2}{2\pi}$ is a closed, integer quantized 3-form. This means that it is invariant under``small'' gauge transformations ($d\varphi\mapsto d\varphi+d\lambda_0$) because $d^2B_2=0$ and it is invariant under ``large'' gauge transformations ($d\varphi\mapsto d\varphi+ \Lambda_1$ for $\frac{\Lambda_1}{2\pi}\in H^1(M;\IZ)$) because the variation 
\eq{
\delta S=\frac{i}{2\pi}\int dB_2\wedge \Lambda_1\in2\pi i\IZ~,
}
is always a  trivial phase.

Now we can perform the Gaussian integral over $\varphi$ by first completing the square
\eq{
 \CL=&\Lambda^2\left(d\varphi-N A_1+ i \frac{\ast dB_2}{4\pi \Lambda^2}\right)\wedge \ast \left(d\varphi - N A_1+ i \frac{\ast dB_2}{4\pi \Lambda^2}\right) +\frac{1}{16\pi^2\Lambda^2} dB_2\wedge \ast dB_2\\
 &+ \frac{iN}{2\pi} dB_2\wedge  A_1 +  \frac{1}{2g^2}F_2\wedge \ast F_2~,
}
and then performing the integral over $\varphi$, which results in the theory 
\eq{
\CL=\frac{4\pi^2}{\Lambda^2}dB_2\wedge \ast dB_2+ \frac{iN}{2\pi} dB_2\wedge  A_1 +  \frac{1}{2g^2}F_2\wedge \ast F_2~.
} 
Note that this dualization can be undone -- much like a Fourier transform.

Now we can take the IR limit $\Lambda \to \infty$. This trivializes the first term and leads to an IR theory which is described by 
\beq
\CL =  \frac{iN}{2\pi} B_2 \wedge d A_1 + \frac{1}{2g^2} F_2 \wedge \ast  F_2 ~.
\eeq
Then, the equation of motion of $B_2$ imposes $d A_1 = 0$ which means that the gauge kinetic term is trivial, and we are left with BF theory.

Since we can track the fields of the BF theory to the UV Abelian Higgs model, we can also identify the defect operators of the BF theory as smooth defects coming from the Abelian Higgs model. 

Following the reduction above, it is clear that the gauge field $A_1$ of the BF theory comes from  the $U(1)$ gauge field of the Abelian Higgs model. This means that  we can directly identify the $\IZ_N$-Wilson line of the BF theory with the Wilson line of the $U(1)$ gauge field of the Abelian Higgs model. Here the $\IZ_N$ reduction occurs because the charge $N$ scalar field $\Phi$ can dynamically screen the Wilson line charge mod$_N$: i.e. there is a $\IZ_N^{(1)}$ 1-form center symmetry of the UV theory which is preserved along the RG flow. 

We also would like to match the vortex of the BF theory, which is the Wilson surface of the $B_2$-field. Since $B_2$ arises as the dual field of $\varphi$, the Wilson surface of $B_2$ corresponds to the winding field configuration of $\varphi$: i.e. the Abelian Higgs model vortex.

From the UV physics, the physical reason that the Wilson line and the BF-vortex have non-trivial winding is that the $\varphi$-vortex is supported by a magnetic flux going through its core. This leads to a non-trivial Aharanov-Bohm phase that is measured by any linking Wilson line.

\subsubsection{Discrete Global Symmetries of BF Theory}
\label{subsubsec:BF_global symmetry}

Now that we understand some aspects of discrete gauge theory, we would like to move on to study discrete global symmetries.  

Unfortunately, there are no currents associated to discrete symmetries. This means that we cannot proceed generally as with continuous global symmetries. Rather, we will have to study discrete global symmetries on a somewhat case-by-case basis. Because of this, one often has to analyze the discrete symmetries of a theory by using a combination of symmetry defect operators and background gauge fields. 
 
Let us consider $d$-dimensional $p$-form BF theory.\footnote{Note that this example contains $3d$ Chern-Simons theory for $d=3$ and $p=1$.}
As we will see, this theory has a $\IZ_N^{(p)}\times \IZ_N^{(d-p-1)}$ higher form global symmetry.\footnote{
These symmetries are the discrete version of center and magnetic 1-form  symmetry for a discrete gauge group. This can be seen by noting  that the center of $\IZ_N$ is $\IZ_N$ (since it is abelian) and hence the center symmetry can be realized as the $\IZ_N^{(p)}$ $p$-form symmetry. Similarly, the  magnetic 1-form  symmetry can be realized by the Pontryagin dual
 $\IZ_N^\vee\cong \IZ_N$ which in this theory we can realize as the $\IZ_N^{(d-p-1)}$ $(d-p-1)$-form symmetry. The Pontryagin dual of a group $G$, is denoted $G^\vee$ and is defined to be the group of homomorphisms from $G^\vee=\{f:G\to U(1)\}$. } These symmetries can be seen by realizing that the BF theory action \eqref{eq:BF theory_action} is invariant under 
\bea
&& \IZ_N^{(p)} \;\;\;\;\;\;\; : \; \; A_p \longmapsto A_p + \frac{1}{N} \epsilon_p, \;\;\; \oint \epsilon_p = 2\pi \IZ \label{eq:BF_ZNp}\\
&& \IZ_N^{(d-p-1)} : \; \; B_{d-p-1} \longmapsto B_{d-p-1} + \frac{1}{N} \epsilon_{d-p-1}, \;\;\; \oint \epsilon_{d-p-1} = 2\pi \IZ \label{eq:BF_ZNdp1}
\eea

This can also be seen explicitly from the equations of motion: 
\eq{
d\frac{N\, A_p}{2\pi}=0\quad, \quad d\frac{N\, B_{d-p-1}}{2\pi}=0~.
}
Here, we can think of these equations as conservation equations for a $(d-p)$-form and $(p+1)$-form conserved ``current'' respectively:
\eq{
\ast j_3=\frac{N\,A_p}{2\pi}\quad, \quad \ast J_2=\frac{N\,B_{d-p-1}}{2\pi}~.
}
However, these currents are clearly not gauge invariant because they are literally gauge fields. 
Let us nevertheless process and define  $\IZ_N$-valued symmetry defect operators as follows.
\eq{
&\CU_{2\pi k/N}(\widetilde\Sigma_p)=e^{\frac{2\pi i k}{N}\oint_{\widetilde\Sigma} \frac{N\,A_p}{2\pi} }=e^{i k \oint_\Gamma A_p} ~,\\
&U_{2\pi k/N}(\Sigma_{d-p-1})=e^{\frac{2\pi i k}{N}\oint_\Sigma \frac{N\,B_{d-p-1}}{2\pi}}=e^{i k \oint_\Sigma B_{d-p-1}}~.
}
The point is that even though the currents are not well-defined, the exponentiation of the associated charge operators, that is, the associated SDOs, are well-defined, but only for $\IZ_N \subset U(1)$.  
We see that the symmetry defect operator for $\IZ_N^{(p)}$ that shifts $A_p$ is the standard Wilson operator constructed out of $B_{d-p-1}$. Similarly, the symmetry defect operator for $\IZ_N^{(d-p-1)}$ that shifts $B_{d-p-1}$ is the Wilson operator of $A_p$. This discussion makes it also clear that charged operators and associated transformation properties are given by
\eq{
& \IZ_N^{(p)} \;\;\;\;\;\;\; : \; \; W_n(\Gamma_p)=e^{i n \int_{\Gamma} A_p} \mapsto e^{\frac{2\pi i n}{N}} W_n (\Gamma_p) \\
& \IZ_N^{(d-p-1)} : \; \;  \CW_m(\widetilde\Gamma_{d-p-1})=e^{i m \int_{\widetilde\Gamma} B_{d-p-1}} \mapsto e^{\frac{2\pi i m}{N}} \CW_m(\widetilde\Gamma_{d-p-1}).
}

Our discussion so far can be summarized as
\begin{center}
\begin{tabular}{|l|l|l|}
\hline
Symmetry & Charged Operator & SDO\\
\hline 
$\IZ_N^{(p)}$ & $W_n(\Gamma_p)=e^{i n \int_{\Gamma} A_p}$ & $U_g(\Sigma_{d-p-1})=e^{i k \int_\Sigma B_{d-p-1}}$ \\
$\IZ_N^{(d-p-1)}$ & $\CW_m(\widetilde\Gamma_{d-p-1})=e^{i m \int_{\widetilde\Gamma} B_{d-p-1}}$ & $\CU_g(\widetilde\Sigma_p)=e^{i k \int_{\widetilde\Sigma} A_p}$\\
\hline
\end{tabular}\begin{tabular}{l}
$g=e^{2\pi i k /N}\in \IZ_N$\\
$k=0,1,...,N-1$\\
\end{tabular}
\end{center}
Notice here that $k$ is constrained to lie in $\IZ_N$ by requiring that the $U_g$ and $\CU_g$ be invariant under $U(1)$ gauge transformations of $B_{d-p-1}$ and $A_p$ respectively. Because of this, there is an identification of operators $W_n\Leftrightarrow \CU_g$ and $\CW_m\Leftrightarrow U_g$. This ``duality'' is a result of the fact that $W_n$ sources fractional charge for the $B$-field and $\CW_m$ sources fractional charge for the $A$-field. 

We would now like to demonstrate the action of the above $\IZ_N$ symmetries. First consider the insertion of $W_n(\Gamma_p)$. Pick a manifold $\Sigma_{d-p-1}$ that links $\Gamma_p$ and a manifold $\Sigma_{d-p-1}'$ that is homotopic to $\Sigma_{d-p-1}$ so that the manifold $\widehat\Sigma_{d-p}$ that is swept out by the deformation intersects $\Gamma_p$. We would now like to compute the correlation function 
\eq{
\big\langle \,U_g(\Sigma_{d-p-1})\,W_n(\Gamma_p)\,\big\rangle=?
}
In fact, we have already determined this in Section~\ref{subsubsec:Wilson_ops_BF} and here we repeat the argument for completeness sake.
Now recall that the insertion of $W_n(\Gamma_p)$ modifies the equation of motion for $A_{p}$ to include a contact term:
\eq{
\twopi{dB_{d-p-1}}=\frac{n}{N}\ \delta^{(d-p)}(x\in \Gamma_p)~.
}
 This means that as we deform $\Sigma_{d-p-1}\to \Sigma'_{d-p-1}$ we pick up a contact term when it intersects $\Gamma_p$ that is proportional to 
\eq{
\big\langle\, U_g(\Sigma_{d-p-1})\,W_n(\Gamma_p)\,\big\rangle&=e^{2\pi i k n /N}\big\langle \,W_n(\Gamma_p)\,U_g(\Sigma_{d-p-1}')\,\big\rangle\\[0.1cm]
&=g^n\big\langle\, W_n(\Gamma_p)\,U_g(\Sigma_{d-p-1}')\,\big\rangle~,
}
where $g=e^{2\pi i k /N}\in \IZ_N$. 

Similarly we can consider the insertion of $\CW_m(\Gamma_{d-p-1})$ and a linking $\Sigma_p$ that can be deformed to $\Sigma_p'$ which does not link $\Gamma_{d-p-1}$. For similar reasons, we see that the insertion of the $\CW_m$ modifies the equation of motion of $A_p$ to have a contact term:
\eq{
\twopi{dA_p}=\frac{m}{N}\ \delta^{(p+1)}(\Gamma_{d-p-1})~, 
}
so that deforming the symmetry defect operator through $\Gamma_{d-p-1}$ generates a phase 
\eq{
\big\langle\, \CU_g(\Sigma_{p})\CW_m(\Gamma_{d-p-1})\,\big\rangle\,&=e^{2\pi i k m/N}\big\langle\, \CW_m(\Gamma_{d-p-1})\,\CU_g(\Sigma_{p}')\,\big\rangle\\[0.1cm]
&=g^m\big\langle\, \CW_n(\Gamma_{d-p-1})\,\CU_g(\Sigma_{p}')\,\big\rangle~,
}
where $g=e^{2\pi i k /N}\in \IZ_N$. 

These symmetries are in a sense ``dual'' because the action is linear in $A,B$ so that the charge of $W_n$ is measured by an integral of $B$ and the charge of $\CW_m$ is measured by an integral of $A$. 

\subsubsection{Anomalies of BF Theory}

The symmetries of BF theory also have a mixed anomaly. Here we consider $\IZ_N$ $p$-form BF theory in $d$-dimensions:
\eq{
S=\twopi{i N} \int B_{d-p-1}\wedge dA_p~.
}
 As discussed, this theory has  $\IZ_N^{(p)}\times \IZ_N^{(d-p-1)}$ global symmetry. Let us introduce the background gauge fields $C_{p+1},\tilde{C}_{d-p}$ for these symmetries which satisfy:
 \eq{
\twopi{ NC_{p+1}}\in H^{p+1}(M_d;\IZ)\quad, \quad \twopi{N\tilde{C}_{d-p}}\in H^{d-p}(M_d;\IZ)~.}
  Now we see that the action is not invariant under background gauge transformations 
 \eq{
&\delta_p C_{p+1}=d\Lambda_{p}\qquad\qquad~, \quad \delta_p A_p=\Lambda_p~,\\
&\delta_{d-p-1} C_{d-p}=d\Lambda_{d-p-1}\quad, \quad \delta_{d-p-1} B_{d-p-1}=\Lambda_{d-p-1}~,
 }
 for $\twopi{d\Lambda_q}\in H^{q+1}(M_d;\IZ)$. We find now that 
 \eq{
\delta S=i N\int\left( \Lambda_{d-p-1}\wedge dA_p+B_{d-p-1}\wedge d\Lambda_p+\Lambda_{d-p-1}\wedge d\Lambda_{p}\right)~. 
 }
 As it turns out, the best we can do is eliminate the first two by adding the couplings:
 \eq{
 S=...-\twopi{iN}\int B_{d-p-1}\wedge C_{p+1}-\twopi{i N}\int \tilde{C}_{d-p}\wedge A_p~,
 } 
 which are gauge invariant under $U(1)$ gauge transformations. However, the final term in the above variation can only be canceled by inflow and hence indicates the anomaly:
 \eq{
 \CA=\twopi{i N}\int \tilde{C}_{d-p}\wedge C_{p+1}~. 
 }

\subsubsection{Other Descriptions of BF Theory}
\label{subsubsec:other descriptions}

There are more than one ways to describe the BF theory and as usual different descriptions are useful to manifest some features which may not appear as clear otherwise. 

First, we dualize the $A_p$ field as described above:
\eq{
\CL =& \frac{iN}{2\pi} B_{d-p-1} \wedge d A_p + \frac{i}{2\pi} d \tilde{A}_{d-p-2} \wedge dA_{p} \\
=& \frac{i}{2\pi} dA_{p} \wedge \left( d \tilde{A}_{d-p-2} + N B_{d-p-1} \right)~.
\label{eq:BF_dual_description_1}
}
 This presentation provides an alternate way to see that the $(d-p-2)$-form $U(1)$ gauge symmetry with the gauge field $B_{d-p-1}$ is broken to $\IZ_{N}^{(d-p-2)}$. Here it is Higgsed by the (dual) matter field $\tilde{A}_{d-p-2}$. In fact, the gauge symmetries are given by
\eq{
& B_{d-p-1} \mapsto B_{d-p-1} + d \Lambda_{d-p-2} \\
& \tilde{A}_{d-p-2} \mapsto \tilde{A}_{d-p-2} + d \lambda_{d-p-3} + N \Lambda_{d-p-2} 
}
Here, we see that the dual field $\tilde{A}_{d-p-2}$ comes with its own $(d-p-3)$-form $U(1)$ gauge invariance, which is often the case when one performs duality transformations.

The equation of motion for $A_{p}$ imposes $d \tilde{A}_{d-p-2} + N B_{d-p-1} = 0$. This can be used to show why the $(d-p-1)$-form global symmetry is $\IZ_N$:
\eq{
\left( \CW_1(\widetilde\Gamma_{d-p-1}) \right)^N = e^{i \oint_{\widetilde\Gamma} N B_{d-p-1}} = e^{i \oint_{\widetilde\Gamma} d \tilde{A}_{d-p-2}} = 1.
}
Also, from \eqref{eq:BF_dual_description_1} we see that $\IZ_N^{(d-p-1)}$ global symmetry acts as
\eq{
&B_{d-p-1} \mapsto B_{d-p-1} + \frac{1}{N} \epsilon_{d-p-1}, \;\; \oint \epsilon_{d-p-1} = 2\pi \IZ \\
& \tilde{A}_{d-p-2} \mapsto \tilde{A}_{d-p-2} - \tilde{\epsilon}_{d-p-2}
}
where locally $d \tilde{\epsilon}_{d-p-2} = \epsilon_{d-p-1}$.

We can also provide a different alternative description by  dualizing the  $B_{d-p-1}$ field:
\eq{
\CL =& \frac{iN}{2\pi} A_p \wedge d B_{d-p-1} + \frac{i}{2\pi} d \tilde{B}_{p-1} \wedge d B_{d-p-1} \\
=& \frac{i}{2\pi} d B_{d-p-1}  \wedge \left( d \tilde{B}_{p-1} + N A_p \right)~.
\label{eq:BF_dual_description_2}
}
Here, we see that the $p$-form $U(1)$ gauge symmetry of $A_p$ is Higgsed, breaking $U(1)^{(p)}\to \IZ_N^{(p)}$ by the (dual) matter field $\tilde{B}_{p-1}$. The gauge symmetries are
\eq{
& A_p \mapsto A_p + d \Lambda_{p-1} \\
& \tilde{B}_{p-1} \mapsto \tilde{B}_{p-1} + d \lambda_{p-2} + N \Lambda_{p-1} 
}
Again, we see the appearance of $(p-2)$-form dual gauge symmetry of $\tilde{B}$. 
The equation of motion for $ B_{d-p-1}$ sets $d \tilde{B}_{p-1} + N A_p = 0$, which can be used to show that $p$-form global symmetry that shifts $A_p$ is indeed $\IZ_N$.
\beq
\left( W_1 (\Gamma_p) \right)^N = e^{i \oint_\Gamma N A_p} = e^{i \oint d \tilde{B}_{p-1}} = 1.
\eeq
From \eqref{eq:BF_dual_description_2} we learn that $\IZ_N^{(p)}$ global symmetry acts as
\eq{
 A_p \longmapsto &A_p + \frac{1}{N} \epsilon_p, \;\; \oint \epsilon_p = 2\pi \IZ \\
 \tilde{B}_{p-1} \longmapsto &\tilde{B}_{p-1} - \tilde{\epsilon}_{p-1}~,
}
where locally $d \tilde{\epsilon}_{p-1} = \epsilon_p$.

\subsubsection{'t Hooft Operators of BF Theory}

So far, we only talked about Wilson operators. In this section, we discuss dual 't Hooft operators. As we describe in detail below, overall all 't Hooft operators of the BF theory are trivial, although they encode non-trivial conditions when a set of Wilson operators can be cut and become trivial. 

\begin{figure}
\center
\includegraphics[scale=0.8,trim=1.5cm 19cm 1.5cm 1cm,clip]{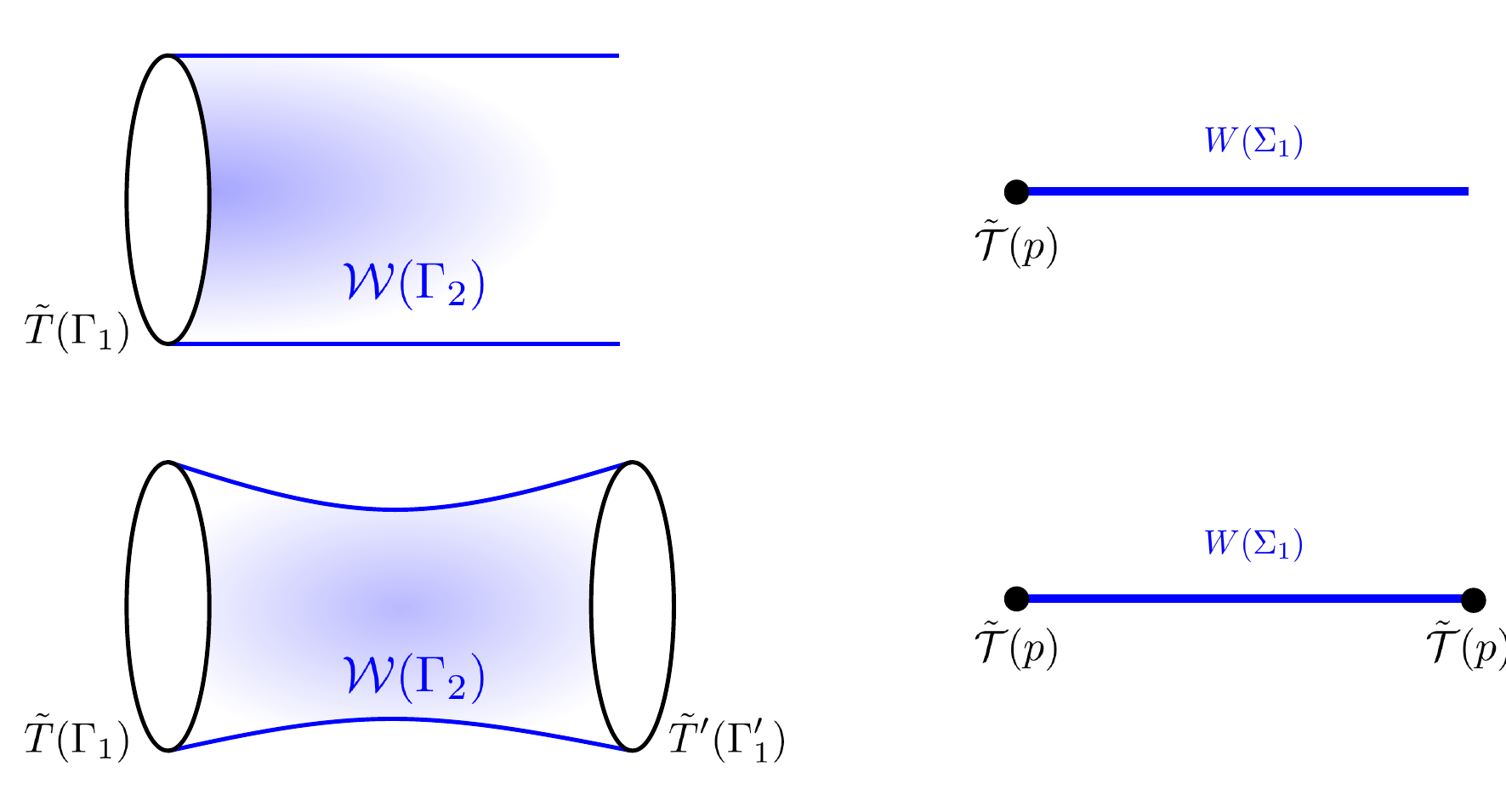}
\caption{Here we illustrate how magnetic/dual defect operators $(\tilde{T}(\Gamma_1)$ and $\tilde{\CT}(p)$) lead to the $\IZ_N$ classification of electric/Wilson operators ($W(\Gamma_1)$ and $\CW(\Gamma_2)$) in $4d$ $\IZN$ BF theory. }
\label{fig:BF_tHooft_op}
\end{figure}

Let us first discuss 't Hooft operators dual to Wilson operators of $A_p$. As described in Section~\ref{subsubsec:other descriptions}, dual field to $A_p$ is $\tilde{A}_{d-p-2}$. Naively, one may write a 't Hooft operator as
\beq
\tilde{T}_n (\Gamma_{d-p-2}) = e^{i n \oint_{\Gamma_{d-p-2}} \tilde{A}_{d-p-2}}.
\eeq
However, as clear from \eqref{eq:BF_dual_description_1}, this operator is not gauge invariant. It can be made gauge invariant by attaching a $(d-p-1)$-dimensional defect to it,
\beq
T_n  = e^{i n \oint_{\Gamma_{d-p-2}} \tilde{A}_{d-p-2}} e^{i n N \int_{\Sigma_{d-p-1}} B_{d-p-1}}
\label{eq:BF_tHooft_op_1}
\eeq
where the boundary of $\Sigma_{d-p-1}$ includes $\Gamma_{d-p-2}$.
To get some intuition, consider $4d$ BF theory with $p=1$. In that case, Wilson operator is a line operator and the naive 't Hooft operator is also a line since the dual field is a 1-form field. There, the gauge invariance of the `t Hooft operator is restored by attaching a charge-$N$ open surface operator to the naive 't Hooft line. The other end of the surface may extend to the infinity or may end on another line of the dual field. See Figure \ref{fig:BF_tHooft_op}. 
 
While the operator in \eqref{eq:BF_tHooft_op_1} is gauge invariant, the equation of motion of $A_{p}$ (see discussion below \eqref{eq:BF_dual_description_1}) makes it a trivial operator. Nevertheless, it encodes a useful information. It shows that a set of $(d-p-1)$-dimensional defects of $B_{d-p-1}$ with total charge $N$ can end on a $(d-p-2)$-dimensional defect of $\tilde{A}_{d-p-2}$. This makes such configuration ``invisible'' within the BF theory since there is no operator that can link the conjoined $B_{d-p-1}$ and $\tilde{A}_{d-p-2}$ defect operator. This ultimately is a restatement of the fact that the Wilson operators of $B_{d-p-1}$ are classified by $\IZ_N$. 

There is an analogous story for the 't Hooft operators dual to Wilson operators of $B_{d-p-1}$. In this case, the dual field is $\tilde{B}_{p-1}$ and one may write down a naive version of the 't Hooft operator
\beq
\tilde{\CT}_m (\widetilde\Gamma_{p-1}) = e^{im \oint_{\widetilde\Gamma} \tilde{B}_{p-1}}~,
\eeq
which similarly is not gauge invariant (see ~\eqref{eq:BF_dual_description_2}). As before, the operator can be made gauge invariant by attaching a $p$-dimensional charge-$N$ defect of $A_p$.
\beq
\CT_m (\widetilde\Gamma_{p-1}) = e^{im \oint_{\widetilde\Gamma_{p-1}} \tilde{B}_{p-1}} e^{i m N \int_{\Sigma_p} A_p},
\label{eq:BF_tHooft_op_2}
\eeq
where the boundary of $\Sigma_p$ includes $\widetilde\Gamma_{p-1}$. This means that the  $A_p$ Wilson  operators with total charge $N$ can be cut, reproducing the $\IZ_N$-classification. 

As an example, consider the $4d$ BF theory with  $p=1$. There, $B_2$ is dual to a 0-form field $\varphi$ and the dual Wilson line is a local operator. This requires us to add a charge-$N$ Wilson lines to make it a gauge invariant. In other words, charge-$N$ Wilson lines are invisible because they can be cut by local operators.  See Figure~\ref{fig:BF_tHooft_op}.

\subsubsection{BF Theory with Torsion}

In this section, we study a BF theory supplemented with a term known as discrete torsion.\footnote{This term is also known as a SPT phase (symmetry protected topological phase) in the condensed matter literature. The reason is that we can dualize the $A_1$ gauge field by introducing a dual gauge field $\tilde{A}_1$ to get 
\eq{
S =&\int \frac{iN}{2\pi} B_2 \wedge d A_1 + \frac{i}{2\pi} d \tilde{A}_1 \wedge d A_1 + \frac{iNp}{4\pi} B_2 \wedge B_2 \\ 
=& \frac{ip}{4\pi N} \int d \tilde{A}_1 \wedge d \tilde{A}_1~, 
}
where here integrating out $A_1$ imposes $NB_2=\tilde{A}_1$ (implying that $\tilde{A}_1$ is $\IZ_N$-valued). 
An action which is a pure (discrete) $\theta$-angle for a discrete gauge field is a ``common'' $\IZ_N$ SPT phase. 
}
 The action in $4d$ is
\beq
S = \frac{iN}{2\pi} \int B_2 \wedge d A_1 + \frac{iNp}{4\pi} \int B_2 \wedge B_2.
\label{eq:BF_SPT_action_1}
\eeq
This theory is still a topological quantum field theory and is interesting to study on its own. What makes it more interesting and relevant to discuss here is that it appears in the discussion of fractional instantons in Section~\ref{subsec:center_YM} and non-invertible symmetries in Section \ref{sec:NIS}. 

This term can be thought of as a sort of ``discrete $\theta$-parameter.'' To make this connection explicit, we can rewrite the action  \eqref{eq:BF_SPT_action_1} in a more suggestive way as 
\beq
S = \frac{iN}{4\pi p} \int \left( dA_1 + p B_2 \right) \wedge \left( dA_1 + p B_2 \right) - \frac{iN}{4\pi p} \int dA_1 \wedge dA_1~.
\label{eq:BF_SPT_action_2}
\eeq
The fact that the torsion term gives rise to the traditional ``$\theta$-term'' for the $U(1)$ gauge field $A_1$, is the reason  why $p$  is sometimes referred to as a discrete $\theta$-parameter. 

Our goal is to understand the effect of this additional term on the generalized global symmetries and associated Wilson operators. To this end, we first show that the integer parameter $p$ must satisfy the conditions:
\beq
\text{(i)} ~ \frac{Np}{2} \in \IZ~,\qquad \text{(ii)} ~  p \sim p + 2N~.
\eeq

To see the first condition, first note that including the discrete torsion term requires that 
 the standard gauge symmetries of the theory \eqref{eq:BF theory_action} be modified to:
\eq{
& B_2 \longmapsto B_2 + d \lambda_1 ~, \hspace{1cm}\oint \lambda_1 = 2\pi \IZ~,\\
& A_1 \longmapsto A_1 + d \lambda_0 - p \lambda_1~, \quad\quad \lambda_0 \sim \lambda_0 + 2\pi~.\label{eq:BF_SPT_Transformations}
}
Here we see that adding the torsion term causes  the 1-form gauge field $A_1$ to transform (with charge $p$) under the 1-form gauge symmetry. 
Note that these modified transformation properties are consistent with the $U(1)$ gauge transformations of $A_1$ when $p$ is an integer. 

Now notice that under a 1-form gauge transformation the action shifts as 
\eq{
\delta S =& \frac{iN}{2\pi} \int d \lambda_1 \wedge d A_1 - \frac{iNp}{2\pi} \int d\lambda_1 \wedge d \lambda_1 \\
=& 2\pi i \int \left( \frac{d \lambda_1}{2\pi} \right) \wedge \left( N \frac{dA_1}{2\pi} \right) - 2\pi i\left( \frac{Np}{2} \right) \int \left( \frac{d \lambda_1}{2\pi} \right) \wedge  \left( \frac{d \lambda_1}{2\pi} \right)~.
}
This means that the theory is only invariant under 1-form gauge transformation provided $\frac{Np}{2} \in \IZ$.\footnote{Note that on a spin manifold 
\beq\label{eq:p2wu}
\int \left( \frac{d \lambda_1}{2\pi} \right) \wedge  \left( \frac{d \lambda_1}{2\pi} \right) \in  2 \IZ~,
\eeq
and thus  $p$ can take any integer value. This follows from the Wu formula which states:
\beq
\int_{M_4} \left[ \frac{\lambda_1}{2\pi} \right] \wedge \left[ \frac{\lambda_1}{2\pi} \right] = \int_{M_4} \nu_2 (TM_4) \cup \left[ \frac{\lambda_1}{2\pi} \right]_{\rm mod 2} \; \text{mod}_2~,
\eeq
where $\nu_2 (TM_4) = \omega_2 (TM_4) + \omega_1^2 (TM_4)$ is the Wu class and $\omega_1$ and $\omega_2$ are the first and second Stiefel-Whitney class of the tangent bundle $TM_4$. 
For spin manifolds $w_2(TM)=w_1(TM)=0$ which proves the identity \eqref{eq:p2wu}. 

However, for more general  orientable manifold $\omega_1 = 0$ and $\nu_2(TM)$ is in general non-zero. Thus, on a general non-spin manifold, we find that $p$ can take any integer value  when $N$ is even, but $p\in 2\IZ$ must be even when $N$ is odd. 
}
To see the second condition, $p \sim p + 2N$,  
first note that the $\IZ_N$ gauge field $B_2$ is normalized 
\eq{
\oint B_2\in \frac{2\pi}{N}\IZ~. 
} 
We can then rewrite  the torsion term as 
\beq
S_{\scriptscriptstyle \rm SPT} = \frac{iNp}{4\pi} \int B_2 \wedge B_2 = \frac{2\pi i p}{2N} \int \left( N  \frac{B_2}{2\pi} \right) \wedge \left( N  \frac{B_2}{2\pi} \right)~, 
\eeq
so that it is clear the term shifts by an multiple of $2\pi$ when 
$p\mapsto p+2N$.  

\bigskip Now let us discuss the effect of the torsion term on the global symmetries. 
First, note that when $p=0$, the theory is simply the standard  BF theory discussed in previous sections. For $p=0$, the global symmetry is given by $\IZ_N^{(1)} \times \IZ_N^{(2)}$ with transformation rules   
\eq{
& A_1 \longmapsto A_1 + \frac{1}{N} \epsilon_1~, \;\;\; \oint \epsilon_1 = 2\pi \IZ~, \\
& B_2 \longmapsto B_2 + \frac{1}{N} \epsilon_2~, \;\;\; \oint \epsilon_2 = 2\pi \IZ~,
}

When $p \neq 0$, symmetry structure changes. 
The equation of motion for $B_2$, $d A_1 + p B_2 = 0$,  implies that the Wilson surface operator satisfies
\beq
\left( e^{i \oint_{\Gamma_2} B_2} \right)^p = e^{-i \oint_{\Gamma_2} d A_1} = 1~.
\label{eq:BF_SPT_W2_cut_p}
\eeq
Additionally, when we introduce a dual gauge field $\tilde{A}_1$, the equation of motion for $A_1$ similarly implies that
\beq
\left( e^{i \oint_{\Gamma_2} B_2} \right)^N = e^{-i \oint_{\Gamma_2} d \tilde{A}_1} = 1~.
\label{eq:BF_SPT_W2_cut_N}
\eeq

Together these conditions mean that the minimal $B_2$ Wilson surface operator $W_1 (\Gamma_2) = e^{i \oint_{\Gamma_2} B_2}$ satisfies\footnote{This can be shown using B\'{e}zout's identity. B\'{e}zout's identity says that given two integers $a, b \in \IZ$ with the greatest common divisor $d = \text{gcd} (a, b)$, there always exist integers $x, y \in \IZ$ such that $ax + by = d$. In our case, any integer $\ell$ can be decomposed as $\ell = n \; \text{gcd} (N,p) + m$ with $n \in \IZ, \;m =0, 1, \cdots , \text{gcd} (N,p) - 1$. Also, by B\'{e}zout's identity $\text{gcd} (N,p)$ can always been written as $\text{gcd} (N,p) = x N + y p$ for integers $x, y$. This immediately implies \eqref{eq:BF_SPT_W2}. }
\beq
\left( W_1 (\Gamma_2) \right)^{{\rm gcd} (N,p)} = 1,
\label{eq:BF_SPT_W2}
\eeq
where $\text{gcd}$ denotes the greatest common divisor.

This means that when $p\neq 0$, the discrete torsion term reduces the number of Wilson surface operators and hence the  
the theory has $\IZ_{\text{gcd}(N,p)}^{(2)}$ 2-form global symmetry with transformation rules 
\eq{
B_2 \longmapsto B_2 + \frac{1}{\text{gcd}(N,p)} \epsilon_2~,\qquad
A_1 \longmapsto A_1 - \frac{p}{\text{gcd}(N,p)} \tilde{\epsilon}_1~,
}
where locally, $d \tilde{\epsilon}_1 = \epsilon_2$. It is a straightforward exercise to show that the action in \eqref{eq:BF_SPT_action_1}  
is indeed invariant under these transformations.

Now let us consider the effect of the torsion term on the 1-form global symmetry. Note that the additional term does not depend on the 1-form field $A_1$ and hence does not directly effect the 1-form symmetry.  
However,  
the naive line operator $e^{i \oint_{\Gamma_1} A_1}$ is not gauge invariant under \eqref{eq:BF_SPT_Transformations}. Rather, it can only be made gauge invariant if we dress it with a surface operator:
\beq
\widetilde{W} = e^{i \oint_{\Gamma_1} A_1} e^{i p \int_{\Sigma_2} B_2}~. 
\eeq
The resulting operator is not a genuine line operator since it is attached to a surface. 
Of course, this is consistent with our previous observation \eqref{eq:BF_SPT_W2_cut_p} that charge-$p$ Wilson surface is ``invisible'' since it can end on a line.  
Note however, that we can use \eqref{eq:BF_SPT_W2_cut_N}
to construct a gauge invariant (genuine) line operator in terms of the line operator
\beq
\widetilde{W}^N = e^{i \oint_{\Gamma_1} \left( N A_1 - p \tilde{A}_1 \right)}~.
\eeq 
This is gcd$(N,p)$ times a minimal (gauge invariant) line operator 
\beq
W (\Gamma_1) = \left( \widetilde{W} \right)^{\frac{N}{\text{gcd} (N,p)}}.
\eeq
which satisfies 
\eq{
\left( W (\Gamma_1) \right)^{\text{gcd} (N,p)} = 1~.}
Therefore, we see that the torsion term reduces the symmetry of both the 1- and 2-form global symmetries:
\eq{
\IZ_N^{(1)}\times \IZ_N^{(2)}~\stackrel{p\neq0}{\longrightarrow}~\IZ_{{\rm gcd}(N,p)}^{(1)}\times \IZ_{{\rm gcd}(N,p)}~. 
}

\subsection{1-Form Center Symmetry in Non-Abelian Yang-Mills Theory}
\label{subsec:center_YM}

Another important class of theories with discrete higher form symmetry is non-abelian Yang-Mills theory \cite{Aharony:2013hda,Gaiotto:2014kfa,Gaiotto:2017yup,Tachikawa:2017gyf,Cordova:2019jnf,Cordova:2019uob}. These theories can also be analyzed by using discrete background gauge fields. 
For our discussion let us take $G$ to be a non-abelian, continuous Lie group that is not simply connected and has a non-trivial center.

In non-abelian Yang-Mills theory, the story of center and magnetic 1-form  symmetry differs from that of $U(1)$ Maxwell theory because non-abelian gauge theories are interacting. 
The self-interacting nature of non-abelian Yang-Mills has many consequences on the dynamics of the theory. 
One consequence of the interactions is that the gluons are charged fields and 
therefore can screen Wilson lines. However, since the gluons are in the adjoint representation -- which is often not the minimal representation of $G$ -- the gluons can only screen certain classes of Wilson lines. In cases where the gluons cannot completely screen Wilson lines there is a (often discrete) 1-form center symmetry that measures the conserved charge modulo screening. This symmetry is classified by the center of the gauge group $Z(G)$. 

Additionally, in non-abelian gauge theories there are smooth, dynamical monopole gauge field configurations that are summed over in the path integral. Because of this, the monopole lines of the theory can also be dynamically screened. This can be seen by a stability analysis of the equations of motion in the monopole background  \cite{Coleman:1982cx}.  
In cases where the monopole lines cannot be screened, we say that they are protected by a magnetic 1-form symmetry which measures their protected magnetic charge. This symmetry is classified by the fundamental group of the gauge group $ \pi_1(G)$.

We will now compare and contrast these 1-form electric and magnetic symmetries  in a pair of connected  examples: $SU(2)$ and  $SO(3)$ non-abelian Yang-Mills theory. 

\subsubsection{$SU(2)$ vs. $SO(3)$}
\label{subsubsec:SU2_vs_SO3}

Here we would like to consider the difference between the center and magnetic 1-form symmetries of $SU(2)$ and $SO(3)$ gauge theories. Since these groups have the same Lie algebra, they have the same perturbative physics. However, since $SO(3)=SU(2)/\IZ_2$, the two groups have different   center and fundamental groups:
\eq{ 
&Z(SU(2))=\IZ_2\qquad \qquad Z(SO(3))=1~,\\
&\pi_1(SU(2))=1~\qquad \qquad \pi_1(SO(3))=\IZ_2~,
}
which means that (according to our proposed classification) the theories will have different higher form symmetries.

First consider $SU(2)$ gauge theory. Here a Wilson line wrapping a curve $\gamma$ is defined by a choice of representation $R_j$ of spin-$j$ where $j=0,\,1/2,\,1,\,3/2,\,...$:
\eq{
W_{R_j}(\gamma)=\Tr_{R_j}\,\CP~{\rm exp}\left\{i \int_\gamma A\right\}~.
}
These Wilson lines can be dynamically screened by the dynamical gluons. However, since gluons are in the adjoint representation (i.e.~the spin-1 representation),
they can only screen Wilson lines of integer spin. We can see this by introducing a charged adjoint scalar field and using it to break $SU(2)\to U(1)$. In this phase, particles charged under the spin-$j$ representation have  abelian charges $q\in 2\IZ+2j$ so that the gluons (charge $\pm2$) can only partially screen Wilson lines of half-integral spin (odd charge). 

Now recall that the center of $SU(2)$ is $Z(SU(2))=\IZ_2$. 
The representations of half-integral spin  are the representations that transform non-trivially under this $\IZ_2$. Therefore, detecting whether or not a Wilson line can be screened is equivalent to measuring its charge under the center $\IZ_2\subset SU(2)$. 
Thus, we see $SU(2)$ Yang-Mills theory has a $\IZ_2^{(1)}$1-form  center symmetry that measures the spin of Wilson lines mod$_\IZ$.  

In $SU(2)$ gauge theory there are no unprotected monopoles (see \cite{Brandt:1979kk,Coleman:1982cx} for details). This implies that  $SU(2)$ gauge theory has no magnetic 1-form  symmetry, matching with the fact that $\pi_1(SU(2))=1$. 
\bigskip

Now consider $SO(3)$ gauge theory. Here the Wilson lines are classified by representations of spin $j=0,1,2,...$. These Wilson lines can all be screened by gluons due to the fact that the minimal Wilson line is the same representation as the gluons. Therefore there is no center symmetry -- matching with the fact that $Z(SO(3))=1$. 

However, due to the fact that $SO(3)=SU(2)/\IZ_2$, the theory has stable  monopole lines. 
In order to see this, let us first consider monopoles in $SU(2)$ gauge theory.  
Here, the monopoles have magnetic charge given by $n\sigma^3$ for $n\in \IZ$. This can be seen by noting that near the monopole, the magnetic field goes like 
\eq{
\vec{B}\underset{r\to 0}{\sim} B_0\frac{1}{2r^2}\hat{r}\quad, \quad B_0=\alpha \sigma^3~.}
An explicit gauge field configuration can be written in two patches (northern/southern-hemisphere) covering the asymptotic 2-sphere:
\eq{A_{N/S}=B_0\frac{\pm 1-\cos(\theta)}{2}d\phi~,
}
where $\pm$ is correlated to the northern/southern hemisphere. In order for this to describe a good monopole configuration, the two patches have to be related by a gauge transformation: $g=e^{i B_0 \phi}$. We require that this gauge transformation be well defined, which implies
\eq{ 
g(\phi+2\pi)=g(\phi)\quad\Rightarrow \quad e^{2\pi i B_0}=\mathds{1}_{SU(2)}~.
}
This leads to the quantization condition $B_0=n \sigma^3$ where $n \in \IZ$. 
As shown in \cite{Brandt:1979kk,Coleman:1982cx}, these monopoles are not stable to emission by $W$-bosons. 

However, in $SO(3)$ gauge theory $-\mathds{1}_{SU(2)}\sim \mathds{1}_{SO(3)}$. This means that $SO(3)$ magnetic monopoles can now have half-integral magnetic charges: $B_0=\frac{n}{2}\sigma^3$. As it turns out,   monopoles with magnetic charge $B_0=\half \sigma^3$  \emph{are} stable to gluon emission. This gauge transformation $g(\phi)=e^{\half \sigma^3 \phi}$ is the non-trivial representative of $\pi_1(SO(3))$. Because  $g(\phi)$ is in a non-trivial topological class, heuristically it cannot ``decay'' by gluon fluctuations and hence the corresponding monopole is stable. Therefore, we see that $SO(3)$ gauge theory has a  $\to \IZ_2^{(1)}$ magnetic  symmetry that measures the magnetic charge $mod_\IZ$.  

\subsubsection{General Center and Magnetic 1-Form Symmetry}

The story for $SU(2)$ and $SO(3)$ gauge theory is in fact   general. Consider a non-abelian gauge group $G$ that has a center $Z$. Let us also define $G_{\rm ad}=G/Z$ -- this is sometimes called the adjoint form of the group  -- and $\hatG$ which is the simply connected cover $G=\hatG/Z$. In general, the   center symmetry of $G$ Yang-Mills theory is described by $Z$ and its 1-form magnetic symmetry is given by $\pi_1(G)$. 

In fact, there is a parallel structure between the center and magnetic 1-form  symmetries:
\begin{center}\begin{tabular}{c|c|c}
Center Symmetry & Gauge Group & Magnetic Symmetry\\
\hline 
``$\pi_1(G)\times Z$'' & $\tG$ & 1\\
$Z$ & $G=\tG/\pi_1(G)$ & $\pi_1(G)$\\
1& $G_{ad}=G/Z$ & ``$\pi_1(G)\times Z$''
\end{tabular}
\end{center}
Here we use `` '' to indicate that this is only true as sets. In general, $Z(G)=Z$ and $\pi_1(G)$ can descend from a simply connected group $\tG$ where $Z(\tG)$ is an extension of $Z$ by $\pi_1(G)$. Similarly, $G_{\rm ad}=G/Z$ will have a fundamental group that is an extension of $\pi_1(G)$ by $Z$.  For more details see Appendix \ref{app:CenterMag}.

\subsubsection{Background Gauge Fields}

\label{centerbackground}
Now we would like to discuss how to use background gauge fields to study the center and magnetic 1-form  symmetry of non-abelian Yang-Mills. 

First, let us discuss center symmetry. Consider the example of $SU(N)$ gauge theory. Center symmetry allows us to shift the dynamical gauge field $a_1$ (with field strength $f_2$) by a $\IZ_N$ gauge field $A_{\IZ_N}$ corresponding to the center $\IZ_N\subset SU(N)$. We can see that this transformation implements a $\IZ_N$ action on Wilson lines: 
\eq{
W_R(\gamma)=Tr_R~{\rm exp}\left\{i \int_\gamma a_1\right\}\longmapsto Tr_R~{\rm exp}\left\{i \int_\gamma a_1+i \int_\gamma A_{\IZ_N}\right\}=e^{i I_N(R)\int_\gamma A_{\IZ_N}}W_R(\gamma)~,
}
where $I_N(R)$ is the $N$-ality of the representation $R$ (i.e.~the charge of $R$ thought of as a $\IZ_N\subset SU(N)$ representation so $N$-ality is the number of boxes in a Young tableau mod $N$). Here we see that the $\IZ_N$-action comes from the fact that $e^{i\int_\gamma A_{\IZ_N}}\in \IZ_N$.

In order to use our knowledge of $\IZ_N$ gauge theory, we can lift the $SU(N)$ gauge theory to a $U(N)$ gauge theory. Here the idea is that we can realize $SU(N)$ gauge theory as a constrained $U(N)$ gauge theory, but in $U(N)$ gauge theory,  
the shift by the $\IZ_N$ gauge field $A_{\IZ_N}$ lifts to a shift by the $U(1)\subset U(N)$ gauge field $\hat{A}_{\IZ_N}$ where we can easily analyze the symmetry using background gauge fields.

So, let us first lift our $SU(N)$ gauge theory to a $U(N)$ gauge theory which has a gauge field and field strength $\tilde{a}_1,\tilde{f}_2$. 
The projection from $U(N)\to SU(N)$ gauge theory can be achieved by introducing a 2-form Lagrange multiplier $\CF_2 \in \Omega^2(M_4)$ so that the full action can be written as  
\eq{
S=\int \frac{1}{g^2}\Tr[\tilde{f}_2\wedge \ast\tilde{f}_2]+\frac{i}{2\pi} \int \CF_2 \wedge \Tr[\tilde{f}_2]~.
}
Here, shifting by the $\IZ_N$ gauge field $A_{\IZ_N}$ is equivalent to shifting by a $U(1)$ gauge field:
 \eq{
\tilde{a}_1\longmapsto \tilde{a}_1 +\hat{A}_{\IZ_N}\quad\Rightarrow \quad \tilde{f}_2\longmapsto\tilde{f}_2+\hatF_{\IZ_N}~,
}
where $\hatF_\IZN$ is the field strength of the $U(1)$ gauge field $\hat{A}_\IZN$. 
However, this shift is not invariant under a choice of lift of $A_\IZN$ to $\hat{A}_\IZN$:
\eq{
\hatF_{\IZ_N}\longmapsto \hatF_{\IZ_N}+\Lambda_2\quad,\quad \tilde{f}_2\longmapsto \tilde{f}_2+\Lambda_2 \mathds{1}_N\quad, \quad \twopi{\Lambda_2}\in H^2(M_4;\IZ)~. 
}
To make the action invariant, we need to introduce a $\IZ_N$ background gauge field $B_2$ that similarly transforms 
\eq{
B_2\longmapsto B_2+\Lambda_2~,
}
and substitute $\tilde{f}_2\longmapsto\tilde{f}_2-B_2\mathds{1}_N$ in the action:
\eq{
S=\int \frac{1}{g^2}\Tr[(\tilde{f}_2-B_2\mathds{1}_N)\wedge \ast (\tilde{f}_2-B_2\mathds{1}_N)]+ \frac{i}{2\pi} \int \CF_2 \wedge\Tr[\tilde{f}_2-B_2\mathds{1}_N]~.
}
Now notice that the Lagrange multiplier term has fixed the $U(1)$ part of the $U(N)$ gauge field to be a $\IZ_N$ gauge field that is fixed by the choice of $B_2$. Due to the fact that we can realize $U(N)=\frac{SU(N)\times U(1)}{\IZ_N}$, we see that the $\IZ_N$ gauge field in the $U(1)$ factor is actually identified with a $\IZ_N$ gauge factor along the center $\IZ_N\subset SU(N)$. 

Mathematically, turning on the background gauge field for $\IZ_N$ center symmetry is equivalent to lifting the $SU(N)$ gauge theory  (where we sum over $SU(N)$ bundles with connection in the path integral) to a $SU(N)^c$ gauge theory where we sum over $SU(N)^c$ bundles with fixed fractional flux $B_2$. Here we use the notation $SU(N)^c$ to denote the $SU(N)$ analog of lifting a $Spin(N)$ bundle to a $Spin(N)^c$ bundle (which is somewhat standard in geometry).\footnote{We would like to thank G. Moore (and N. Seiberg) for comments clarifying this point.} See  Appendix \ref{app:twisted_bundle} for more details.

Now we can understand how to introduce a background gauge field for the magnetic 1-form  symmetry of $G$.  Let us start with $\tG$ gauge theory where $\pi_1(\tG)=1$. $G$ gauge theory differs by taking the quotient of the gauge group $\tG$ by $\pi_1(G)\subset Z(\tG)$. At the level of the gauge theory, taking the quotient of the gauge group is equivalent to gauging 1-form center symmetry $\pi_1(G)\subset Z(\tG)$. This can be seen by noting that gauging $\Gamma\subset Z$ projects out states that are electrically charged under $Z$ while adding states that have fractional magnetic charge. This is analogous to the story of orbifolding 2D CFT's where we project out charged states while adding twisted sectors. See \cite{DiFrancesco:1997nk,Ginsparg:1988ui} for related reviews.   

Formally, gauging $\pi_1(G)\subset Z(\tG)$ can be achieved by summing over all possible $B_2$ corresponding to $\pi_1(G)\subset H^2(M_d;Z(\tG))$. 
At the level of the path integral, we can express this as
\eq{
Z_G=\sum_{[B_2]}Z_{\tG}[B_2]~,
}
where $Z_{\tG}[B_2]$ is the path integral of the $\tG$ gauge theory with background field $B_2$ turned on. 

We can now formally introduce a new background gauge field $B_2^{(m)}$  as the discrete gauge field that is ``dual'' to $B_2$:
\eq{
Z_G=\sum_{[B_2]}Z_{\tG}[B_2]e^{\twopi{in} \int B_2^{(m)}\cup B_2}=\sum_{[B_2]}Z_{\tG}[B_2]e^{\twopi{in }\int \hat{B}_2^{(m)}\wedge \hat{B}_2}~,
}
where we have written $\hat{B}_2$ as an integral lift of the discrete cohomology class $B_2$ and $n=|\pi_1(G)|$. Here, $B_2^{(m)}$ becomes the 1-form background gauge field for the new magnetic 1-form symmetry associated to $\pi_1(G)$. 
Here we see explicitly that taking the quotient of the gauge group exchanges center symmetry for magnetic 1-form  symmetry. \\ 

\rmk.~
Note that including charged matter fields can ``break'' center symmetry. The physical reason is that in addition to gluons, dynamical charged matter fields can also screen Wilson lines. Thus, adding matter fields that are charged under the center of the gauge group  can lead to additional screening of Wilson lines, thereby reducing the dynamically conserved charge. 

\subsubsection{Instantons and Center Symmetry}

An important consequence of center symmetry   is that turning on a background gauge field  has a dramatic effect on instanton physics  \cite{Aharony:2013hda,Gaiotto:2017yup,Cordova:2019uob}. 

Let us consider the $\theta$-term in $SU(N)$ Yang-Mills theory. Here we can turn on a background gauge field $B_2$ for the $\IZ_N$ center symmetry as discussed in the previous section. The $\theta$-term now reads
\eq{
S_\theta=\frac{i\theta}{8\pi^2}\int \Tr[(\tilde{f}_2-B_2\mathds{1}_N)\wedge (\tilde{f}_2-B_2\mathds{1}_N)]~.
}
We can now expand this out to separate out the background gauge field component from the dynamical field:
\eq{
S_\theta=\frac{i \theta}{8\pi^2}\int \left(\Tr[\tilde{f}_2\wedge \tilde{f}_2]-N B_2\wedge B_2\right)~,
}
where we used the constraint $\Tr[\tilde{f}_2-B_2\mathds{1}_N]=0$. Now we can express the first term as an integral class (i.e.~the second Chern class):
\eq{
c_2(\tilde{f}_2)=\frac{\Tr[\tilde{f}_2\wedge \tilde{f}_2]-\Tr[\tilde{f}_2]\wedge \Tr[\tilde{f}_2]}{8\pi^2}~,
}
since $\tilde{f}_2$ is  simply a $U(N)$ gauge field that we are interpreting in a special way.
After trivial manipulation, we find that the $\theta$-term can be written as 
\eq{
S_\theta&=\frac{i \theta}{8\pi^2}\int \left(\Tr[\tilde{f}_2\wedge \tilde{f}_2]-\Tr[\tilde{f}_2]\wedge \Tr[\tilde{f}_2]\right)+\frac{i \theta}{8\pi^2}N(N-1)\int B_2\wedge B_2\\
&=i  \theta\left(\int c_2(\tilde{f}_2)+N(N-1)\int\frac{B_2\wedge B_2}{8\pi^2}\right)~.
}
Now, since $N\oint \twopi{B_2}\in \IZ$, we see that turning on a background gauge field for $\IZ_N$ center symmetry shifts the instanton number by a $1/N$-fractional amount. In fact, it is often standard to define the integral class $w_2= N \frac{B_2}{2\pi}\in H^2(M_d;\IZ)$ to emphasize the fact that the new term is fractional
\eq{
S_\theta =i  \theta\left(n+\frac{(N-1)}{N}\int\frac{w_2\wedge w_2}{2}\right)\quad, \quad \int\frac{w_2\wedge w_2}{2}\in \IZ~.
}

As we discussed, mathematically, turning on a $\IZ_N$ background gauge field is equivalent to lifting the $SU(N)$ gauge theory to a $SU(N)^c$ bundles with fixed fractional flux: $B_2$. Our computation here shows that lifting to $SU(N)^c$ bundles in this way results in  summing over gauge fields that have a fixed fractional flux and hence a fractional instanton number:  
\eq{
I_{inst}=\frac{(N-1)}{N}\int\frac{w_2\wedge w_2}{2}~{\rm mod}_\IZ~.
}
See  \cite{Gaiotto:2014kfa} for more details. 

There is also a generalization of 1-form center symmetry story to the case where there are combined actions of gauge and global symmetries that act trivially on all the matter in the theory. In this case, one can only turn on combined fractional fluxes in the gauge and global symmetry groups which is sometimes called BCU-, or CFU- background 
\cite{Anber:2021iip, Anber:2019nze}. This idea is is important for the analysis of anomalies in gauge theories as well as for the idea of charge fractionalization \cite{Wang:2018qoy,Brennan:2022tyl,Delmastro:2022pfo,Barkeshli:2014cna}.

\subsection{Discrete $\theta$-Angles}

Here we would like to demonstrate a way in which discrete symmetries can be implemented to further probe non-abelian gauge theories with center symmetry. 

We consider gauge theories with gauge groups that have a trivial center. These theories, while defined in their own right, can also be obtained by taking the gauge theory of the simply connected cover   and then gauging the center symmetry. 

A simple case is $G=SU(N)$ where $Z(SU(N))=\IZ_N$. The centerless form of the group is then $PSU(N)=SU(N)/\IZ_N$. If we write the partition function of $SU(N)$ gauge theory with $\theta$-term and background gauge field $B_2$ for the center symmetry as $Z_{SU(N)}[B_2;\theta]$, then we can write down the partition function of $PSU(N)$ gauge theory as 
\eq{
Z_{PSU(N)}[\theta]=\sum_{[B_2]\in H^2(M_d;\IZ_N)} Z_{SU(N)}[B_2;\theta]~. 
}
In general, one is allowed to include $B_2$-dependent (gauge-invariant) phases to the partition function that differentiate between different topological sectors. In general, these phases are simply added by 
\eq{
Z_{PSU(N)}=\sum_{[B_2]\in H^2(M_d;\IZ_N)} Z_{SU(N)}[B_2;\theta] \; e^{\twopi{i p N} \int \frac{B_2\wedge B_2}{2}}~,
}
where $p=0,...,N-1$.   These terms are called \emph{discrete theta angles}. 

Note that this term is the only possible phase we can add to the partition function that is only dependent on $B_2$ and is gauge invariant. Note that the quantization of $p$ also comes from the demanding that the phase is gauge invariant (which in this case means independent of choice of integer lift $B_2\to B_2+\Lambda_2$ where $\twopi{\Lambda_2}\in H^2(M_d;\IZ)$). 

However, we know that there is an intimate connection between the instanton number and the proposed phase we wish to add. In particular, due to the identification 
\eq{I=
\frac{1}{8\pi^2}\int \Tr[F\wedge F]=N(N-1)\int \frac{B_2\wedge B_2}{8\pi^2}~{\rm mod}_\IZ~,
}
we see that shifting $\theta\to \theta+2\pi n$ shifts the phase 
\eq{
Z_{SU(N)}[B_2;\theta+2\pi n]=Z_{SU(N)}[B_2;\theta]e^{-\twopi{i n N}\int \frac{B_2\wedge B_2}{2}}~. 
}
So, we see that in the case of $PSU(N)$ gauge theory, we can generate all of the possible phases in the partition function by simply shifting $\theta$. 

However, this is not the case for a generic gauge group. For other gauge groups, shifting $\theta$ does not generate all such possible phases. In the following table we  enumerate the relations for the classical Lie groups ( (6.13) in \cite{Aharony:2013hda}):

\begin{center}
\begin{tabular}{l|l|l|l|l}
$\widetilde{G}$ & Center & Background Fields & $I$ mod$_\IZ$ & Extra Phases\\
\hline 
$SU(N)$ &$\IZ_N$& $\twopi{B_2}=\frac{1}{N}w_2$ & $\frac{N-1}{N}\int \frac{w_2\wedge w_2}{2}$ & none\\[0.2cm]
$Sp(2N)$ & $\IZ_2$&$\twopi{B_2}=\half w_2$& always $\in \IZ$ & $\half \int \frac{w_2\wedge w_2}{2}$\\[0.2cm]
$Sp(2N+1)$& $\IZ_2$&$\twopi{B_2}=\half w_2$& $\half \int \frac{w_2\wedge w_2}{2}$& none\\[0.2cm]
$Spin(2N+1)$ & $\IZ_2$ &$\twopi{B_2}=\half w_2$& always $\in\IZ$ & $\half \int \frac{w_2\wedge w_2}{2}$ \\[0.2cm]
$Spin(8N)$ &$\IZ_2\times \IZ_2$ & $\twopi{B_2}=\half w_2$, $\frac{\tB}{2\pi}=\half \tilde{w}_2$ & $\half \int w_2\wedge \tilde{w}_2$ & $\half \int \frac{w_2\wedge w_2}{2}$, $\half \int \frac{\tilde{w}_2\wedge \tilde{w}_2}{2}$\\[0.2cm]
$Spin(8N+4)$  &$\IZ_2\times \IZ_2$ & $\twopi{B_2}=\half w_2$, $\frac{\tB}{2\pi}=\half \tilde{w}_2$ & $\half \int \left( \frac{w_2\wedge w_2}{2}+ \frac{\tilde{w}_2\wedge \tilde{w}_2}{2}\right)$ & $\half \int  \frac{w_2\wedge w_2}{2} $, $\half \int w_2\wedge \tilde{w}_2$\\[0.2cm]
$Spin(8N+2)$ & $\IZ_4$ & $\twopi{B_2}=\frac{1}{4} w_2$ & $\frac{1}{4} \int \frac{w_2\wedge w_2}{2}$&none\\[0.2cm]
$Spin(8N+6)$ & $\IZ_4$ & $\twopi{B_2}=\frac{1}{4} w_2$ & $-\frac{1}{4} \int \frac{w_2\wedge w_2}{2}$&none\\[0.2cm]
$\widetilde{E_6}$ & $\IZ_3$ & $\twopi{B_2}=\frac{1}{3}w_2$ & $ \frac{2}{3}\int \frac{w_2\wedge w_2}{2}$ & none\\[0.2cm]
$\widetilde{E_7}$ & $\IZ_2$ & $\twopi{B_2}=\half w_2$ & $\half \int \frac{w_2\wedge w_2}{2}$ & none\\[0.2cm]
$E_8 $ & trivial & none & always $\in \IZ$& none
\end{tabular}
\end{center} 
This tells us that pure gauge theories with gauge group $Sp(2N)$, $SO(2N+1)$, $SO(8N)$, $SO(8N+4)$ all have phases we can turn on in the path integral that are independent of the usual $\theta$-angle. These extra phases are in a sense independent  discrete $\theta$-angles. For more details see \cite{Aharony:2013hda,Hsin:2020nts}.

\subsection{Comment on Dual Higher Form Global Symmetries}

We would now like to comment on the role of dual higher form global symmetries and their relation to the anomaly of $p$-form gauge theories. Here we will compare and contrast the dual symmetry for continuous vs. discrete symmetries. 

We say that a pair of continuous higher form symmetries are dual when their currents are related by Hodge duality. This occurs in the following situation. Let us consider the case of a conserved (co-closed) $(p+1)$-form current $J_{p+1}$ that is associated to a $p$-form global symmetry. We say that there is a dual $(d-p-1)$-form current $\tJ_{d-p-1}$ if $J_{p+1}$ is also closed:  
\eq{
d J_{p+1}=0\quad\Rightarrow\quad d\ast\tJ_{d-p-1}=0\quad, \quad \tJ_{d-p-1}=\ast J_{p+1}~. 
}
This current corresponds to a $(d-p-2)$-form global symmetry which we call the ``dual'' global symmetry. It follows that dual continuous global symmetries are related as:
\eq{
\text{Continuous:~~}p\text{-form symmetry}~\longleftrightarrow ~(d-p-2)\text{-form symmetry}~.
} 

Due to the fact that $dJ_{p+1}=d\ast J_{p+1}=0$, we can locally express $J_{p+1}$ in terms of a $p$-form field $A_{p}$ that satisfies:
\eq{\label{discreteanom}
J_{p+1}=dA_p\quad, \quad d\ast d A_{p}=0~,
}
which are simply the equations of motion for $p$-form electromagnetism. 
We then see that $J_{p+1}$ and $J_{d-p-1}$ are the currents corresponding to dual electric and magnetic higher form symmetries and thus carry an anomaly of the type  \eqref{mixedEManom}.

In this setting, EM duality can be implemented by 
coupling $A_p$ to a dynamical  $(d-p-2)$-form $U(1)^{(d-p-3)}$ gauge field $\tilde{A}_{d-p-2}$ via 
\eq{
S=...+\twopi{i} \int dA_{p}\wedge d\tilde{A}_{d-p-2}~. 
}
This process is allowed and well defined since it is gauge invariant. 
Now we can see that the duality is induced at the level of the equations of motion:
\eq{
d\left(\ast dA_{p}-\twopi{ d\tilde{A}_{d-p-2}}\right)=0~\Longrightarrow~ \twopi{d\tilde{A}_{d-p-2}}= \ast dA_{p}~.
}
Then by integrating out the gauge field $A_p$, we find a theory of the dual gauge field $\tilde{A}_{d-p-2}$.\footnote{
In the case of $4d$  $U(1)$ Maxwell theory, this duality relates the field strength $dA_1$ to the field strength of a dual gauge field $d\tilde{A}_1= \ast dA_1$ and integrating out $A_1$ in exchange for $\tilde{A}_1$ implements $U(1)$ electric-magnetic duality. }
Now the duality relating the $p$- and $(d-p-2)$-form global symmetries can be understood as the exchanging the symmetry that measures the charges of Wilson lines of $A_p$ for the symmetry that measures the charges of Wilson lines of $\tilde{A}_{d-p-2}$.

There is a similar story for discrete global symmetries. Here, pairs of dual symmetries are related by 
\eq{
\text{Discrete:~~} p\text{-form symmetry}~\longleftrightarrow~ (d-p-1)\text{-form symmetry}~.
} 
Note that the degree of the dual symmetry is different from that of the continuous case. The reason is because continuous gauge theories have 2-derivative actions while discrete gauge theories have 1-derivative actions. 
To see this, we can identify dual symmetries as in the continuous case by performing the discrete version of EM-duality. 

Let us consider a $\IZ_N$ $p$-form discrete gauge field which we lift to a $U(1)$ gauge field $\hat{A}_p$. Now let us introduce a dual $\IZ_N$ $(d-p-1)$ form gauge field that we lift to $U(1)$: $\hat{B}_{d-p-1}$. We can again implement the duality between these two by adding the gauge invariant coupling\footnote{
Here we see that this is gauge invariant under $\IZ_N$ gauge transformations: $A_p\to A_p+d\lambda_{p-1}$ and $B_{d-p-1}\to B_{d-p-1}+d\Lambda_{d-p-2}$. Since $\twopi{d\Lambda},\ \twopi{d\lambda}$, $\twopi{N A_p}$, and $\twopi{N B_{d-p-1}}$ are all integer quantized, the shift 
\eq{
\delta S=\twopi{i N}\int d\lambda_{p-1}\wedge dB_{d-p-1}\in 2\pi i \IZ~,
}
means that the exponentiated action is gauge invariant.}
\eq{
S=...+\twopi{i N} \int \hat{A}_p\wedge d\hat{B}_{d-p-1}~.
} Note that in contrast to the case of continuous gauge symmetries, the discrete quantization of discrete gauge fields allows us to write down a gauge invariant  coupling between a $p$-form and a $(d-p-1)$-form which means that the dual to a $\IZN^{(p)}$ discrete $p$-form symmetry is a $\IZN^{(d-p-1)}$ $(d-p-1)$-form global symmetry.  
  As in the case of continuous electric-magnetic duality, dual discrete symmetries will carry the anomaly of discrete gauge theory \eqref{discreteanom}.

\subsection{Mixed 0-Form/1-Form Anomaly}
 
We would now like to discuss a general class of mixed anomaly between 0-form and 1-form global symmetries in QCD-like gauge theories. 

Consider a $4d$  $SU(N)$ non-abelian gauge theory coupled to some collection of charged fermions $\{\psi_i\}$. Now let us consider an abelian flavor symmetry $G_0$ that rotates the fermions:
\eq{
\psi_i \longmapsto e^{i q_i \alpha} \psi_i~. 
}
Here we can take the rotation parameter $\alpha$ to be either discrete or continuous. In the case where there is an ABJ anomaly for $G_0$, the chiral rotation above generates a term in the action 
\eq{
\frac{i k \alpha}{8\pi^2}\int \Tr[F\wedge F]~,
}
for some $k\in \IZ$. 

Now we see that the action is invariant under shifts of $\alpha$ if $\alpha\in \frac{2\pi}{k}$. Because of this, we say that the symmetry is broken to $\IZ_k\cap G_0$ by the ABJ anomaly. In the case where $\IZ_k\subset G_0$, it simply breaks to $\IZ_k$. Let us assume that this is indeed the case so that the ABJ anomaly breaks $G_0\to \IZ_k$. 

Let us take the fermions to transform in in representations of $G$ so that the center symmetry is preserved. In this case we can turn on a background gauge field for the center symmetry $\twopi{B_2}=\frac{1}{N}w_2$, where $w_2\in H^2(M_4;\IZ)$. As we have discussed, turning on this background gauge field for the center symmetry of $SU(N)$ changes the quantization of the instanton number:
\eq{
\frac{1}{8\pi^2}\int \Tr[F\wedge F]= \frac{N-1}{N}\int \frac{w_2\wedge w_2}{2}\quad {\rm mod}_\IZ~.
}
This fractional quantization changes the standard periodicity of the $\theta$-angle from $2\pi$ to  to $2\pi N$. 

This has a nontrivial implication on the $G_0$ ABJ anomaly.  As with the $\theta$-angle, when the background gauge field $B_2$ is turned on, the effective periodicity of $\alpha$ is modified to $2\pi N/k$. This means that under the $\IZ_k$ transformation, the partition function shifts by a phase given by:
\eq{
\delta_\alpha Z[w_2]={\rm exp}\left\{ik\alpha \frac{N-1}{N}\int \frac{w_2\wedge w_2}{2}\right\}\,Z[w_2]~.
\label{eq:mixed 0form 1form anom}
}%
Explicitly, this means that the $\IZ_k$ symmetry is broken in the fractional instanton background.  
Since this phase is only dependent on the background gauge field and transformation parameters of a global symmetry, we can interpret the phase as a mixed 0-form/1-form 't Hooft anomaly between $\IZ_k$ and 1-form center symmetry. If the gauge group were $PSU(N)$ instead, then the fractional instantons would be summed over in the path integral and the anomalous phase  \eqref{eq:mixed 0form 1form anom} becomes an ABJ anomaly which breaks $\IZ_k$ to its anomalous subgroup $\IZ_k\to \IZ_{\text{gcd} (k, N)}$.

\subsection{The Time Reversal/Center Symmetry Anomaly in Yang-Mills Theory}

Let us consider $SU(N)$ Yang-Mills theory in $4d$ . We take the action to include a $\theta$-angle:
\eq{
S=\int \frac{1}{2g^2}\Tr[F\wedge \ast F]+\frac{i \theta}{8\pi^2}\Tr[F\wedge F]~. 
}
One important  feature of this theory is that the $\theta$-term generically breaks time reversal symmetry $T$. Effectively, $T$ acts on $\theta$ as
\eq{
T:~~\theta\longmapsto -\theta~.
}
Now, because the instanton number is integer quantized in $SU(N)$ gauge theory, $\theta$ is $2\pi$-periodic, $\theta\sim \theta+2\pi$. This implies that  $T$-symmetry is  preserved when $\theta=0,\pi$ since $\pi=-\pi+2\pi$. 

As we have discussed many times by this point, 
$SU(N)$ Yang-Mills additionally has a 1-form center symmetry that measures $N$-ality of the representation of Wilson lines. As in previous discussions, we can turn on a background gauge field $\frac{B_2}{2\pi}=\frac{1}{N}w_2$ for the center symmetry. Again this changes the quantization of the instanton number
\eq{
\frac{1}{8\pi^2}\int \Tr[F\wedge F]=\frac{N-1}{N}\int \frac{w_2\wedge w_2}{2}\quad {\rm mod}_\IZ~. 
}
The effect of this fractionalization of the instanton number is that it changes the periodicity of $\theta$ to $2\pi N$. Now we see that $\theta=\pi$ is no longer symmetric under $T$ (i.e.~$\pi\neq -\pi $ mod$_{2\pi N}$). 
Rather, we now see that the partition function now transforms  under $T$-symmetry as
\eq{
\delta_TZ[w_2]={\rm exp}\left\{2\pi i\frac{N-1}{N}\int \frac{w_2\wedge w_2}{2}\right\}\,Z[w_2]~.
}
Since this phase is only dependent on background gauge fields, we can interpret this phase as a mixed 't Hooft anomaly between time reversal and center symmetry \cite{Gaiotto:2017yup}. 

Because 't Hooft anomalies are preserved under RG flows, this implies that non-abelian Yang-Mills theory cannot flow to a trivially gapped theory at $\theta=\pi$. Much of the analysis of \cite{Gaiotto:2017yup} is then dedicated to investigating the implications of this anomaly on what we know about the phase structure of the low energy effective theory. For related topics, also see \cite{Komargodski:2017keh,Gaiotto:2017tne,Gomis:2017ixy,Cordova:2017kue,Cordova:2018acb}.  

\section{Higher Group Symmetries}
\label{sec:highergroup}

As it turns out, higher form global symmetries fit into a more general  symmetry structure called higher group symmetry. Formally, higher groups can be realized as an $n$-category -- generalizing the structure of a group \cite{Baez:2004in,Baez:2010ya}. However, in physics it  is perhaps more useful to think of higher group global symmetry as the symmetry structure that arises when higher form symmetries of different degrees mix together. 

Let us consider a theory with a collection of $p_i$-form global symmetries where 
\eq{
p_1<p_2<...<p_n~.
}
Now introduce background gauge fields for the $p_i$-form global symmetries, $A_i$, which transform under the standard background gauge transformations 
\eq{\label{eqtrans}
\delta_i A_i=d\Lambda_{p_i}~. 
}

We say that the theory has a \emph{higher group global symmetry} -- in particular a $(p_n+1)$-group global symmetry, which we denote $\IG^{(p_n+1)}$ -- if the theory is  not invariant under \eqref{eqtrans}, but rather is only invariant under a more general type of background gauge transformation:
\eq{\label{highergrouptrans}
\delta  A_i=d\Lambda_{p_i}+\sum_{j\leq i}\Lambda_{p_j}\wedge \alpha_j^{(i)}(\{A_j\})+\{{\rm Schwinger~terms}\}~,
}
where $\alpha_j^{(i)}$ is a $(p_i-p_j)+1$-form that is dependent on background gauge fields $A_j$ for $j<i$ and Schwinger terms are terms that are non-linear in the background gauge transformation parameters (but again we demand that these only depend on $\{A_j,\Lambda_{p_j}\}$ for $j<i$). 

It is important to note that the background gauge fields $A_i$ only transform non-trivially under the background gauge transformations of the $p_j$-form global symmetries of lower degree ($p_{j}<p_i$). This fact leads to two important consequences \cite{Cordova:2018cvg,Brennan:2020ehu}. \\

\noindent\textbf{Breaking Patterns for $\IG^{(p_n+1)}$}

First let us consider the possible breaking patterns of the above higher group symmetry. Let us consider the (spontaneous) breaking of the top $p_n$-form global symmetry. Since the theory only preserves the modified transformations as in \eqref{highergrouptrans}, any background transformation that couples to a non-trivial $\alpha^{(n)}_j$ will generate a non-trivial $A_n$. However, since the $p_n$-form global symmetry is broken, this implies that \emph{all} background gauge transformations that couple to $A_n$ and their associated symmetries are broken. Therefore,  breaking the $p_n$-form symmetry generically breaks the entire $p_n+1$-group $\IG^{(p_n+1)}$  symmetry. 

A simple generalization of the above analysis shows that breaking the $p_i$-form global symmetry breaks all $p_j$-form symmetries for $p_j<p_i$. This leads to a nested structure of higher group symmetries 
\eq{
\IG^{(p_n+1)}=\IG^{(p_n+1)}_{p_1}\subset \IG^{(p_n+1)}_{p_2} \subset ... \subset \IG^{(p_n+1)}_{p_n}~,
}
where $\IG^{(p_n+1)}_{p_i}$ is the $(p_n+1)$-group whose lowest degree symmetry is a $p_i$-form global symmetry. 

Because of this nested $(p_n+1)$-group structure, the allowed breaking patterns are of the form 
\eq{\label{highergroupnested}
\IG^{(p_n+1)}\longrightarrow \IG^{(p_n+1)}_{p_i}\longrightarrow \IG^{(p_n+1)}_{p_{j<i}}\longrightarrow \IG^{(p_n+1)}_{p_{k<j}}\longrightarrow ... 
}

\noindent\textbf{Emergent $\IG^{(p_n+1)}$ Symmetry }

 The nested higher group structure also has an effect on the emergence of higher group symmetries along RG flows. 
 To see this, let us first consider a related phenomenon with ordinary group symmetries.  
  
  Consider a theory with  a 0-form  global symmetry $G$ which is a group extension of $H$ by an abelian group $A$:
\eq{
1\longrightarrow A\longrightarrow G \longrightarrow H\longrightarrow 1~.
}
Such an extension is determined by a function $c:H\times H\to A$ with certain nice properties (classified by $c\in H^2(H,A)$).\footnote{Technically, an extension of $H$ by an abelian group $A$ (called a central extension) is classified by two pieces of data: 1.) an action of $H$ on $A$ given by $\omega:H\to Aut(A)$ and 2.) a twisted cohomology class $c\in H^2_\omega(H,A)$. In the case where $\omega$ is trivial, $H^2_\omega(H,A)=H^2(H,A)$ and the extension is given by the data above. See \cite{GmooreGroupTheory} for more details. 
}  See \cite{Dijkgraaf:1989pz,GmooreGroupTheory} for details. 

Let us realize $G$ as a set $G=A\times H$. The non-triviality of $c$ implies that the multiplication of $h_1,h_2\in H$ generates an element of $A$:
\eq{
(1,h_1)\cdot (1,h_2)=(c(h_1,h_2)\,,\, h_1h_2)~. 
}
This means that $H$ is not a subgroup of $G$. 

Therefore, along an RG flow, a theory cannot have $H$-symmetry without also (or first) developing $A$-symmetry. This is clear because if $A$ is not a symmetry, the multiplication of two elements of $H\subset G$ can generate an element of $A$ which implies $H$ also cannot be a symmetry. Therefore, along an RG flow we find a hierarchy of scales 
\eq{
 E_A\gtrsim  E_H~,
}
where $ E_{A,H}$ are the scales at which $A,H$-symmetries emerge. 

There is a completely analogous story for higher group symmetries. Let us consider an RG flow which starts at a UV symmetry with no higher form symmetry that flows in the IR to a theory with a $p_n$-group global symmetry in the IR. Since the $\IG^{(p_n)}_{p_i}$ obey a nested structure as in \eqref{highergroupnested}, we see that along an RG flow there must also be a hierarchy of scales
\eq{
...\gtrsim E_{p_{i+1}\text{-}form}\gtrsim  E_{p_{i}\text{-}form}\gtrsim  E_{p_{i-1}\text{-}form}\gtrsim...~,
}
where $E_{p_i\text{-}form}$ is the energy scale at which the $p_i$-form symmetry emerges. 

The hierarchy of energy scales clearly gives one a great deal of information about the structure of the theory along the RG flow. Since  symmetries  often emerge when dynamical fields are either integrated out or when dynamical extended objects (such as strings or domain walls) become stable, the hierarchy of energy scales can often be translated into a statement about different coupling constants in theories that develop higher group global symmetry in the IR.  

Note that since the above bounds are not strict, the proper physical interpretation is that the bound cannot be parametrically violated:
\eq{
 E_{p_{i+1}\text{-}form}\not{\ll}E_{p_{i}\text{-}form}~.
}

We will now give a couple of examples of higher group global symmetries. For more uses of higher group global symmetries in phenomenological settings, see \cite{Cordova:2022qtz,Hidaka:2020iaz,Hidaka:2020izy,Brennan:2020ehu,Brennan:2023kpw}.

\subsection{Example: 2-Groups}

A 2-group is the simplest example  of a higher group global symmetry. They consist of a 0-form global symmetry that mixes with a 1-form global symmetry \cite{Baez:2004in,Baez:2010ya,Gukov:2013zka,Kapustin:2013uxa,Cordova:2018cvg,Benini:2018reh}. Here we will take an example that is a continuous 2-group (i.e. made up of continuous 0- and 1-form global symmetries) although more generally we can also consider 2-groups that are composed of discrete higher form symmetries. 

Let us consider the example of a $U(1)$ 0-form global symmetry that mixes with a $U(1)$ 1-form global symmetry in $4d$ . Here we have a background gauge field $A_1$ (with curvature $F_2$) for the 0-form global symmetry and $B_2$ (with curvature $H_3)$ for the 1-form global symmetry. Their transformation properties obey 
\eq{\label{2grouptrans}
\delta A_1=d\lambda_0\quad, \quad \delta B_2=d\Lambda_1-\twopi{\kappa} \lambda_0 F_2~. 
}

Theories that have 2-group global symmetry are often realized as theories that have a 1-form global symmetry and an anomalous 0-form global symmetry whose 6-form anomaly polynomial can be realized as 
\eq{
\CI_6=i\kappa \int \frac{F_2\wedge F_2\wedge \ast J_2}{4\pi}~, 
}
where $J_2$ is the current for the 1-form global symmetry. In this case, the anomalous transformation of the partition function is of the form 
\eq{
\delta_{\lambda_0}Z=\exp\left\{\twopi{i\kappa }\int \lambda_0 F\wedge  \ast J_2\right\}Z~. 
}
However, this anomalous phase can be canceled by modifying the standard transformation of the background gauge field $B_2$ to \eqref{2grouptrans} due to the coupling 
\eq{
S=...+i \int B_2\wedge  \ast J_2~.
}
Note that this cancellation is simply an application of the Green-Schwartz mechanism \cite{Green:1984sg}.

\subsubsection{Concrete Example: QED}

\label{sec2groupex}

Let us consider a concrete theory that has a 2-group global symmetry. Let us take $U(1)_g$ gauge theory with dynamical gauge field $a_g$ (with curvature $f_g$) coupled to 4 Weyl fermions $\{\psi_i^\pm\}$. In addition to being charged under the gauge group $U(1)_g$, the fermions also transform under a global flavor symmetry $U(1)_f$:

\begin{center}
\begin{tabular}{l|l|l}
Field & $U(1)_g$ & $U(1)_f$\\
\hline
$\psi_1^+$& +1 & +1\\
$\psi_2^+$& +1 & $-1$\\
$\psi_1^-$& $-1$ & +$q$\\
$\psi_2^-$& $-1$ & $-q$
\end{tabular}
\end{center}

The most general possible anomaly is of the form 
\eq{
\CI_6=\frac{1}{(2\pi)^2}\int \Big[\frac{\kappa_{g^3}}{3!}f_g\wedge f_g\wedge f_g+\frac{\kappa_{g^2f}}{2!}f_g\wedge f_g\wedge F_2+\frac{\kappa_{gf^2}}{2!}f_g\wedge F_2\wedge F_2+\frac{\kappa_{f^3}}{3!}F_2\wedge F_2\wedge F_2\Big]~.
}
By a standard computation we can compute the coefficients to be 
\eq{
\kappa_{g^3}=\kappa_{f^3}=\kappa_{g^2f}=0\quad, \quad \kappa_{gf^2}=2(1-q^2)~.
}
Let us assume that $|q|\neq 1$ so that $U(1)_f$ has an 't Hooft anomaly:
\eq{
\CI_6=\frac{2(1-q^2)}{8\pi^2}\int  f_g\wedge F_2\wedge F_2
~.
}
This leads to the anomalous variation of the path integral by inflow 
\eq{
\delta_{\lambda_0}Z=\exp\left\{\frac{2(1-q^2)i}{(2\pi)^2}\int \lambda_0 F_2\wedge f_g\right\}Z~.
}

 In addition to the (anomalous) $U(1)_f$ 0-form global symmetry, the theory also has a 1-form magnetic global symmetry $\ast J_2=\frac{1}{2\pi}f_g.$ This allows us to rewrite the anomalous variation as 
 \eq{
\delta_{\lambda_0}Z=\exp\left\{\twopi{2(1-q^2)i}\int \lambda_0 F_2\wedge \ast J_2\right\}Z~.
 }
 This symmetry can be coupled to a  background gauge field $B_2$ by adding to the action 
\eq{
S=...+i\int B_2\wedge\ast  J_2~,
}
which allows us to cancel the anomaly by modifying the transformation properties of $B_2$:
\eq{
\delta B_2=d\Lambda_1-\twopi{2(1-q^2)}\lambda_0 F_2~.
}
This is the 2-group global symmetry of the theory.

\subsubsection{Spontaneous Symmetry Breaking}

Now we can ask what happens when we spontaneously break 2-group global symmetry \cite{Cordova:2018cvg}. Let us consider the case where we have a 0-form global symmetry $G^{(0)}$ and a $U(1)^{(1)}$ 1-form global symmetry that make up a 2-group that is spontaneously broken. From what we know, the spontaneous breaking of $G^{(0)}$ gives us a $G^{(0)}$-valued Goldstone boson and the spontaneously broken $U(1)^{(1)}$ 1-form symmetry gives us a $U(1)$ gauge field. 

Let us denote the Goldstone boson $\chi_a$ and the $U(1)$ gauge field $a_1$. This theory now has a 0-form global shift symmetry of $\chi_a$ and a dynamical Goldstone gauge field $a_1$ (with curvature $f_2)$. The action can then be written as 
\eq{
S=\int \Tr[d\chi\wedge \ast d\chi]+\frac{1}{2g^2}f_2\wedge \ast f_2~.
}
This theory now has a 0-form shift symmetry of $\chi$ and a 1-form magnetic symmetry of $f_2$. Let us introduce background gauge fields $A_1$ and $B_2$ respectively. Now the background gauge fields couple to the theory as 
\eq{
S=\int \Tr[(d\chi-A_1)\wedge \ast(d \chi -A_1)]+\frac{1}{2g^2}f_2\wedge \ast f_2+2\pi i B_2\wedge f_2-\frac{i \kappa}{4\pi^2} \Tr[\chi F_2]\wedge f_2~. 
}
Now we see that the theory is invariant under the 2-group transformations 
\eq{
\chi\to \chi+\lambda \quad, \quad A_1\to A_1+d\lambda\quad, \quad B\to B+d\Lambda_1-\twopi{\kappa} \lambda F_2~.
}
This indeed reproduces the current equation 
\eq{
d j_3=\twopi\kappa F_2\wedge J_2~,
}
where now 
\eq{
j_3=\ast (d\chi-A_1)-\frac{i \kappa}{4\pi^2} d\chi \wedge f_2~. 
}

\subsubsection{Implication on IR Theory}

Now let us consider the case where the above theory is the IR limit of some UV theory which we \emph{assume has no 2-group global symmetry}. By assumption, such a UV completion must break the 0-form flavor symmetry and 1-form magnetic symmetry by dynamical objects. In particular, the breaking of 1-form magnetic symmetry requires a theory with dynamical monopoles (as in a non-abelian gauge theory). 

So let us take the UV theory to be an $SU(2)$ gauge theory (which has no magnetic 1-form symmetry) coupled to an adjoint Higgs field. We can now embed the fermions in the UV theory as two Weyl fermions in the fundamental representation. If we introduce a potential for the Higgs field so that it develops a vev, the theory will flow in the IR to a $U(1)$ gauge theory with 2 Weyl fermions of charge +1 and 2 Weyl fermions of charge $-1$. Now there is an emergent symmetry $U(1)_f\subset SU(4)$ as considered in Section \ref{sec2groupex} which mixes with the 1-form global symmetry to make up the 2-group. This symmetry is broken in the IR by the fact that it does not commute with the $SU(2)$ gauge  symmetry.
\eq{
\xymatrix{
\text{UV: }SU(2)~\text{ gauge theory, }\Phi\text{ adjoint scalar } \Psi_1,\Psi_2\text{ fund Weyl}\ar[d]\\
\big\langle Tr[\Phi]^2\big\rangle= v^2\Rightarrow SU(2)\to U(1)~,~\Psi_{1,2}\to \psi_{1,2}^\pm\ar[d]\\
\text{IR: } U(1)\text{ gauge theory with }\psi_{1,2}^\pm\text{ Weyl with 2-group}
}
}
The fact that 2-group symmetry emerges in the IR implies that constraints on the above RG flow (and indeed any RG flow that ends at our IR theory in question). The constraint is 
\eq{
E_{1\text{-}form}\gtrsim E_{0\text{-}form}~,}
where $E_{p\text{-}form}$ is the energy scale where the $p$-form symmetry emerges.

 In our above proposed UV completion $E_{1\text{-}form}$ is the scale where monopoles become stable which is determined by the scale at which non-abelian gauge symmetry is broken. This is the energy scale needed to ``unwind'' the gauge field, which is the scale necessary to activate dynamical non-abelian degrees of freedom: 
 \eq{
 E_{1\text{-}form}=m_W=gv~,
 }
 where $g$ is the gauge coupling. 
 
The scale $E_{0\text{-}form}$ is the scale where the $U(1)_f$ flavor symmetry emerges which also occurs when we spontaneously break gauge symmetry. Here, the flavor symmetry is violated by non-abelian interactions at tree level -- i.e. scattering with $W$-bosons. This means that   the energy scale $E_{0\text{-}form}$ is also given by 
 \eq{
 E_{0\text{-}form}=m_W=gv~.
 }

We then see that the hierarchy constraint from emergent 2-group symmetry is trivially satisfied since 
\eq{
E_{1\text{-}form}=E_{0\text{-}form}=m_W=gv~. 
}

\subsection{Example: 3-Groups}

A 3-group is the next simplest example of a higher group global symmetry. It consists of 0-form, 1-form, and 2-form global symmetries that mix together. Let us consider the case where the three symmetry groups $G^{(0)},G^{(1)},G^{(2)}$ are all $U(1)$. Now introduce background gauge fields $A_1,B_2,C_3$. The most general transformations of these fields are of the form\footnote{Here we restrict to terms that are gauge invariant (i.e.~not dependent on the gauge fields except through their field strength).}
\eq{
\delta A_1&=d\lambda_0~,\\
\delta B_2&=d\Lambda_1+k \lambda_0 F_1~,\\
\delta C_3&= d\Lambda_2+\alpha \, d\lambda_0 \wedge F_1+\beta_1\,  \Lambda_1\wedge B_2+\beta_2\,  \Lambda_1\wedge d\Lambda_1+\kappa_1\,  \Lambda_1\wedge F_2+...
}
Due to the complexity of these transformation properties, it is perhaps clearer to illustrate the structure of the 3-group with an  example. 

\subsubsection{Axion-Yang-Mills}

Here we will take a slightly different example in which $G^{(0)}$ is trivial, and $G^{(1)}=\IZN^{(1)}$. 

Let take $4d$  $SU(N)$ axion-Yang-Mills \cite{Brennan:2020ehu,Cordova:2019uob,Hidaka:2020iaz,Hidaka:2020izy}. This theory is described by the action 
\eq{
S=\half \int da\wedge \ast da+\frac{1}{g^2}\int \Tr[F\wedge \ast F]- \frac{i}{8\pi^2 f_a}\int a\,\Tr[F\wedge F]~,
}
where $a\sim a+2\pi f_a$. 

This theory has two higher form global symmetries: $\IZ_N^{(1)}$ 1-form center symmetry and $U(1)^{(2)}$ 2-form axion string symmetry. Here $\IZ_N^{(1)}$ is the standard center symmetry that shifts the gauge field by a $\IZ_N$ gauge field.  The $U(1)^{(2)}$ magnetic string symmetry has  a current given by $\ast J_3=\frac{1}{2\pi f_a}da$. 
As discussed in Section \ref{periodicscalar}, this symmetry counts the winding charge of the axion field. 

We can now couple the theory to background fields $B_2$, $C_3$ for $\IZ_N^{(1)}$ and $U(1)^{(2)}$:
\eq{
S=&\half \int da\wedge \ast da+\frac{i}{2\pi f_a} \int a\, dC_3+\frac{1}{g^2}\int \Tr[(F-B_2)\wedge \ast (F-B_2)]\\
&- \frac{i}{8\pi^2 f_a}\int a\,\Tr[(F-B_2)\wedge (F-B_2)]+\int \varphi\, \Tr[F-B_2]~. 
}
Note that here we have chosen to integrate the coupling of $C_3$ to $\ast J_3$ by parts.

 Now we see that turning on the background gauge field $B_2$ causes the axion coupling to be ill-defined. In other words, the fact that the instanton number is now fractional 
\eq{
\frac{N}{8\pi^2}\int \Tr[(F-B_2)\wedge (F-B_2)]
\in \IZ~,
}
implies that the action is no longer invariant under $a\to a+2\pi f_a$:
\eq{
\delta_a S=-\frac{2\pi N(N-1)}{8\pi^2}\int B_2\wedge B_2\neq 2\pi i n ~,
}
where $n \in \IZ$. Since this is a gauge transformation that makes the axion a periodic scalar, we would say that the 1-form symmetry is explicitly broken.  
 
However, the offending variation of the action can be eliminated  by modifying the background gauge transformations for $C_3$. In particular, we can elevate its field strength
\eq{
G_4=dC_3+\frac{N(N-1)}{4\pi}B_2\wedge B_2~,
}
which we demand is gauge invariant under gauge transformations of $\IZ_N^{(1)}$ and $U(1)^{(2)}$. This implies that the fields transform under the 3-group background as transformations 
\eq{
B_2&\longmapsto B_2+d\Lambda_1~,\\
C_3&\longmapsto C_3+d\Lambda_2-\frac{N(N-1)}{2\pi}\Lambda_1\wedge B_2-\frac{N(N-1)}{4\pi}\Lambda_1\wedge d\Lambda_1~.
}
These transformations indicate a 3-group symmetry structure  without a 0-form global symmetry. 

This theory has several variations that include modifying the axion coupling and adding charged matter fields which all have more interesting/complicated 3-group symmetries. For further details see \cite{Brennan:2020ehu}. 

\subsubsection{Constraints on RG Flow}

We can now infer information about UV completions of our theory by using the constraints from emergent 3-group symmetry along RG flows that end at axion-Yang-Mills:
\eq{\label{3groupconst}
E_{2\text{-}form}\gtrsim E_{1\text{-}form}~.
} 

Let us examine this hierarchy of scales in a simple UV completion. Take the UV theory to be a $SU(N)$ gauge theory with a pair of Weyl fermions $\psi_\pm$ in conjugate (fundamental/anti-fundamental) representations and an uncharged complex scalar field $\varphi(x)$. Let us take the action to be 
\eq{
S=\int d^4x\left\{\frac{1}{g^2}\Tr[F\wedge \ast F]+i \bar\psi_\pm \slashed{D}\psi_\pm +\half d\varphi \wedge \ast d\bar\varphi-V(\varphi)+\lambda \bar\varphi \psi_+\psi_-+\lambda \varphi \bar\psi_-\bar\psi_+\right\}~,
}
where we take the potential 
\eq{
V(\varphi)=m^2(|\varphi|^2-f_a^2)^2~.
}

This theory has a classical $U(1)$ global symmetry which we will denote $U(1)_{PQ}$  under which the matter fields have charges
\begin{center}
\begin{tabular}{l|l}
Field & $U(1)_{PQ}$ charge\\
\hline
$\psi_\pm $ & +1\\
$\varphi $ & +2
\end{tabular}
\end{center}

\noindent This symmetry suffers from an ABJ anomaly so that the rotation 
\eq{
\psi_\pm \to e^{ i \alpha} \psi_\pm~,
}
generates a term 
\eq{
\frac{i \alpha}{8\pi^2}\int \Tr[F\wedge F]~,
}
in the action. 

Now let us flow to the IR. Along the flow, the scalar field will obtain a vev $\big\langle |\varphi|\big\rangle=f_a$. Below the scale $\sqrt{m}f_a$, the scalar field is classically unable to reach the local maximum of the scalar potential so that the field is effectively valued in an interval$\times S^1$ and the axion strings become stable. Thus, in the effective theory at energies $E<\sqrt{m}f_a$, it is appropriate to describe the field as 
\eq{\label{radialexp}
\varphi=(f_a+\rho)e^{i a/f_a}~,
}
where $\rho,a$ are dynamical degrees of freedom. 

Now the Yukawa term becomes 
\eq{
\CL_{Y}=\lambda f_a e^{i a/f_a}\psi_+\psi_-+...
}
We can then perform a chiral rotation of the $\psi_\pm$ to move the coupling to $a(x)$ to an axion term
\eq{
S_{ax}=\frac{i}{8\pi^2f_a}\int a\,\Tr[F\wedge F]~. 
}
Now, the Yukawa term has generated a mass term for the fermions where $m_\psi=\lambda f_a$. Thus, if we flow to the IR below the mass scale of the radial mode $\rho$ and the fermions $\psi_\pm$, we can trivially integrate out the massive fields $\rho,\psi$. Then, the theory flows to pure axion-Yang-Mills which has an emergent 3-group symmetry. 

Now we can again consider the  3-group constraint \eqref{3groupconst}.  
Here the 2-form symmetry is emergent when the axion strings becomes stable. As we discussed above, this occurs when the scalar field $\varphi$ becomes effectively interval$\times S^1$-valued  at  $E_{2\text{-}form}\sim \sqrt{m}f_a$.  
Additionally, the 1-form center symmetry emerges when the fundamental fermion fields are integrated out which allows us to identify $E_{1\text{-}form}=m_\psi=\sqrt{m}f_a$. 

Now by putting these together, we see that the bound from emergent 3-group symmetry implies
\eq{
\sqrt{m} \gtrsim \lambda ~.
}
As it turns out this bound coincides with the bound from EFT. As shown in \cite{Brennan:2020ehu} when $\sqrt{m}\ll\lambda$ the fermion contribution to the scalar potential is large and outside of the control of EFT. 

\section{Non-Invertible Symmetries}
\label{sec:NIS}

As we have stressed, the notion of global symmetries in quantum theories is best phrased in terms of the topological operators of the theory. Thus far, we have focused on the theories where the topological operators have additional special properties such as obeying a group multiplication law. However, topological operators more generally are described by the  story of ``categorical'' global symmetries. In general, categorical symmetries do not obey a group like structure. Because of this, we often call the symmetries which are not group-like or higher group-like \emph{non-invertible} global symmetries. 

There is a lot of mathematical structure that is used to describe the general categorical symmetries -- this structure gets more and more complicated in higher and higher dimensions. Here it is our goal to avoid category theory, but the subject is large and incredibly rich/interesting! We direct the interested reader to \cite{Bhardwaj:2023ayw,Bartsch:2023wvv,Schafer-Nameki:2023jdn,Bhardwaj:2022lsg,Bhardwaj:2022yxj,Bartsch:2022ytj,Bartsch:2022mpm,Freed:2022qnc,Freed:2022iao,Kaidi:2022cpf,Choi:2022jqy,Choi:2022zal,Choi:2022fgx,Choi:2022rfe,Cordova:2022ieu,Bhardwaj:2022kot,Bhardwaj:2022maz} for more details.

Here we would like to describe a couple constructions of non-invertible symmetries that arise in Lagrangian theories. The main idea of a non-invertible symmetry is the following. Consider a set $\CC$  to which we associate topological defect operators $\CD_x$ for $x \in \CC$. As we have discussed, if $\CD_x$ were group-like symmetry defect operators, their multiplication would obey some group structure: $\CD_{x_1}\times \CD_{x_2}=\CD_{x_1x_2}$. However, the fusion of two defect operators can obey a more general categorical structure which can take the form 
\eq{
\CD_{x_1}\times \CD_{x_2}=\begin{cases}
\sum_i \CD_{x_i}\\
\CT\otimes \CD_{x_1x_2}\\
\quad\vdots
\end{cases}
}
for example, where here $\CT$ is the partition function of some $3d$ TQFT.

The linchpin for constructions  in Lagrangian theories of non-invertible symmetries is the construction of what are called condensation defects.  The idea and terminology name for condensation defects comes from condensed matter.  The general idea is that in a $d$-dimensional QFT, we can gauge a global symmetry along a fixed $p$-dimensional hypersurface.  This allows us to define a larger class of topological defects in the theory -- and therefore allows us to construct a larger class of global symmetries. 

This idea of condensation defects is a bit esoteric, so let us relate this idea to something that is perhaps more familiar. The idea of condensation defects is somewhat similar to the  idea of having a defect with a non-trivial world volume theory such as axion strings or monopoles. However, for condensation defects the defect theory is a gauge theory that couples to the bulk via a conserved current. In particular, we can think of condensation defects as theories which flow to a TQFT in the IR that couples  to the bulk theory only via a gauge-current interaction.

In this section, we plan to first discuss the construction of condensation defects. Then we will go on to discuss two examples of non-invertible symmetries in Lagrangian theories: non-invertible chiral symmetries and non-invertible 1-form symmetries in axion-electromagnetism.

\subsection{Condensation Defects}

Let us consider a $d$-dimensional theory with a $p$-form global symmetry $G^{(p)}$. For the moment let us assume that the symmetry has no obstructions to being gauged -- i.e. no ABJ anomaly or `t Hooft anomaly only involving itself. The symmetry $G^{(p)}$ can be gauged in the full space time. As we have discussed this can be formally treated by summing over all possible insertions of the associated symmetry defect operators in the path integral. 

In our theory, we can consider removing a co-dimension $q$ hypersurface $\Sigma^{(q)}$. We can then introduce a ``defect'' QFT that lives on this $\Sigma^{(q)}$ that couples to the bulk theory. This theory by virtue of coupling to the bulk theory admits an action of the symmetry $G^{(p)}$. This can be seen by taking charged operators in the bulk theory and dragging them to the defect -- this leads to operators in the defect theory that must also necessarily be charged under the action of the bulk $G^{(p)}$. However, we are also allowed to couple to QFTs where $G^{(p)}$ is gauged on $\Sigma^{(q)}$. In this case, the result of dragging a charged operator to $\Sigma^{(q)}$ will lead to charged operator in the defect theory which will in general be dressed by a Wilson operator to make a gauge invariant operator. 
When we are allowed to gauge the symmetry $G^{(p)}$ on any $\Sigma^{(q)}$, we say that the theory is \emph{$q$-gaugeable}. 

Formally, we can think about gauging a $q$-gaugeable $p$-form symmetry $G^{(p)}$ along a co-dimension-$q$ hypersurface $\Sigma^{(q)}$ by summing over the insertion of all $(d-q-p-1)$-dimensional  symmetry defect operators restricted to $\Sigma^{(q)}$. Practically, we can think of gauging $G^{(p)}$ along $\Sigma^{(q)}$ by introducing an associated gauge field $B_{p+1}$ that only lives on $\Sigma^{(q)}$. Since these $G^{(p)}$ symmetry defect operators restricted to $\Sigma^{(q)}$ are not the same as the bulk $G^{(p)}$ symmetry defect operators, the restricted $G^{(p)}$ can have fewer `t Hooft anomalies than the bulk $G^{(p)}$. This allows some symmetries to be $q$-gaugeable for $q>p+1$ even though the bulk $G^{(p)}$ symmetry may have obstructions to its gauging. In essence, restricting to a co-dimension-$q$ hypersurface can sometimes trivialize anomalies of the bulk $G^{(p)}$ symmetry. 
 
As in the ordinary procedure of gauging of any symmetry, we are allowed to add a phase that weights the different gauge sectors -- i.e. add a non-trivial action. 
For our purposes, we will restrict to the case where we only consider phases  that are topological (this will be important for constructing non-invertible symmetry defect operators). This class of local actions are given by gauge-invariant Chern-Simons-type actions of $B_{p+1}$. These Chern-Simons-type actions are also referred to as $(d-q)$-dimensional SPT phases that are described by cohomology classes $\eta\in H^q(B^pG,U(1))$. 

Formally, we can think of the condensation defects as operators we insert into the path integral. Formally, we can write these as 
\eq{
\CC_\eta(\Sigma^{(q)})=\frac{1}{\CN}\sum_{B_{p+1}}\exp\left\{i \int \eta[B_{p+1}]\right\}~,
}
where here the sum enforces the gauging, $\eta[B_{p+1}]$ is the Chern-Simons-type action, and $\CN$ is a normalization factor that captures the space of gauge inequivalent $B_{p+1}$. 

\subsubsection{Example: 1-Gauging $\IZ_N^{(1)}$ in 4d}

Let us illustrate these ideas by showing how to 1-gauge a 1-form $\IZ_N^{(1)}$ global symmetry in a $4d$ theory. Here, we will have in mind the case where we gauge a $\IZ_N^{(1)}$ magnetic symmetry of a  gauge theory.

Let us fix a $3d$ hypersurface $\Sigma^{(1)}$, and consider 1-gauging $\IZ_N^{(1)}$ on $\Sigma^{(1)}$. On $\Sigma^{(1)}$, the magnetic 1-form global symmetry is generated by line operators (the same way that $3d$ Chern-Simons theory has a 1-form symmetry whose lines are both the charged operators and symmetry defect operators). We can then 1-gauge the $\IZ_N^{(1)}$  by coupling to a $\IZ_N$ 1-form gauge field $v_1$ that only lives on $\Sigma^{(1)}$. In the case where we are 1-gauging a $\IZ_N^{(1)}$ magnetic symmetry, we can couple this gauge field to the bulk symmetry as 
\eq{\label{CCfirstminimalex}
\CC_{0}(\Sigma^{(1)})=\int [dv_1] \,e^{ i  \int v_1\wedge \frac{f}{2\pi}}~, 
}
where here the path integral over $v_1$ is restricted to gauge fields on $\Sigma^{(1)}$.  

We can then couple the condensation defect by a SPT phase -- i.e. Chern-Simons term -- so that we get an expression of the form 
\eq{\label{CNcondensation}
\CC_{N}(\Sigma^{(1)})=\int [dv_1]~{\rm exp}\left\{\frac{iN}{4\pi}\int v_1\wedge dv_1+\frac{i }{2\pi} \int v_1\wedge f\right\}~.
}
In the case where $p=1$, this particular choice of condensation defect is an example of a minimal $3d$ $\IZ_N$ TQFT which in the literature is often denoted $\CA^{N,1}(f)$  \cite{Hsin:2018vcg}. This also describes the $3d$ fractional quantum Hall state at filling fraction $\nu=1/N$.

\subsection{Example: Non-Invertible Chiral Symmetry in $4d$}

In this section, we discuss non-invertible chiral symmetries appearing in $4d$ gauge theories with fermions. As we will see, these symmetries arise in the study of the Standard Model  and may have meaningful applications to BSM model building. These symmetries were first introduced in \cite{Cordova:2022ieu,Choi:2022jqy} where it was used in the study of pions and axions.  Subsequently, non-invertible symmetries have also been 
applied to explain the smallness of the neutrino mass \cite{Cordova:2022fhg} and further study axion theories in \cite{Choi:2022rfe}.

Let us take $4d$ QED: $U(1)$ gauge theory with a Dirac fermion $\psi$ of charge $1$. This has an action
\eq{\label{CANonTop}
S=\int \frac{1}{2g^2}f\wedge \ast f+i \bar\psi \slashed{D}\psi~,
}
where $f = dA$ is the dynamical field strength. 
This theory has a $U(1)_A$ chiral symmetry that acts as\footnote{In the limit where the gauge coupling vanishes, $g \to 0$, the free Dirac theory has $U(1) \times U(1)$ global symmetry: $U(1)$ rotation for each Weyl fermions in the Dirac fermion $\psi = (\chi, \epsilon \psi^*)^T$. The vector-like combination is free of cubic anomaly and is the gauged $U(1)$ in $4d$ QED. The orthogonal combination is $U(1)_A$.}
\eq{
\psi\longmapsto e^{ i \alpha \gamma_5}\psi\quad, \quad j_A^\mu(x)=\bar\psi \gamma^\mu \gamma^5\psi(x)~.
}
However, the $U(1)_A$ chiral symmetry has an ABJ anomaly which is given by: 
\eq{\label{eq:NIS_ABJ_anomaly}
d\ast j_A=\frac{2}{8\pi^2}f\wedge f~.
}
This means that if we try to construct a symmetry defect operator for $U(1)_A$
\eq{\label{eq:NIS_naive_SDO}
U_g(\Sigma_3)={\rm exp}\left\{i \alpha \oint_{\Sigma_3} \ast j_A\right\}~,
}
we see that the operator is not topological 
\eq{\label{NISDOphase}
U_g(\Sigma_3)=U_g(\Sigma_3^\prime)\times {\rm exp}\left\{2i \alpha \int_{\Gamma_4} \frac{f\wedge f}{8\pi^2}\right\}~,
}
where $\Gamma_4$ is a 4-manifold that fills in the space between $\Sigma_3,\Sigma_3^\prime$: $\partial \Gamma_4=\Sigma_3\cup \overline{\Sigma_3^{\prime}}$ where $\overline{\Sigma_3^{\prime}}$ is the orientation reversal of $\Sigma^\prime$. 

Starting from the anomaly equation \eqref{eq:NIS_ABJ_anomaly}, one may try to construct a conserved current 
\beq
* \widehat{j}_5 = * j_A - \frac{2}{8\pi^2} A \wedge d A~,
\eeq
and a associated symmetry defect operator
\beq\label{SDOattempt}
\widehat{U}_g (\Sigma_3) = \text{exp} \left\lbrace i \alpha \oint_{\Sigma_3} \ast \widehat{j} \right\rbrace= \text{exp} \left\lbrace i \alpha \oint_{\Sigma_3} \left(\ast j_A - \frac{2}{8\pi^2} A \wedge d A\right) \right\rbrace~.
\eeq
The conservation of the new current $d * \widehat{j}_5 = 0$ ensures that this new symmetry defect operator is in fact topological. The issue with this construction is that neither the current $\widehat{j}_5$ nor the symmetry defect operator $\widehat{U}_g (\Sigma_3) $ is gauge invariant and hence they are ill-defined. This exercise, however, provides a useful viewpoint. 

The above operator is ill-defined because  of the Chern-Simons term which is in general not gauge invariant for arbitrary $\alpha$. To see this more explicitly, we recall that the $3d$ Chern-Simons action
\beq
S_{\rm \scriptscriptstyle CS} = \frac{iK}{4\pi} \int A \wedge dA~,
\eeq
is only gauge-invariant for $K \in 2 \IZ$ (or for $K \in \IZ$ on a spin manifold). This can be seen by noting that under a large gauge transformation $A \to A + \Lambda_1$, the action shifts as
\beq
\delta S_{\rm \scriptscriptstyle CS} = \frac{iK}{4\pi} \int \Lambda_1 \wedge d \Lambda_1 + \frac{iK}{4\pi} \int 2 \Lambda_1 \wedge d A~.
\eeq
Using $\oint \Lambda_1 = 2\pi \IZ$, one obtains the quantization condition for $K$. 

Naively, the fact that generically the modified (topological) defect operator \eqref{SDOattempt} has improperly quantized Chern-Simons level for generic $\alpha \in U(1)_A$ would be interpreted as an explicit violation of the classical $U(1)_A$ symmetry by quantum mechanical effects (reflecting the  ABJ anomaly). One would not expect any useful selection rules imposed by $U(1)_A$ to hold in any quantum process. On the other hand, one may remember that the abelian instanton does not exist on $\IS^4$ (since $\pi_3 (U(1)) = \emptyset$) and wonder if above described effect is completely absent. This observation, suggests that there may be more going on.  

To investigate, let us consider a rational angle $\alpha = \frac{2\pi}{N}$. The Chern-Simons term in $\widehat{U}_g (\Sigma_3)$ is given by
\beq
\tilde{S}_{\rm \scriptscriptstyle FQH} = - \frac{i}{4\pi N} \oint_{\Sigma_3} A \wedge dA.
\label{eq:FQH_1}
\eeq
This is known in condensed matter literature as the action for the fractional quantum hall state in $(2+1)d$ at filling fraction $\nu = \frac{1}{N}$.  
As it turns out, there is a gauge-invariant description which is equivalent to this naively ill-defined action. 

The gauge-invariant action for the fractional quantum hall state in $(2+1)d$ at filling fraction $\nu = \frac{1}{N}$ is 
\beq
S_{\rm \scriptscriptstyle FQH} = \int_{\Sigma_3} \left(\frac{iN}{4\pi} a \wedge da + \frac{i}{2\pi}  a \wedge dA\right)~,
\label{eq:FQH_2}
\eeq
as we have discussed above. 
The theory of fractional hall state with $\nu = 1/N$   is also known as $\CA^{N,1}(f,\Sigma_3)$ which is the ``minimal $\IZ_N$ TQFT'' \cite{Hsin:2020nts}. More generally, for $\alpha=\frac{2\pi p}{N}$ we can construct a co-dimension 1 topological operator by dressing the naive symmetry operator a fractional quantum hall state with $\nu=p/N$  (where $\text{gcd} (N,p) = 1$) which is denoted by $\CA^{N,p}(f,\Sigma_3)$:
\eq{\label{eq:NIS_Dq}
\CD_q(\Sigma_3)={\rm exp}\left\{2\pi i\, q \oint \ast j_A\right\}\times \CA^{N,p}(f,\Sigma_3)\quad, \quad q=\frac{p}{N}\in \IQ/\IZ~.
}
The general $\CA^{N,p}(f,\Sigma_3)$ are somewhat complicated define as an intrinsically $3d$ theory. It is easier to describe as the boundary state of a $4d$ TQFT which we will do in the next section.

\subsubsection{Half-Space Gauging Construction}
\label{subsubsec:NIS_half_space}

In this section, we will discuss the construction of the non-invertible chiral symmetry defect operator \eqref{eq:NIS_Dq} by means of ``half-space gauging.''  
This will clarify the topological nature of the non-invertible SDO and also illustrate the role of magnetic 1-form symmetry. This latter point is especially important when one thinks about the \emph{breaking} of non-invertible symmetry as discussed in \cite{Cordova:2022ieu,Cordova:2022fhg}. 

We start by noting that QED has a $U(1)$ magnetic 1-form symmetry as implied by the Bianchi identity $d f = 0$. We imagine dividing the space time $M_4$ into two pieces by a 3-manifold $\Sigma_3$ and let us take $x\in \IR$ to be a normal a coordinate direction.  

We define the insertion of the fractional quantum Hall state $\CA^{N,p}(f,\Sigma)$ by gauging $\IZ_N^{(1)}$ subgroup of $U(1)$ 1-form magnetic symmetry on the ``half'' of $M_4$ given by $x \geq 0$ where we add to the action  
\beq
\CA^{N,p}(f,\Sigma)~\Longleftrightarrow~
\Delta S =
 \int_{x\geq0}\left( \frac{i}{2\pi} b_2 \wedge f + \frac{iN}{2\pi} b_2 \wedge d c_1 + \frac{Nk}{4\pi} b_2 \wedge b_2 \right)~.
\label{eq:NIS_half_space_gauging}
\eeq
Here $pk=1~{\rm mod}_N$ (i.e.~$k$ is the multiplicative inverse of $p$ modulo $N$) and we require $\text{gcd} (k, N) = 1$ (which is equivalent to $\text{gcd} (p, N) = 1$). We also impose a topological boundary condition $b_2 \vert_{x=0} = 0$. This condition is topological because $b_2 \vert_{x=0} - b_2 \vert_{x'=0} = d b_2 = 0$.

In the half-space TQFT action, the first term describes the coupling of the 1-form magnetic symmetry current to a 2-form background gauge field $b_2$, while the last two terms are the $4d$ $\IZ_N$ BF theory with a SPT term (see Section~\ref{sec:discrete} for details). Here, $b_2$ and $c_1$ are dynamical $U(1)$ 2- and 1-form gauge field which, due to the BF action, effectively gauge $\IZ_N^{(1)} \subset U(1)_m^{(1)}$.

We will now show that this gauging has an effect of shifting the $\theta$ parameter of QED by
\beq
\theta \longmapsto \begin{cases} \theta - \frac{2\pi p}{N}&x\geq0\\
\theta&x<0
\end{cases}
\label{eq:NIS_theta_shift}
\eeq
To see this, let us consider the gauge field part of the action with $\theta$ angle:
\eq{
S = \int \left(\frac{1}{2g^2} f \wedge * f + \frac{i\theta}{8\pi^2} f \wedge f \right)+
 \int_{x\geq0}\left( \frac{i}{2\pi} b_2 \wedge f + \frac{iN}{2\pi} b_2 \wedge d c_1 + \frac{Nk}{4\pi} b_2 \wedge b_2 \right)~.
 }
  First note that integrating out $c_1$ constrains $b_2$ to be a $\IZ_N$ field and leads to a theory described by the action
\eq{
S =& \int\left(\frac{1}{2g^2} f \wedge * f + \frac{i\theta}{8\pi^2} f \wedge f\right) + \int_{x\geq0}\left(\frac{i}{2\pi} b_2 \wedge f + \frac{Nk}{4\pi} b_2 \wedge b_2\right) \\
=& \int\left(\frac{1}{2g^2} f \wedge * f + \frac{i\theta}{8\pi^2}    f \wedge f \right)+ \int_{x\geq0}\left( \frac{i Nk}{4\pi} \left( b_2 + \frac{p}{N} f \right)^2 - \frac{2\pi ip}{8\pi^2 N}  f \wedge f\right)~.
}
We can then perform the Gaussian integral over $b_2$ and we arrive at the QED action with additional term defined on $x\geq0$:
\eq{
S 
=& \int\left(\frac{1}{2g^2} f \wedge * f + \frac{i\theta}{8\pi^2}    f \wedge f \right) - \frac{2\pi ip}{8\pi^2 N} \int_{x\geq0}  f \wedge f~.}
We can rewrite this action as 
\eq{
S=&\int\frac{1}{2g^2} f \wedge * f + \frac{i}{8\pi^2}\int \left(\theta-  \frac{2\pi ip}{ N} \Theta(x)\right)  f \wedge f ~,
}
where $\Theta(x)$ is the Heaviside step function. Therefore, we see that inserting the operator $\CA^{N,p}(f,\Sigma_3)$ effectively shifts the $\theta$-parameter across $\Sigma_3$ as in \eqref{eq:NIS_theta_shift}.

From this, we can see that the effect of the half-space gauging is not independent of the choice of $\Sigma_3$ (i.e. the boundary dividing the two halves of space time). Because of the  difference of the (spatially dependent) effective $\theta$-angle between the two disconnected regions, we can see explicitly that the effect of deforming the operator $\CA^{N,p}(f,\Sigma_3)$ (or in other words the effect on contribution to the action \eqref{eq:NIS_half_space_gauging} of deforming the boundary $\Sigma_3$) is:
\eq{\label{MinimalAVar}
\CA^{N,p}(f,\Sigma_3)=\CA^{N,p}(f,\Sigma_3^\prime)\times {\rm exp}\left\{
-\frac{ip}{4\pi N}\int_{\Gamma_4} f\wedge f
\right\}~, 
}
where $\Gamma_4$ is a 4-manifold that fills in the space between $\Sigma_3,\Sigma_3^\prime$: $\partial \Gamma_4=\Sigma_3\cup \overline{\Sigma_3^{\prime}}$ where $\overline{\Sigma_3^{\prime}}$ is the orientation reversal of $\Sigma^\prime$. 

We now see by comparing the non-topological phase of the naive symmetry defect operator \eqref{NISDOphase} to that of \eqref{MinimalAVar}, we see that the operator constructed by dressing the naive symmetry defect operator by $\CA^{N,p}(f,\Sigma_3)$
\eq{
\CD_q(\Sigma_3)={\rm exp}\left\{2\pi i\, q \oint \ast j_A\right\}\times \CA^{N,p}(f,\Sigma_3)\quad, \quad q=\frac{p}{N}\in \IQ/\IZ~,
}
is indeed topological and hence defines a (generalized) symmetry for our theory.

\subsubsection{Non-Invertibility and Fusion Algebra}

While non-invertible symmetry defect operators are topological and hence define a symmetry, they do not obey the group structure of $\IQ/\IZ\subset U(1)$ and therefore define a non-invertible symmetry. We can demonstrate this explicitly by computing the product of $\CD_{1/N}$ and $\CD_{-1/N}$. 

If the $\CD_q$ obeyed the multiplicative group law of $\IQ/\IZ$, we would expect that $\CD_{1/N}\times \CD_{-1/N}=\mathds{1}$. 
First note that the factors of ${\rm exp}\left\{ \frac{2\pi i\,p}{N} \oint\ast j_5\right\}$ obey an invertible multiplication law following from the computations of Section \ref{sec:continuous higher-form}. We then need to compute the product of the $3d$ minimal TQFTs. 

In the case of $\CD_{\pm 1/N}$, we can write down $\CA^{N,1}(f,\Sigma_3)$ as a simpler condensation operator as in \eqref{CNcondensation}:
\eq{
\CA^{N,1}(f,\Sigma_3)=
\CC_N(\Sigma_3)=\int [dv_1]~{\rm exp}\left\{
\frac{iN}{4\pi}\int v_1\wedge dv_1+\frac{i}{2\pi}\int v_1\wedge f
\right\}~.
}
 We can use this expression to  explicitly compute the product 
 \eq{
\CD_{1/N}\times \CD_{-1/N}=\int [dv_1][dv_2]~{\rm exp}\left\{
\frac{iN}{4\pi}\int (v_1\wedge dv_1-v_2\wedge dv_2)+\frac{i}{2\pi}\int (v_1-v_2)\wedge f
\right\}~,
}
which simplifies by performing a change of variables $v_\pm=\half(v_1\pm v_2)$:
\eq{\label{onlycomputableCD}
\CD_{1/N}\times \CD_{-1/N}=\int [dv_\pm]~{\rm exp}\left\{
\frac{iN}{2\pi}\int v_+\wedge dv_-+\frac{i}{2\pi}\int v_-\wedge f
\right\}\neq \mathds{1}~,
}
which is equal to a type of condensation operator. 

We can also see that the $\IQ/\IZ$ group structure is also broken by computing $\CD_{1/N}\times \CD_{1/N}\neq \CD_{2/N}$:
\eq{
\CD_{1/N}\times \CD_{1/N}= e^{\frac{4\pi i}{N}\int \ast j_5}\, \int [dv_1][dv_2]~{\rm exp}\left\{
\frac{iN}{4\pi}\int (v_1\wedge dv_1+v_2\wedge dv_2)+\frac{i}{2\pi}\int (v_1+v_2)\wedge f
\right\}~,
}
which again simplifies by the same change of variables
\eq{
\CD_{1/N}\times \CD_{1/N}&=e^{\frac{4\pi i}{N}\int \ast j_5}\,\int [dv_\pm]~{\rm exp}\left\{
\frac{iN}{4\pi}\int (v_+\wedge dv_++v_-\wedge dv_-)+\frac{2i}{2\pi}\int v_+\wedge f
\right\}\\
&=\CA^{N,2}(f,\Sigma_3)\times \CD_{2/N}~,
}
for $N$ odd. Here the equality follows from some non-trivial identities of the $\CA^{N,p}(f)$ TQFTs. We refer the interested reader to \cite{Hsin:2018vcg} for more details on the general computation. 

While we have not given all of the details to fully compute the products of the $\CD_q$ operators, it should be clear even from \eqref{onlycomputableCD} that the topological operators $\CD_q$, while labeled by elements of a set $\IQ/\IZ$ (which can be given a group structure), define a non-invertible symmetry. 

\subsection{Example: Non-Invertible Symmetries in Axion-Electromagnetism}

Let us now demonstrate another example of a non-invertible symmetry in $4d$: axion electromagnetism. 
This theory also  contains several types of non-invertible global symmetries. 
Let us take the action 
\eq{
S=\int \frac{1}{2g^2}f\wedge \ast f+(\partial a)^2+\frac{i}{8\pi^2 f_a}\int a\,f\wedge f~,
}
where $a\sim a+2\pi f_a$ and $f=dA$ is the $U(1)$ field strength. It is commonly said that the axion, being a periodic boson, has a continuous shift symmetry that is broken by the axion interaction term. Under a shift $\delta a=\alpha\, f_a$ for $\alpha\in U(1)$, we see that the action shifts
\eq{
\delta S=i \alpha \int \frac{f_2\wedge f_2}{8\pi^2}~. 
}
This variation of the action is identical to that of a chiral symmetry with an ABJ anomaly. 
Given our previous discussion of the non-invertible chiral symmetry, we see that the axion shift symmetry can be restored to a $\IQ/\IZ$ non-invertible shift symmetry in exactly the same way. If we define the shift symmetry current as 
\eq{
\ast J_1=\ast da~,
}
then we can define the non-invertible symmetry defect as 
\eq{
\CD_q=e^{ 2\pi i q\int \ast J_1}\times  \CA^{N,p}(f,\Sigma_3)\quad, \quad q=\frac{p}{N}\in \IQ/\IZ~.
}
This result is perhaps obvious if we note that we can realize the axion as the pseudo-Goldstone mode for an ABJ-anomalous chiral symmetry as in the KSVZ model:\footnote{
The KSVZ model flows to axion-electromagnetism by noting that $\phi$-condenses and gives a mass to the fermions. Then, when integrating out the fermions, the chiral phase mode $\psi\mapsto e^{i \alpha \gamma_5/2}\psi$ gives rise to the axion by the fact that the associated symmetry is an ABJ-anomalous symmetry (note that this symmetry also rotates $\phi\mapsto e^{-i \alpha}\phi$). In modern language, the non-invertible chiral symmetry enforces a sort of Goldstone theorem so that the symmetry which is spontaneously broken by the $\phi$-condensation so that the IR theory has an axion \cite{GarciaEtxebarria:2022jky}. 
}
\eq{
S_{\rm KSVZ}=\int \frac{1}{2g^2}f\wedge \ast f+(\partial \phi)^2+i \bar\psi\slashed{D}\psi-m(|\phi|^2-f_a^2)^2+ i\lambda \phi \psi \psi+c.c.
}
Note that in this example the non-invertible chiral symmetry of the UV KSVZ theory is matched by the non-invertible shift symmetry of the IR axion theory. 

Axion electrodynamics also has a non-invertible 1-form symmetry \cite{Choi:2022fgx}. Let us consider the electric 1-form global symmetry of free electromagnetism. It has an associated current 
\eq{
\ast J_2^{(m)}=\frac{1}{g^2}\ast f~. 
}
When coupling to the axion theory, this group-like symmetry is also broken by the interaction:
\eq{
d\ast J_2^{(e)}=\frac{1}{4\pi^2 f_a}da\wedge f~.
}
As we have discussed previously, this theory also has a conserved $U(1)^{(1)}$ 1-form magnetic symmetry and $U(1)^{(2)}$ 2-form axion string symmetry which have associated conserved currents:
\eq{
d\ast J_2^{(m)}=d\frac{f}{2\pi}=0\quad, \quad d\ast J_3=d\frac{da}{2\pi f_a}=0~.
}
Now, because the violation of the conservation equation for $J_2^{(e)}$ can be written in terms of currents:\footnote{For the case of axion shift/chiral symmetry, the conservation equation can also be written as 
\eq{
d\ast J_1=\half \ast J_2^{(m)}\wedge \ast J_2^{(m)}~,
} which is the reason that we are allowed to ``save'' the ABJ-anomalous symmetry. }
\eq{
d\ast J_2^{(e)}=\ast J_3\wedge \ast J_2^{(m)}~, 
}
we can ``save'' the broken symmetry by going to a non-invertible global symmetry. In particular, we can construct a topological (non-invertible) defect operator by modifying the standard $U(1)^{(1)}_e$ 1-form electric symmetry defect operator by coupling to a $2d$ TQFT:
\eq{
\CD^{(1)}_q(\Sigma_2)=e^{ 2\pi i q \int \ast J_2^{(e)}}\times \int [D\phi\,Dc_1] ~{\rm exp}\left\{ \frac{i}{2\pi} \int \left(N \phi dc_1+ p\frac{a}{f_a}dc_1+\phi f\right)\right\}~, 
}
where $q=\frac{p}{N}\in \IQ$, $\phi\sim \phi+2\pi$ is a periodic scalar field and $c_1$ is a dynamical $U(1)$ 1-form gauge field. We can explicitly check that the multiplication of this topological defect is non-invertible:
\eq{
\CD^{(1)}_{1/N}\times \CD^{(1)\dagger}_{1/N}&=\int [D\phi\,Dc_1\,D\phi^\prime\, Dc_1^\prime] ~{\rm exp}\left\{ \frac{i}{2\pi} \int \left(N (\phi dc_1-\phi^\prime dc_1^\prime+ \frac{a}{f_a}(dc_1-dc_1^\prime)+(\phi-\phi^\prime)f\right)\right\}\\
&= \int [D\phi_\pm\,Dc_1^\pm]~{\rm exp}\left\{\frac{i}{2\pi}\int \left(N\phi_+dc_1^+ +N\phi_-dc_1^-+2\frac{a}{f_a}dc_1^-+\phi_- f
\right)\right\}\\
&\neq 1
} 
By comparison to the 0-form non-invertible symmetry, the 1-form non-invertible symmetry is emergent along the RG flow from the KSVZ model. 

The power of the non-invertible symmetries in this theory is that they impose constraints on UV models. In particular, since the non-invertible symmetries are good symmetries of the IR theory, they must either be matched in the UV theory, or they must be emergent. In the case they are emergent, the structure of the non-invertible symmetries requires that the symmetries associated to the currents that arise in the conservation current for the non-invertible symmetry emerge before the ABJ-anomalous current. This implies: 
\begin{itemize}
\item Non-invertible shift symmetry: This requires that the 1-form magnetic symmetry emerge at energy scales higher than the classical shift symmetry
\item Non-invertible electric symmetry: This requires that the 1-form magnetic symmetry and winding symmetry emerge at energy scales that are higher than the scale of emergent classical electric symmetry
\end{itemize}
Note that the non-invertible 1-form symmetry gives a model-independent constraint on the UV physics that is compatible with those derived from the 3-group global symmetry. And indeed, the energy heirarchy that results from the analysis of non-invertible symmetries is more restrictive than that coming from the 3-group global symmetry:\footnote{Note that when the axion interaction is of the form 
\eq{
S_{axion}=\frac{iK}{8\pi^2f_a}\int a\, f\wedge f\quad, \quad K\in \IZ~,
}
for general $K\neq \pm1$, the 3-group symmetry structure is enhanced to include a $\IZ_{|K|}^{(0)}$ shift symmetry. This further constrains ${\rm min}\left\{E_{\rm shift},E_{\rm electric}\right\}\lesssim E_{\rm magnetic}$. This inequality is still consistent (and weaker) with the one obtained from non-invertible symmetries. 
}
\eq{
{\rm 3-group:}&~E_{\rm electric}\lesssim E_{\rm winding}~,\\
{\rm Non-Invertible:}&~E_{\rm electric}\lesssim {\rm min}\left\{E_{\rm winding},E_{\rm magnetic}\right\}\quad, \quad E_{\rm shift}\lesssim E_{\rm magnetic}~.
}

\section*{Acknowledgements}

DB would like to thank the University of Chicago Particle Physics group for the invitation to give a series of group meetings as well as for numerous comments. SH would like to thank the organizers and attendees of the series of two ``Snail Lectures'' supported by the APCTP. 
 We would especially like to thank Liantao Wang, Seth Koren, Christian Ferko, and Anthony Ashmore for detailed comments and insight. We would also like to thank Clay Cordova, Thomas Dumitrescu, Liantao Wang, Ken Intriligator, Kantaro Ohmori, and Seth Koren for fruitful collaborations on related topics and to Clay Cordova, Thomas Dumitrescu, and Gregory Moore for explaining many of the finer points of generalized global symmetries. 

TDB is supported by Simons Foundation award 568420 (Simons
Investigator) and award 888994 (The Simons Collaboration on Global Categorical Symmetries) and SH is supported by  the DOE grants DE-SC-0013642 and DE-AC02-06CH11357.  

\appendix  

\section{Differential Forms and Gauge Theory}
\label{app:differentialforms}

In this section we will give a very brief review of the mathematical background required for the study of generalized global symmetries. For a more complete review see \cite{Nakahara:2003nw}. 

\subsection{Crash Course on Differential Forms}

One mathematical tool we will use again and again is that of differential forms. Differential forms are a mathematical tool to describe geometric properties of manifolds and bundles in a coordinate independent and index-free language. 

Differential forms are labeled by a degree $n$ which is a non-negative integer. We will denote the set of differential $n$-form on a manifold $X$ as $\Omega^n(X)$. The $n$-forms are made out of basic components which are differential volume elements which themselves can be constructed from differential line elements $dx$ for some set of coordinates $\{x\}$. The differential 1-forms are of the form 
\eq{
\omega_1=\sum_i f_i(x) dx_i\in \Omega^1(X)~.
} 
These are the ``dual'' of vector fields. By this, we mean that there is a natural inner product 
\eq{
\langle \vec{v},\omega\rangle=\sum_i v^i(x) f_i(x)\quad, \quad \vec{v}=\sum_i v^i(x)\frac{\partial}{\partial x^i}~. 
}
And indeed, there is a map between vector fields and 1-forms (and more generally anti-symmetric $n$-tensors and $n$-forms) by using the metric
\eq{\label{vecduality}
\vec{v}\longmapsto v_1=\sum_{ij}g_{ij}v^j(x) dx_i~. 
}

Differential $n$-forms can be constructed out of differential $1$-forms by the \emph{wedge product}.
The wedge product acts on differential line elements as the anti-symmetric (tensor) product 
\eq{
dx_i\wedge dx_j=-dx_j\wedge dx_i~.
}
Analogously, the wedge product of $n$ differential line elements is also their anti-symmetric (tensor) product. The reason for this is that the $n$-dimensional differential polytope (i.e. fundamental $n$-volume) has a volume which is given by the anti-symmetrization of the differential line elements. For example, the 3-dimensional differential parallelogram has a volume element which is given by the triple product (i.e. anti-symmetric product) 
\eq{
d{\rm Vol}=d\vec{x}_1\cdot(d\vec{x}_2\times d\vec{x}_3)=dx_1\wedge dx_2\wedge dx_3~.
}
 This is because the anti-symmetrization projects out any  redundancies that arises from coordinate systems that are not orthogonal. For example, the change of coordinates $(x,y,z)\mapsto (x,y+f(x),z+g(x))$ leads to the differential line elements 
\eq{
&(dx,dy,dz)\longmapsto (dx\,,\,dy+f'(x)dx\,,\,dz+g'(x)dx)~.
}
However, due to the anti-symmetry of the wedge product, the differential volume 3-form is invariant under the coordinate transformation
\eq{
dx\wedge dy\wedge dz\longmapsto dx\wedge dy\wedge dz~.
}

Because of this, we can describe the generic $n$-form from the wedge product of $n$ differential line elements:
\eq{\label{nform}
\omega_n=\sum_{i_1,...,i_n}f_{i_1...i_n}(x)~dx_{i_1}\wedge ...\wedge dx_{i_n}~.
}
Notice that the structure of differential forms allows us to take the arbitrary wedge product of a $p$-form $\alpha_p$ with a $q$-form $\beta_q$:
\eq{
\alpha_p=\sum_{i_1,...,i_p}\alpha_{i_1...i_p}(x)~dx_{i_1}\wedge...\wedge dx_{i_p}\quad, \quad \beta_p=\sum_{j_1,...,j_q}\beta_{j_1...j_q}(x)~dx_{j_1}\wedge...\wedge dx_{j_q}~,\\
\alpha_p\wedge \beta_p=\sum_{i_1,...,i_p}\sum_{j_1,...,j_q}\alpha_{i_1...i_p}(x)\beta_{j_1...j_q}(x)dx_{i_1}\wedge...\wedge dx_{i_p}\wedge dx_{j_1}\wedge...\wedge dx_{j_q}~.
}
One can show from the anti-symmetry of wedge-product on differential line elements that 
\eq{
\alpha_p\wedge \beta_q=(-1)^{pq}\beta_q\wedge \alpha_p~. 
}

As we have mentioned, differential forms  are naturally integrable quantities. By construction, we can integrate an $n$-form along  a closed $n$-manifold $\Sigma_n$. Given $\omega_n$ as in \eqref{nform}, the integral over $\Sigma_n$ is given by 
\eq{ 
\oint_{\Sigma_n} \omega_n= \sum_{i_1,...,i_n}\oint_{\Sigma_n} f_{i_1...i_n}(x)~dx_{i_1} ... dx_{i_n}~,
}
where here the restriction (i.e. pullback) of $dx_{i_1}...dx_{i_n}$ to $\Sigma_n$ vanishes except when $dx_{i_1}...dx_{i_n}$ is proportional to the volume element on $\Sigma_n$. 
Note additionally, that we can integrate an $n$-form along a $p$-dimensional manifold when $n>p$ to get an $n-p$ form. This can be done straight-forwardly by decomposing the differential form $\omega_n$ as 
\eq{
\omega_n=\left(\sum_{i_1,...,i_{n-p}}f_{i_1,...,i_{n-p}}(x)dx_{i_1}\wedge...dx_{i_{n-p}}\right)\wedge d{\rm Vol}_p(\Sigma_p)~.
}
Then the integral is given by 
\eq{
 \oint_{\Sigma_p}\omega_n=\sum_{i_1,...,i_{n-p}}\left(\oint f_{i_1,...,i_{n-p}}(x)d{\rm Vol}_p(\Sigma_p)\right)dx_{i_1}\wedge...dx_{i_{n-p}}
}

\bigskip
Another fundamental concept in differential forms is the \emph{exterior derivative}. It is the natural generalization of the partial derivative to the exterior algebra of differential forms -- in fact, we have already implicitly used the exterior derivative. Consider a general differential $n$-form $\omega_n$:
\eq{
\omega_n=\sum_{i_1,...,i_n}f_{i_1...i_n}(x)~dx_{i_1}\wedge ...\wedge dx_{i_n}~. 
}
We can compute the exterior derivative by:
\eq{
d\omega_n=\sum_{i,i_1,...,i_n}\frac{\partial f_{i_1...i_n}(x)}{\partial x_i}~dx_i\wedge dx_{i_1}\wedge ...\wedge dx_{i_n}~. 
}
Importantly, due to the anti-symmetry of the wedge product, the exterior derivative squares to zero:
\eq{
d^2\omega_n=\sum_{i,j,i_1,...,i_n}\frac{\partial^2 f_{i_1...i_n}(x)}{\partial x_i\partial x_j}~dx_i\wedge dx_j\wedge dx_{i_1}\wedge ...\wedge dx_{i_n}~,
}
since partial derivatives commute. Note that these properties hold for general curved manifolds since all of the machinery of differential forms only uses partial derivatives -- the ``magic'' comes from the exterior algebra.  

One can show from the definition that the exterior derivative obeys the product rule:
\eq{\label{extderivproduct}
d(\alpha_p\wedge \beta_q)=d\alpha_p\wedge \beta_q+(-1)^p \alpha_p\wedge d\beta_q~.
}

\bigskip The exterior derivative is especially useful for differential forms as it leads to a simple statement of the \emph{Generalized Stokes' Theorem}. \\

\noindent \textbf{Generalized Stokes' Theorem:} Let $\Sigma$ be an oriented $(n+1)$-manifold with boundary $\partial \Sigma$ and let $\omega_n$ be a differential $n$ form. Then the integral of $d\omega_n$ is given in terms of the integral of $\omega_n$ on $\partial \Sigma$:
\eq{
\int_\Sigma d\omega_n=\oint_{\partial \Sigma}\omega~. 
}
~\\

\noindent Here will not prove this theorem. The interested reader can read the details in \cite{Nakahara:2003nw}.\\

Together with the Stokes' Theorem, the product law for exterior derivative \eqref{extderivproduct} allows us to integrate differential forms by parts:
\eq{
\int_M d\alpha_p\wedge \beta_q=\int_{\partial M}\alpha_p\wedge \beta_q+(-1)^p\int_M\alpha_p\wedge d\beta_q~. 
}

\bigskip 

Another important manipulation involving differential forms is the \emph{Hodge star} operation. In $d$-dimensions on a Riemannian manifold $M$ (i.e. a manifold with metric) this is a map: 
\eq{
\ast : \Omega^n(M)\longmapsto \Omega^{d-n}(M)~. 
}
This depends on a choice of orientation (i.e. a choice of ordering of the coordinates $\{x^1,...,x^d\}$ and the metric and acts on the infinitesimal $n$-form  as 
\eq{
\ast\Big(dx^{i_1}\wedge... \wedge dx^{i_n}\Big)=\frac{\sqrt{{\rm det}\,g}}{(d-n)!}g^{i_1j_1}...g^{i_nj_n}\epsilon_{j_1...j_n} dx^{j_{n+1}}\wedge dx^{j_d}~,
}
where $\epsilon_{1,...,n}$ is the fully antisymmetric Levi-Civita tensor. 
It is important that acting on an $n$-form, the Hodge star squares to $\pm$ the trivial operation:
\eq{
\ast \ast \omega_n=(-1)^{n(d-n)}s\,\omega_n~,
}
where $s=\pm 1$ is the signature of the space time (i.e. $s=1$ Euclidean and $s=-1$ for Lorentzian). The Hodge star also allows us to compute divergences of tensors using their associated differential forms. For example, if we take $v_1$ to be the 1-form associated to the vector field $\vec{v}$ as in \eqref{vecduality}, then the divergence of $\vec{v}$ is a function (i.e. 0-form) which can be computed in terms of $v_1$ as:
\eq{
\vec\nabla\cdot \vec{v}=\ast d\ast v_1~. 
}
We leave this as an exercise for the interested reader.

\bigskip

Often, we are interested in understanding which differential forms $\omega_n$ can be written as an exterior derivative $\omega_n=d\alpha_{n-1}$. Due to the fact that $d^2=0$, one may be tempted to try to solve this by classifying $d\omega_n=0$. However, on manifolds with non-trivial topology, there exist solutions to the equation $d\omega_n=0$ where no $\alpha_{n-1}$ such that $\omega_n=d\alpha_{n-1}$. These differential forms are classified by \emph{cohomology} classes called \emph{de Rahm Cohomology}. 

The space of $n$-forms $\omega_n$ on a manifold $M$ that are \emph{closed} -- i.e. that obey $d\omega_n=0$ -- that are not \emph{exact} -- i.e. there is no $(n-1)$-form $\alpha_{n-1}$ s.t. $\omega_n=d\alpha_{n-1}$ -- are classified by the cohomology class $H^n_{dR}(M)$. This cohomology class is often written in terms of the set $Z^n(M)$ of closed $n$ co-chains (closed differential $n$-forms) modulo the set $C^{n-1}(M)$ of exact $(n-1)$ co-chains (i.e. exact $n$-forms):
\eq{
H^n_{dR}(M)=Z^n(M)/C^{n-1}(M)~.
}
What is useful about these cohomology classes is that they are equivalent to other constructions of ``cohomology classes'' that are easier to compute:
\eq{
H^n_{dR}(M)=H^n(M;\IR)~.
}
We will not construct these cohomology classes here -- the interested reader can see \cite{Nakahara:2003nw}. 

Something we will use for our discussion here is that we can restrict to the set of differential $n$-forms with \emph{integral periods} $\Omega^n_\IZ(M)$: i.e. they integrate to an integer on \emph{all} $n$-dimensional sub-manifolds of $M$. We can then construct an integral-valued cohomology class 
\eq{
H^n(M;\IZ)=Z^n_\IZ(M)/C^{n-1}_\IZ(M)~,
}
where the definition follows analogously from before where we have now restricted to only differential forms in $\Omega_\IZ^n(M)$.

\subsection{Gauge Theory}
Gauge theory is the theory of principal $G$ bundles with connection where $G$ is some Lie group\footnote{Note that discrete groups such as $\IZ_N$ are also technically Lie groups.}. Principal $G$ bundles are fibrations of $G$ over some base manifold 
\eq{
\xymatrix{
G\ar[r] & P\ar[d]^\pi\\
& X
}
}
where $\pi$ is the projection map onto the base manifold. 

Here we will describe a construction of $G$-bundles $P$ with connection. 
Let us fix $G$ to be a semi-simple Lie group. A $G$-bundle $P$ over $X$ can be locally as follows. Take a ``nice'' covering of $X$ by simply connected open sets $\{\CU_\alpha\}$ so that $\bigcup_\alpha \CU_\alpha=X$. The bundle $P$ locally can be trivialized over each $\CU_\alpha$ so that the fiber of $P$ over $\CU_\alpha$ looks like a product: $\pi^{-1}(\CU_\alpha)=P_\alpha\cong G\times \CU_\alpha$. The fibers of $P$ are related between different patches $\CU_\alpha, \CU_\beta$ by the gauge transformation map:
\eq{\label{overlap}
g_{\alpha\beta}:\CU_{\alpha\beta}\to G\quad, \quad (x,p_\alpha )=(x,g_{\alpha\beta}(x)\cdot p_\beta )\quad, \quad x\in \CU_{\alpha\beta}~,
}
where $\CU_{\alpha\beta...}=\CU_\alpha\cap\CU_\beta\cap...$ All fiber bundles obey the ``triple intersection'' condition:
\eq{\label{tripleintersect}
g_{\alpha\beta}(x)\cdot g_{\beta\gamma}(x)\cdot g_{\gamma \delta}(x)=\mathds{1}\qquad x\in \CU_{\alpha\beta\gamma}~.
}
Note that this will differ for ``twisted'' fiber bundles. 

A connection on $P$ is formally a rule for comparing the fibers $P_x=\pi^{-1}(x)$ with $P_y$ for $x\neq y$. The point is that parallel transport along around the base manifold can be identified with a tangent vector field to $X$ and for non-trivial $P$, motion along such a vector field can induce a translation in the fiber. This translation is itself described by a vector that is tangent to the fiber $G$  that is a function of the tangent vector field $X$. Thus, the translation in $G$ is (locally) described by a $\fg$-valued 1-form which is called the connection $A$. 

The intersection condition \eqref{overlap}  implies that the connection $A_\alpha$ defined on $\CU_\alpha$ can be related to the connection on $\CU_\beta$ by the gauge transformation:
\eq{
A_\beta=g_{\alpha\beta}^{-1}A_\alpha g_{\alpha\beta}+g_{\alpha\beta}^{-1}dg_{\alpha\beta}~.
} 
From this connection we can define the covariant derivative which is given by 
\eq{
D_A=d+A\wedge ~.
}
This connection can have a non-trivial curvature (in physics called the field strength) which is given by the commutator of covariant derivatives (or alternatively as the covariant derivative of the connection $A$):
\eq{
F=dA+A\wedge A~.
}

\subsection{Characteristic Classes}

It is a result from mathematics that for a fixed $G$, any principal $G$ bundle over a fixed base manifold $X$ can be constructed by pulling back a $G$-bundle from a universal space called the classifying space: $BG$.\footnote{
The classifying space $BG$ can be constructed as follows. Pick a contractible space $EG$ that has a free $G$ action. $BG$ can then be constructed as the quotient $EG/G$. Now any principal $G$ bundle  $P\to X$ can be constructed by pulling back the bundle $EG\to BG$: 
\eq{
\xymatrix{
G\ar[r]& P=f^\ast(EG)\ar[d]& EG\ar[d]& G\ar[l]\\
&X\ar[r]^f&BG&
}
}}
As a consequence of this construction, all characteristic classes of the principal $G$ bundle $P$ are given by the pullback of cohomology classes of $BG$. So in particular, the cohomology of $BG$ classifies the possible topology of principal $G$ bundles. 

The characteristic classes classify the possible topological invariants of an associated gauge bundle. As such, they must be measured by gauge invariant differential forms that are written in terms of the field strength and gauge field. When $G$ is a simply connected Lie group, the cohomology classes are given by the gauge-invariant polynomials in $F$ and their powers (called Chern-classes):\footnote{The Chern-characters are given by the simple polynomials $\frac{\Tr[F\wedge...\wedge F]}{n!(2\pi)^n}$. Chern-classes are the reduction of Chern characters by products of lower-dimensional Chern-classes -- they agree for $ch_1(F)=c_1(F)=\frac{\Tr[F]}{2\pi}$. For example, $c_2 (F) = ch_2(F) - \frac{1}{2} \left[ c_1(F) \right]^2$.}
\eq{
\frac{\Tr[F\wedge...\wedge F]}{n!(2\pi)^n}\in H^{2n}(BG;\IZ)~. 
}
Other important cases include the case of $U(1)$ which is completely generated by polynomials of $F$: 
\eq{
H^{2n}(BU(1);\IZ)=\IZ\left[\frac{F\wedge...\wedge F}{n!(2\pi)^n}\right]~,
}
and $\IZ_N$:\footnote{Here the cohomology classes are $\IZ_N$ valued because the ``field strength'' is the $\IZ_N$ Bockstein of the gauge field $\beta_N(A)$ which is itself $\IZ_N$-valued. The cohomology classes are then generated by polynomials of $\beta_N(A)$. }
\eq{
H^{2n}(B\IZ_N;\IZ)=\IZ_N~. 
}
For our purposes, it will be important to consider the case of non-simply connected Lie groups such as $SO(N)$ or $PSU(N)$. In these cases, we have the relation  
\eq{
H_1(G;\IZ)\cong H^2(BG;\IZ)~,
}
due to the Universal Coefficient Theorem \cite{Hatcher}. This means that $G$-gauge theory has discrete 2-form fluxes that correspond to the first homotopy group of $G$ which encodes the 1-form magnetic flux sectors.

\subsection{Twisted Bundles and Center Symmetry} 
\label{app:twisted_bundle}

Center symmetry of a $G$-bundle with (abelian) center $Z=Z(G)$ acts by shifting the gauge field by a $Z$-gauge field. The mathematical description of this process is described by the ``twisting'' of the $G$ bundle which is classified by an element $w_2\in H^2(BG;Z)$.

Let us pick $G$ to be a Lie group with non-trivial center $Z=\IZ_N$. As we discussed in the previous section, the standard triple intersection constraint for $G$-bundles is given by:
\eq{
g_{\alpha\beta}\circ g_{\beta\gamma}\circ g_{\gamma \alpha}=\mathds{1}_G\quad, \quad g_{\alpha\beta}:U_\alpha\cap U_\beta\to G~.
}
In the case where we are considering principal $G$-bundles that couple to matter (i.e. associated vector bundles) that is uncharged under $\IZN\subset G$, the triple overlap condition can be modified (i.e. twisted) to 
\eq{
g_{\alpha\beta}\circ g_{\beta\gamma}\circ g_{\gamma \alpha}=\omega\quad, \quad \omega\in \IZN~.
}
The standard problem with twisting a bundle is that it will lead to field configurations with branch cuts. However, since the matter fields are uncharged under $\IZN$, they do not see the twist, there are no induced branch cuts, and the path integral is still well defined. The effect of performing this twist is that the ``standard'' curvature is now shifted by the discrete flux $w_2\in H^2(BG;\IZN)$:
\eq{
F_G\longmapsto F_G+\frac{1}{N}w_2\mathds{1}_G~.
}
This procedure of passing to twisted $G$-bundles is similar to lifting the $G$-gauge theory (path integral over $G$-bundles) to a $G^c$-gauge theory (path integral over $G^c$-bundles). $G^c$-bundles can be defined analogously to $Spin^c$-bundles as follows (see \cite{GMspinc} for a  review of $Spin^c$ structures).
 A $G^c$-bundle is a principal $\frac{G\times U(1)}{\IZ_N}$-bundle where the triple overlap condition is now  modified to 
\eq{
g_{\alpha\beta}\circ g_{\beta\gamma}\circ g_{\gamma \alpha}=(z,z^{-1})\sim (1,1)\quad, \quad g_{\alpha\beta}:U_\alpha\cap U_\beta\to \frac{G\times U(1)}{\IZ_N}~,
}
where $z\in \IZ_N \subset U(1)$. Here we see that the $G^c$-bundle is composed of a twisted $G$-bundle and a twisted $U(1)$-bundle which, combined together, form an untwisted total bundle (due to the $\IZ_N$-quotient):
\eq{
\xymatrix{
\frac{G\times U(1)}{\IZ_N}\ar[r]&G^c\ar[d]\\
&X
}
}
We can also introduce a connection on $G^c$. The effective result of performing this lift is that the curvature is restricted 
\eq{
F_{G^c}={\omega}_{U(1)}+F_G,	
}
where ${\omega}_{U(1)}$ is the curvature of the twisted $U(1)$  and $F_G$ is the curvature of the twisted $G$-bundle. Since $G^c$ is a well defined bundle, $F_{G^c}$ is an integer class. This means that the twisted-$U(1)$ and $G$ bundles will have opposite fractional fluxes.  

For applications to center symmetry of gauge theories, we can replace the $U(1)$ factor with $\IZ_N$. The reason is that  with a factor of $\IZ_N$, we are studying a $\frac{G\times \IZ_N}{\IZ_N}\cong G$-bundle which does not introduce any new degrees of freedom, but merely turns on a fractional flux along the center of $G$. 
 This can be achieved at the level of the path integral or at the level of the action by projecting $U(1)\to \IZN$ by including some Lagrange multiplier.

Let us review how this works in an example. Consider the case of $G=SU(N)$ and let us turn on a $\IZ_N^{(1)}$ flux -- i.e. twist the $SU(N)$ bundle by a $\IZ_N$-valued $b_2$-class. To do this, we will go to a $G^c$ bundle where
\eq{
G^c=\frac{SU(N)\times U(1)}{\IZ_N}\cong U(N)~.
}
Because of this isomorphism, we can rewrite
\eq{
F_{SU(N)}=F_{U(N)}-F_{U(1)}~,
}
where again $F_{U(1)}$ is the field strength where we have twisted by $b_2$:
\eq{
\oint\frac{F_{U(1)}}{2\pi}=\oint \frac{b_2}{2\pi}~{\rm mod}_\IZ\quad, \quad \oint \frac{b_2}{2\pi}\in \frac{1}{N}\IZ~. 
}

For a $SU(N)$ gauge theory, we can make the substitution above in the action if we also impose the constraint 
\eq{\label{UNConstraint}
S=...+\frac{i}{2\pi}\int \varphi_2\wedge \Tr[F_{U(N)}-B_2\mathds{1}_N]\quad\Longrightarrow \quad \Tr[F_{U(N)}]=NB_2~.
}
where here we have fixed $F_{U(1)}$ to be a background gauge field given by a representative $B_2$ of $b_2$. This constraint projects $U(1)\to \IZN$ and therefore $U(N)\to SU(N)$ which is twisted by $b_2$. 

Using this, we can show that the instanton number  becomes fractional when we turn on a center symmetry flux. Plugging in the shifted field strength, we can compute
\eq{
&\int\frac{\Tr[F_{SU(N)}\wedge F_{SU(N)}]}{8\pi^2}=\frac{1}{8\pi^2}\int\Tr[(F_{U(N)}-B_2\mathds{1}_N)\wedge (F_{U(N)}-B_2\mathds{1}_N)]\\
&
=\int\frac{\Tr[F\wedge F]-\Tr[F]\wedge \Tr[F]}{8\pi^2}+N\int\frac{B_2\wedge B_2}{8\pi^2}+\int\frac{\Tr[F]\wedge \Tr[F]}{8\pi^2}-\int\frac{B_2\wedge \Tr[F]}{4\pi^2}~.
}
Here, this first term is the second Chern class of the $U(N)$ connection which is integer quantized. We then want to impose the constraint \eqref{UNConstraint}:
\eq{
\int\frac{\Tr[F_{SU(N)}\wedge F_{SU(N)}]}{8\pi^2}&=N(N-1)\int\frac{B_2\wedge B_2}{8\pi^2}~{\rm mod}_\IZ=\frac{N-1}{2N}\int w_2\wedge w_2 ~{\rm mod}_\IZ~. 
}
Here we have rescaled $B_2=\frac{2\pi}{N}w_2$ to make the fractional part explicit: 
\eq{
\oint w_2=\IZ\quad \Longrightarrow \quad \oint w_2\wedge w_2\in \IZ~.
}

 \section{Classifying Center/Magnetic  Symmetry}
 \label{app:CenterMag}

In this appendix we more carefully discuss the relation between the center and magnetic 1-form symmetries in non-abelian gauge groups and the center, fundamental group of the gauge group. 

 Consider a non-abelian gauge group $G$ that has a center $Z$. Let us also define $G_{\rm ad}=G/Z$ -- this is sometimes called the adjoint form of the group   -- and $\hatG$ which is the simply connected cover $G=\hatG/Z$. As we have discussed, the   center symmetry of $G$ Yang-Mills theory is described by $Z$ and its 1-form magnetic symmetry is given by $\pi_1(G)$. The physical picture is the following. 
 
All finite dimensional representations of a compact Lie group $G$ are ``highest weight'' representations. This means that every representation $R$ corresponds to an element of (i.e. a vector in) the weight lattice $\Lambda_{\rm wt}(G)$.\footnote{Technically, the weight lattice is defined for a Lie algebra which captures all representations of $\tildeG$. Here, what we are calling the ``$G$-weight lattice''  
$\Lambda_{\rm wt}(G)=\Lambda_{\rm char}(G)$ is the character lattice which classifies representations of $G$: 
\eq{
\Lambda_{\rm char}(G)={\rm Hom}(T,U(1))=\left\{\mu \in \ft^\ast ~|~\langle\mu,H\rangle\in \IZ~,~\forall H\in \ft~{\rm s.t.}~e^{2\pi i H}=\mathds{1}_G\right\}~,
}
where $T$ (with Lie algebra $\ft$) is a Cartan torus of $G$ (with Lie algebra $\fg$).  
We hope that this introduction of slightly incorrect terminology is helpful and not confusing. 
} The weight lattice sits inside of the weight lattice of the simply connected group $\Lambda_{\rm wt}(G)\subseteq \Lambda_{\rm wt}(\tildeG)$ and it further contains the weight lattice of the adjoint form of the group: $\Lambda_{\rm wt}(G_{\rm ad})\subseteq \Lambda_{\rm wt}(G)$. This is the statement that every representation of $G$ is also a representation of $\tildeG$ and that every representation of $G_{\rm ad}$ is a representation of $G$. 
The lattice classifying the representations of $G$, $\Lambda_{\rm char}(G)$, fits into a hierarchy \cite{Moore:2014jfa}: 
\eq{
\Lambda_{\rm rt}(\fg)=\Lambda_{\rm wt}(G_{\rm ad})~~\subseteq~~ \Lambda_{\rm wt}(G)~~\subseteq~~  \Lambda_{\rm wt}(\widetilde{G})~,
}
where $\Lambda_{\rm rt}$ is the root lattice.

By definition the adjoint representation of $G$ is always uncharged under $Z\subset G$, which means that it is a representation of $G_{\rm ad}$. As it turns out, the minimal/defining representation of $\Lambda_{\rm wt}(G_{\rm ad})$ is the adjoint representation. This means that 
the set of representations that can be screened by gluons are those classified by $\Lambda_{\rm wt}(G_{\rm ad})$.  
 Thus the set of unscreenable Wilson lines are classified by $\Lambda_{\rm wt}(G)/\Lambda_{\rm wt}(G_{\rm ad})=\Lambda_{\rm wt}(G)/\Lambda_{\rm rt}(\fg)$. As it turns out, the hierarchy of lattices is related to their center and fundamental group:
\eq{
\Lambda_{\rm wt}(G)/\Lambda_{\rm wt}(G_{\rm ad})=Z(G)\quad&, \quad \Lambda_{\rm wt}(\tildeG)/\Lambda_{\rm wt}(G)=\pi_1(G)~.}
Therefore, we see that for the general gauge theory that $Z(G)$ classifies the charge of unscreened Wilson lines. The fact that these are related by the center/fundamental group means that passing from $G$ to $G/Z$ is equivalent to taking the quotient of the lattice $\Lambda_{\rm wt}(G)$ by $Z$. 

There is a similar story for monopoles. The lattices classifying the allowed (not necessarily stable) monopole charges also fit into a hierarchy of lattices:
\eq{
\Lambda_{\rm cr}(\fg)=\Lambda_{\rm co-char}(\widetilde{G})~~\subseteq~~ \Lambda_{\rm co-char}(G)~~\subseteq~~  \Lambda_{\rm co-char}(G_{\rm ad})=\Lambda_{\rm mw}(\fg)~.
}
where $\Lambda_{\rm cr}(\fg)=\Lambda_{\rm rt}(\fg)^\ast$ is the co-root lattice. Additionally, $\Lambda_{\rm mw},\Lambda_{\rm co-char}(G)$ are  the  ``magnetic weight'' and co-character lattices respectively. They are defined as  \cite{Moore:2014jfa}:
\eq{
&\Lambda_{\rm mw}(\fg)=\left\{H\in \ft ~|~\big\langle \alpha, H\big\rangle\in \IZ~,~\forall \alpha \in \Lambda_{\rm rt}\right\}~,\\
&\Lambda_{\rm co-char}(G)=\left\{H\in \ft~|~{\rm exp}\{2\pi i H\}=\mathds{1}_G\right\}~. \label{eq:def_cochar}
}

We know from our previous discussion that stable monopoles are classified by $\pi_1(G)$ due to the existence of topologically non-trivial winding field configurations for a monopole field configuration. It is then clear that the monopole charges in $\Lambda_{\rm cr}(\fg)$ all correspond to unstable monopoles since $\tG$ is simply connected. As it turns out, these are all of the unstable monopoles. On the other hand, all possible magnetic charges in $G$ gauge theory is captured by $\Lambda_{\rm co-char}$ as can be seen from its definition in \eqref{eq:def_cochar}. Recall the discussion of monopoles given in Section~\ref{subsubsec:SU2_vs_SO3} which shows that a well-defined monopole charge (stable or not) requires the existence of a gauge transformation $g=e^{iH\phi}$ to patch the $A_N$ and $A_S$ and such a map must satisfy $g (\phi + 2\pi) = g (\phi)$. The set of such map is equivalent to choosing an element in $\Lambda_{\rm co-char}$. 
Therefore, the set of stable monopoles of $G$ is classified by $\Lambda_{\rm co-char}(G)/\Lambda_{\rm cr}(\fg)$. 
Again, the hierarchy of lattices turns out to be related to the center/fundamental group of $G$ as:
\eq{ 
\Lambda_{\rm co-char}(G)/\Lambda_{\rm cr}(\fg)=\pi_1(G)\quad, \quad\Lambda_{\rm mw}(\fg)/\Lambda_{\rm co-char}(G)=Z(G) ~. 
}

If we now compare the two hierarchies of lattices, we see that there is a parallel structure between the center and magnetic 1-form  symmetries:
\begin{center}\begin{tabular}{c|c|c}
Center Symmetry & Gauge Group & Magnetic Symmetry\\
\hline 
``$\pi_1(G)\times Z$'' & $\tG$ & 1\\
$Z$ & $G=\tG/\pi_1(G)$ & $\pi_1(G)$\\
1& $G_{0}=G/Z$ & ``$\pi_1(G)\times Z$''
\end{tabular}
\end{center}
Here we use `` '' to indicate that this is only true as sets. In general, $Z(G)=Z$ and $\pi_1(G)$ can descend from a simply connected group $\tG$ where $Z(\tG)$ is an extension of $Z$ by $\pi_1(G)$. Similarly, quotienting $G$ by $Z$ can create a group $G_0$ whose fundamental group is an extension  of $Z(G)$ by $\pi_1 (G)$. 

From this table, we see that quotienting the gauge group by subgroups of the center ``removes'' center symmetry and ``produces'' magnetic 1-form  symmetry. We will now explain this mechanism.

\bibliographystyle{utphys}
\bibliography{GGSbib}

\end{document}